\documentclass[prd,aps,superscriptaddress,twocolumn,floatfix,nofootinbib]{revtex4}
\usepackage{grffile}
\usepackage{graphicx}
\usepackage{amssymb}
\usepackage{float}
\usepackage{amsmath}
\usepackage{hyperref}
\usepackage[nameinlink]{cleveref}
\usepackage[dvipsnames]{xcolor}
\newcommand{\ket}[1]{\ensuremath{\left|#1\right\rangle}}

\newcommand{\braket}[1]{\ensuremath{\left\langle #1 \right\rangle}}
\newcommand{\bra}[1]{\ensuremath{\left\langle#1\right|}}
\def\simge{
    \mathrel{\rlap{\raise 0.511ex
        \hbox{$>$}}{\lower 0.511ex \hbox{$\sim$}}}}
\def\simle{
    \mathrel{\rlap{\raise 0.511ex 
        \hbox{$<$}}{\lower 0.511ex \hbox{$\sim$}}}}
\let\ifcomments\iftrue
\def\commentsoff{\global\let\ifcomments\iffalse}

\let\commentsize\small
\def\tinycomments{\global\let\commentsize\footnotesize}
\newcommand{\Tr}{{\rm{Tr}}}

\allowdisplaybreaks[1]

\begin{document}

\begin{abstract}
{A study of two-pion scattering for the isospin channels, $I=0$ and $I=2$, using lattice QCD is presented. M\"obius domain wall fermions on top of the Iwasaki-DSDR gauge action for gluons with periodic boundary conditions are used for the lattice computations which are carried out on two ensembles of gauge field configurations generated by the RBC and UKQCD collaborations with physical masses, inverse lattice spacings of 1.023 and 1.378~GeV}, and spatial extents of $L=4.63$ and 4.58 fm, respectively. 
The all-to-all propagator method is employed to compute a matrix of correlation functions of two-pion operators.
The generalized eigenvalue problem (GEVP) is solved for a matrix of correlation functions to extract phase shifts with multiple states, two pions with a non-zero relative momentum as well as two pions at rest.
Our results for phase shifts for both $I=0$ and $I=2$ channels are consistent with and the Roy Equation and chiral perturbation theory, though at this preliminary stage our errors for $I=0$ are large. An important outcome of this work is that we are successful in extracting two-pion excited states, which are useful for studying $K\to\pi\pi$ decay, on physical-mass ensembles using GEVP.
\end{abstract}

\title{Isospin 0 and 2 two-pion scattering at physical pion mass using all-to-all propagators with periodic boundary conditions in lattice QCD}

\newcommand{\Columbia}{
Physics Department,
Columbia University,
New York, NY 10027, USA
}
\newcommand{\UConn}{
Physics Department,
University of Connecticut,
Storrs, CT 06269, USA
}
\newcommand{\BNL}{
Physics Department,
Brookhaven National Laboratory,
Upton, NY 11973, USA
}
\newcommand{\RBRC}{
RIKEN-BNL Research Center,
Brookhaven National Laboratory,
Upton, NY 11973, USA
}
\newcommand{\CSIBNL}{
Computational Science Initiative,
Brookhaven National Laboratory,
Upton, NY 11973, USA
}
\newcommand{\Bern}{
Albert Einstein Center,
Institute for Theoretical Physics,
University of Bern,
CH-3012 Bern, Switzerland
}
\newcommand{\Regensburg}{
Fakult\"at f\"ur Physik, Universit\"at Regensburg,
Universit\"atsstra{\ss}e 31,
93040 Regensburg, Germany
}
\newcommand{\UCB}{
Department of Physics,
University of California,
Berkeley, CA 94720, USA
}
\newcommand{\LBNL}{
Nuclear Science Division,
Lawrence Berkeley National Laboratory,
Berkeley, CA 94720, USA
}
\newcommand{\Milano}{
Dipartimento di Fisica,
Universit\'a di Milano-Bicocca,
Piazza della Scienza 3,
I-20126 Milano, Italy
}
\newcommand{\INFN}{
INFN, Sezione di Milano-Bicocca,
Piazza della Scienza 3,
I-20126 Milano, Italy
}

\author{Thomas~Blum}
\affiliation{\UConn}
\affiliation{\RBRC}

\author{Peter~A.~Boyle}
\affiliation{\BNL}

\author{Mattia~Bruno}
\affiliation{\Milano}
\affiliation{\INFN}

\author{Daniel~Hoying}
\affiliation{\Bern}

\author{Taku~Izubuchi}
\affiliation{\RBRC}
\affiliation{\BNL}

\author{Luchang~Jin}
\affiliation{\UConn}
\affiliation{\RBRC}

\author{Chulwoo~Jung}
\affiliation{\BNL}

\author{Christopher~Kelly}
\affiliation{\CSIBNL}

\author{Christoph~Lehner}
\affiliation{\Regensburg}

\author{Aaron~S.~Meyer}
\affiliation{\UCB}
\affiliation{\LBNL}

\author{Amarjit~Soni}
\affiliation{\BNL}

\author{Masaaki~Tomii}
\affiliation{\UConn}

\collaboration{RBC and UKQCD Collaborations}

%

\maketitle

\section{Introduction}
\label{sec:intro}

Understanding the interactions of two pions is an interesting endeavor for practitioners of non-perturbative QCD. Not only do we learn how the fundamental interactions of quarks and gluons give rise to the observable properties of hadrons, these two particle systems play an important role in Standard Model processes under intense investigation, such as $K\to\pi\pi$ decays~\cite{Blum:2015ywa,Bai:2015nea,RBC:2020kdj} and the muon's anomalous magnetic moment $g-2$~\cite{RBC:2018dos,Bruno:2019nzm}. Our focus in this study is primarily on isospin $I=0$ and 2 for the former while $I=1$ is important for the latter. Isospin symmetry and Bose-Einstein statistics constrain the states that appear in these processes.

With L\"uscher's technique \cite{Luscher:1990ux} that relates two-pion energy in a finite box with the corresponding scattering phase shift, there have been many studies of two-pion scattering in lattice QCD at unphysical pion masses \cite{Kuramashi:1993ka,CP-PACS:2004dtj,CP-PACS:2005gzm,Beane:2005rj,Beane:2007xs,Feng:2009ij,NPLQCD:2011htk,Dudek:2012gj,Fu:2013ffa,Bulava:2016mks,Briceno:2016mjc,Liu:2016cba,Fu:2017apw,Culver:2019qtx}.  For these studies, a chiral extrapolation was needed to obtain physical results.  
The analytic evaluation of two-pion scattering in chiral perturbation theory \cite{Bijnens:1995yn,Bijnens:1997vq} (ChPT) was employed in these works to perform the extrapolation of important parameters of two-pion scattering such as the scattering length.
While we can expect the extrapolation is reasonable for the scattering length, which can be extracted near two-pion threshold, the extrapolation of the scattering amplitudes or phase shifts might not be accurate at high energies.

Now it is possible to perform a lattice calculation at the physical pion mass so that we can directly compute the two-pion phase shifts at relatively large energies without a chiral extrapolation.
There was a study where the $I=2$ scattering length was computed including the physical pion mass for the first time~\cite{Fischer:2020jzp}.  The $I=0$ channel is challenging already at unphysical pion masses~\cite{Kuramashi:1993ka,Fu:2013ffa,Briceno:2016mjc,Liu:2016cba,Fu:2017apw} because of the presence of disconnected diagrams and two-pion operators coupling with the vacuum state. 
This paper is part of a series of studies of two-pion scattering undertaken
by the RBC and UKQCD collaborations~\cite{RBC:2021acc,Bruno:2023pde}, where the challenging $I=0$ channel is examined at physical pion mass. 
Here we present results for phase shifts at various energy levels for $I=2$ and $I=0$ at physical pion mass using $2+1$ flavors of M\"obius domain wall fermions (MDWF) with periodic boundary conditions.

The RBC and UKQCD collaborations have reported results for CP violation in kaon decays and pion phase shifts at the physical point for the corresponding $I=0$ and 2 final states~\cite{Bai:2015nea,RBC:2020kdj,RBC:2021acc}. Because the physical kinematics for such decays require pions with back-to-back relative momenta, which is not the ground state achieved in ordinary lattice calculations where the pions are at rest \cite{Maiani:1990ca}, G-parity spatial boundary conditions (GPBC) were employed in the simulations~\cite{Christ:2019sah}. GPBC forbid pions with zero momentum, and if the box size is adjusted appropriately, then the two-pion ground state computed on the lattice will have physical momenta satisfying $E_{\pi\pi} \approx M_K$. In the GPBC two-pion scattering work \cite{RBC:2021acc}, we obtained the phase shifts of the $I=0$ channel as well as the $I=2$ channel at various two-pion energies with non-zero pion momenta that are consistent with the prediction from the dispersion theory \cite{Ananthanarayan:2000ht,Colangelo:2000jc,Colangelo:2001df,Garcia-Martin:2011iqs,Garcia-Martin:2011nna} based on the Roy Equation \cite{Roy:1971tc} with inputs obtained by a combination of chiral perturbation theory for the scattering lengths and experimental data for the high energy regime.
GPBC are not implemented without extra cost, however. They are at least twice as expensive for measurements compared with periodic boundary conditions because the G-parity Dirac operator is explicitly two-flavor, with mixing between the flavors occurring at the boundary, and they also require gauge ensembles with the same boundary conditions to be generated. In addition, an important future step is to include isospin breaking effects in the calculation of $\varepsilon^\prime$ which is expected to be significant but GPBC may not be suitable due to the intrinsic role of the isospin symmetry.

The long-term aim of this study is to explore the use of periodic boundary conditions (PBC) to answer the question whether the decay amplitudes with physical kinematics can be extracted reliably from an {\it excited state} computed on the lattice.  As a first step, we investigate pion scattering in this setup.

Two-pion states with a definite total momentum can vary their total energy not only by a standard excitation of a single pion but also by changing the momenta of individual pions, or equivalently the relative momentum.  
A finite box forces the momenta to be quantized in units of $2\pi/L$ for PBC, implying that the typical interval among two-pion energies is in general of $O(2\pi/L)$.  At the same time simulations are carried out with typical values of $m_\pi L$ of approximately 3.3--4 to keep exponential finite volume effects under control.
Therefore as we lower the pion mass towards the physical value, the box sizes grows and it may become increasingly challenging to extract the signals of an excited state with the statistical and systematic errors under control.  This is the case especially for $I=0$, where there are disconnected diagrams and corresponding operators couple with the vacuum state.
In fact, we learned from our earlier works~\cite{Bai:2015nea,RBC:2020kdj,RBC:2021acc} with GPBC that there is significant higher-state contamination in two-pion correlation functions. 

The generalized eigenvalue (GEVP) method~\cite{Luscher:1990ck,Blossier:2009kd} provides us with a systematic procedure to decompose correlation functions into contributions from the several lightest states with the same quantum numbers and have been widely used for hadron spectrum studies.
In our particular case, it turned out from earlier works~\cite{Bai:2015nea,RBC:2020kdj,RBC:2021acc,Briceno:2016mjc} that introducing a $\sigma$ operator for $I=0$ in the measurements plays a crucial role in removing the contamination from excited states and that the introduction of the $\sigma$ operator significantly reduces the statistical error.
In this work, we introduce a $\sigma$ operator as well as four two-pion operators with various pion momenta for our measurements and GEVP analysis to extract the ground and excited states.

In addition, we propose a variant of the GEVP approach which we call the {\it re-based} GEVP (RGEVP).  The eigenvectors of GEVP obtained at a time slice give us a new basis of operators.  In principle, each operator in the new basis couples well with one of the lowest energy states considered in the GEVP.  With limited statistics, since we could lose the signal from one or more of those states at large time separations, it may be reasonable to exclude such noisy states by removing the corresponding operators from the basis so that all the states included in the GEVP analysis have good statistical precision.  The RGEVP is to reduce the size of GEVP by using fewer operators that couple well with states. 
We find that this approach gives us an improvement on statistical precision for the ground and first excited states of the $I=0$ channel. 

We perform lattice calculation for two-pion scattering with 258 configurations on the $24^3\times64$ lattice with the lattice cutoff $a^{-1} = 1.023$~GeV and 107 configurations on the $32^3\times64$ lattice with $a^{-1} = 1.378$~GeV~\cite{RBC:2014ntl,Tu:2020vpn}.  Both ensembles are generated with $2+1$-flavor M\"obius domain wall fermions and Iwasaki plus DSDR (dislocation suppressed determinant ratio) gauge action.  See Tab.~\ref{tab:ensembles} for more detail.
We employ several cutting-edge lattice methods: all-mode-averaging (AMA)~\cite{Blum:2012uh,Shintani:2014vja} and all-to-all (A2A) propagators~\cite{Foley:2005ac} as well as the GEVP method~\cite{Luscher:1990ck,Blossier:2009kd} to compute correlation functions and extract energy eigenvalues. 
While better statistical precision is desired and we will update our results in the near future, we take the continuum limit of the scattering phase shifts and scattering lengths at this point.  Our determination of the scattering length does not need a chiral extrapolation, which assumes the leading order as an input from ChPT and hence gives a precise value.  Our results are meaningful as a pure lattice determination, though they have larger uncertainty.
A companion paper~\cite{Bruno:2023pde} using distillation~\cite{HadronSpectrum:2009krc} will also be available soon.

Although the core of this study is the application to $K \to \pi \pi$ and the direct CP violation parameter, $\varepsilon'$, the experience gained here will provide impetus to other applications 
of $\pi$-$K$ scattering phases. Examples that we have in mind so far are 
direct CP violation in charm decays~\cite{LHCb2019,AS_LAT19}, possible CP violation in $\tau \to \nu K \pi$~\cite{Babar_tauCP}, and 
three body proton decays.

This paper is organized as follows. Section~\ref{sec:th} describes the theoretical framework underlying the calculation. In Section~\ref{sec:lat details} we give the lattice details. Section~\ref{sec:results} gives results for the computed two-pion energies and the corresponding phase shifts. Here we also compare our results to recent data-driven studies~\cite{Ananthanarayan:2000ht}. In Section~\ref{sec:comparison} we compare the PBC calculation to the GPBC one~\cite{RBC:2021acc}.
Section~\ref{sec:conclusion} summarizes the present work and future prospects.


\section{Theoretical framework}
\label{sec:th}

\subsection{Operator Construction}
\label{sec:operator def}

In this subsection we describe the operators and states used in this work.

We start with pion operators with definite spatial momentum 
\begin{align}
    \pi^a(t,\vec p) 
    = \sum_{\vec x,\vec y}\,
    &e^{-i(\vec p_1\cdot\vec x + \vec p_2\cdot\vec y)}
    f_r(||\vec x - \vec y||) 
    \notag\\
    &\times\overline\psi(t,\vec x)i\gamma_5F^a\psi(t,\vec y),
    \label{eq:pionOp}
\end{align}
which is defined with the Coulomb gauge fixing and the momentum $\vec p = \vec p_1 + \vec p_2$ of the pion operator.
In this work we consider pion operators whose momentum for a direction is zero or one unit.  Thus a natural way of assigning the inner momenta when $\vec p$ is non-zero is that $\vec p_1$ or $\vec p_2$ carries one unit of momentum and the other zero.
We assign $\vec p_1 = \vec 0$ and $\vec p_2 = \vec p$ in this work and add only $\vec p$ as the momentum argument of the single-pion operators.  
While there is no dependence on the relative momentum $\vec p_1 - \vec p_2$ without smearing,
we introduce the exponential smearing function 
\begin{equation}
    f_r(||\vec x-\vec y||) = \exp(-||\vec x-\vec y||/r), 
\end{equation}
with smearing radius $r$ and the periodic modulus $||\vec x -\vec y||$, the length of the shortest straight path from $\vec y$ to $\vec x$ in the periodic box. This hydrogen-like wave function has been used in our earlier works ~\cite{RBC:2020kdj,RBC:2021acc}.
While the single-pion operator of course depends on the smearing radius, we drop $r$ from the pion operator on the left hand side of Eq.~\eqref{eq:pionOp} for simplicity.
The quark and anti-quark isospin doublets are defined as
\begin{equation}
    \psi 
    = \left(
    \begin{array}{c}
    u \\ d
    \end{array}
    \right),\hspace{10mm}
    \overline\psi = (
    \begin{array}{cc}
    \bar u & \bar d
    \end{array}
    ),
\end{equation}
and 
\begin{align}
    F^+ &= \frac{1}{2}(\sigma_1 + i\sigma_2),\\
    F^- &= \frac{1}{2}(-\sigma_1 + i\sigma_2),\\
    F^0 &= \frac{1}{\sqrt2}\sigma_3,
\end{align}
with the Pauli matrices $\sigma_{1,2,3}$.

The two-pion operators are constructed by multiplying two single-pion ones:
\begin{align}
\widetilde O_{\pi\pi}^{I,I_z}(t_1, t_2,\vec P, \vec p/2)
= \sum_{a,b}c^{I,I_z}_{ab}\, &\pi^a(t_1,(\vec P+\vec p)/2)
\notag\\
\times\, & \pi^b(t_2, (\vec P-\vec p)/2),
\end{align}
where $\vec P$ and $\vec p$ are the center-of-mass and relative momenta of the two pseudoscalar operators, respectively, and $a,b\in \{+,-,0\}$.
The coefficients $c^{I,I_z}_{ab}$ project the two-pion operator to an isospin-definite channel labeled by $(I,I_z)$.  Appendix~\ref{sec:contractions} gives the explicit forms of the $(I,I_z) = (2,0)$ and $(0,0)$ two-pion operators.

The discrete, finite lattice breaks the continuum rotational symmetry of angular momentum SO(3) down to a discrete subgroup, which depends on the center-of-mass momentum.  The irreducible representations of such a discrete subgroup do not give rise to angular momentum eigenstates that appear as irreducible representations of SO(3). Instead, they are mixtures which can be classified in terms of the continuum irreducible representations.

It is fairly straightforward to make this classification based on fundamental group theory for one and two particle systems, both moving and at rest~\cite{Moore:2005dw,Dudek:2012gj}. 
Our main target is the $s$-wave two-pion states and their phase shifts. The corresponding interpolating operators are defined as
\begin{equation}
O_{\pi\pi}^{I,I_z}(t_1, t_2, \vec P, \vec p/2)
= \sum_{\hat T\in G}
\chi_{A_1}(\hat T)
\widetilde O_{\pi\pi}^{I,I_z}(t_1, t_2,\vec P, \hat T[\vec p/2]),
\label{eq:pipi_swave}
\end{equation}
where we sum over all elements $\hat T$ in the finite-volume symmetry group $G$ and the normalization factor $\chi_{A_1}(\hat T)$ is the character of the group element $\hat T$ in the representation $A_1$~\cite{Moore:2005dw,Dudek:2012gj}.

We can consider two-pion operators composed of two bilinear operators located at different time slices $t_1$ and $t_2$.  As long as there is no other operator placed in between, the time-non-local two-pion operators still play a role in creating and annihilating two-pion states with corresponding quantum numbers and we can discuss the spectrum at time slices outside the operator.
It has been shown that placing the two bilinear operators on slightly different time slices is advantageous for reducing statistical noise especially for $I=0$~\cite{Bai:2015nea,RBC:2020kdj,RBC:2021acc}, where the overlap of the two-particle operator with the vacuum state can be suppressed exponentially by the separation $\Delta\equiv |t_2 - t_1|$. 

In addition to these two-pion operators, we introduce a $\sigma$ operator, or iso-singlet scalar bilinear operator, for $I=0$:
\begin{equation}
    \sigma(t, \vec p)
    = \sum_{\vec x,\vec y}\,
    e^{-i(\vec p_1\cdot\vec x + \vec p_2\cdot\vec y)}
    f_r(||\vec x - \vec y||) 
    \overline\psi(t,\vec x)\psi(t,\vec y),
\end{equation}
with $\vec p = \vec p_1 + \vec p_2$.
Again, we set $\vec p_1 = \vec0$ and $\vec p_2 = \vec p$ in this work.
This operator has been found to play an important role in controlling the contamination from excited states~\cite{RBC:2020kdj,RBC:2021acc}.

In this work, we concentrate on the rest frame $\vec P = \vec 0$ and $s$-wave operators and states.
For $I=2, I_z=0$, we consider four values of relative pion momenta and use the following operator basis:
\begin{equation}
    O^{2,0}(t)
    = \left(
    \begin{array}{c}
    O_{\pi\pi}^{2,0}(t,t+\Delta,\vec 0,(0,0,0)\times2\pi/L) \\
    O_{\pi\pi}^{2,0}(t,t+\Delta,\vec 0,(0,0,1)\times2\pi/L) \\ 
    O_{\pi\pi}^{2,0}(t,t+\Delta,\vec 0,(0,1,1)\times2\pi/L) \\ 
    O_{\pi\pi}^{2,0}(t,t+\Delta,\vec 0,(1,1,1)\times2\pi/L) 
    \end{array}
    \right).
    \label{eq:basisI2}
\end{equation}
Similarly for $I=0,I_z=0$, we define
\begin{equation}
    O^{0,0}(t)
    = \left(
    \begin{array}{c}
    O_{\pi\pi}^{0,0}(t,t+\Delta,\vec 0,(0,0,0)\times2\pi/L) \\
    \sigma(t, \vec 0) \\
    O_{\pi\pi}^{0,0}(t,t+\Delta,\vec 0,(0,0,1)\times2\pi/L) \\ 
    O_{\pi\pi}^{0,0}(t,t+\Delta,\vec 0,(0,1,1)\times2\pi/L) \\ 
    O_{\pi\pi}^{0,0}(t,t+\Delta,\vec 0,(1,1,1)\times2\pi/L) 
    \end{array}
    \right).
    \label{eq:basisI0}
\end{equation}
The order of the operators in these bases matters when it comes to $N\times N$ GEVP as described in Section~\ref{sec:GEVP}.
When we do not need to specify the isospin, we simply call the four two-pion operators $\pi\pi(000)$, $\pi\pi(001)$, $\pi\pi(011)$ and $\pi\pi(111)$, respectively.

\subsection{Correlation Functions}

For the GEVP a matrix correlation function is defined in the basis of operators given in the previous sub-section,
\begin{equation}
    C_{ij}^{I,I_z}(t)
    = \left\langle
    O^{I,I_z}_i(t) O^{I,I_z}_j(-\Delta_j)^\dag
    \right\rangle,
    \label{eq:def_2ptCorr}
\end{equation}
where 
\begin{equation}
\Delta_j = \Bigg\{
\begin{array}{ll}
\Delta & (I=2)
\\[2mm]
(1-\delta_{2,j})\Delta & (I=0)
\end{array},
\end{equation}
translates the source two-pion operators by $\Delta$ but does nothing for the $\sigma$ operator so that the time variable $t$ always indicates the minimum time separation between a bilinear of the source operator and that of the sink operator.  While the measured correlator matrix is not exactly symmetric with finite statistics, we symmetrize it by averaging with the transposed partner.

\begin{figure}[tbp]
    \centering
    \includegraphics[width=.22\textwidth]{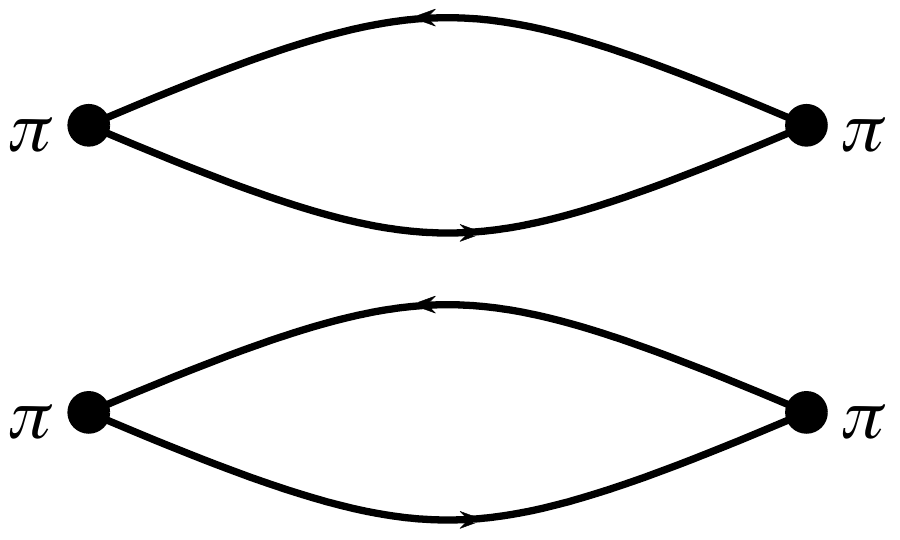}~~~~~~~
    \includegraphics[width=.22\textwidth]{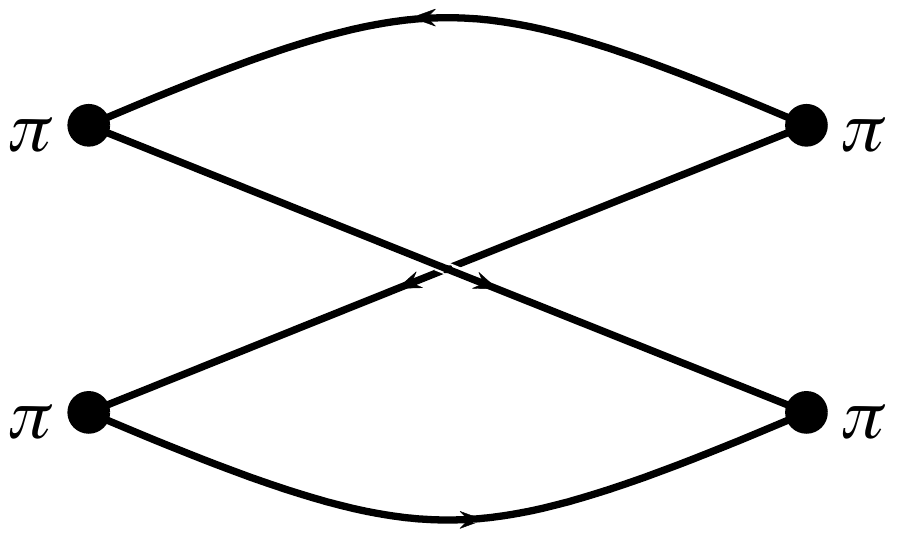}
    \\[6mm]
    \includegraphics[width=.22\textwidth]{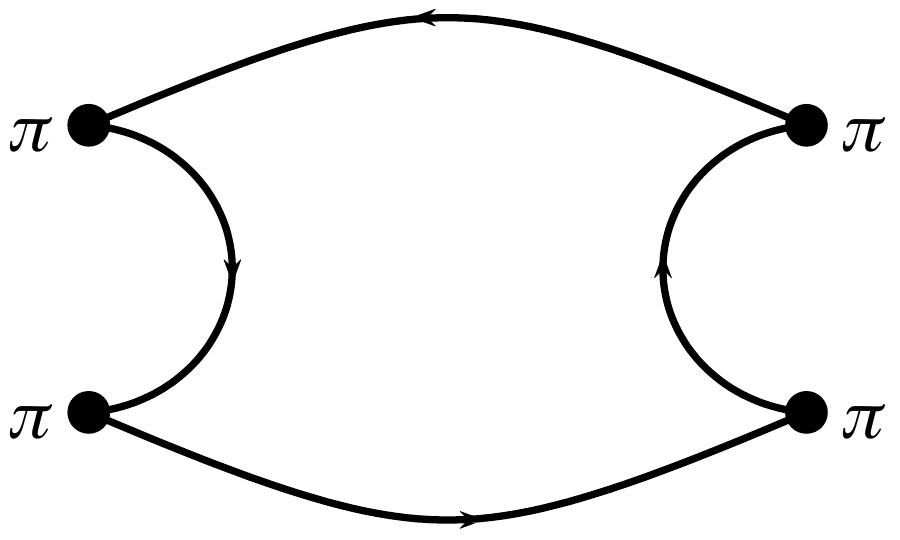}~~~~~~~
    \includegraphics[width=.22\textwidth]{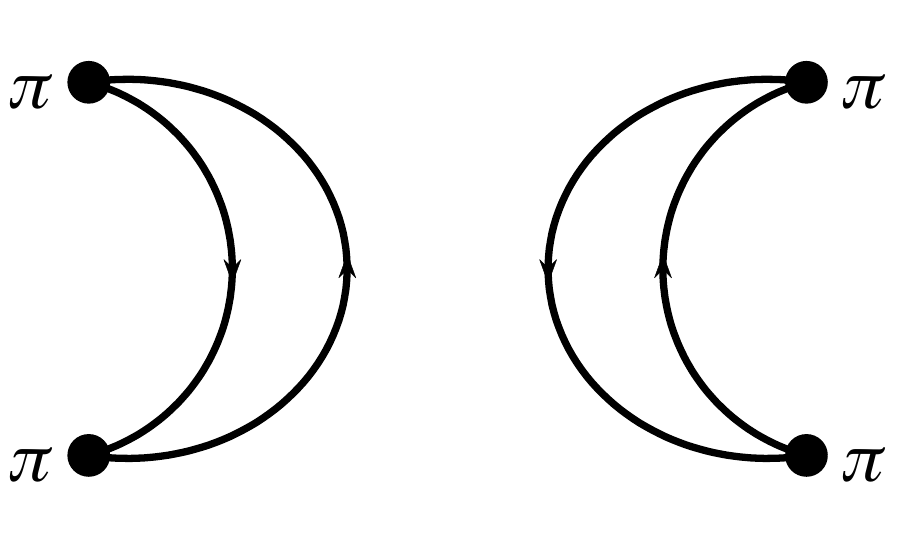}
    \caption{Elemental Wick contractions. Clockwise from top-left: D (direct), C (cross), V (vacuum), and R (rectangle). Linear combinations of the four diagrams are used to construct pion scattering correlation functions for $I=0$, 1, and 2 states with definite lattice hypercubic symmetry.}
    \label{fig:Wick contractions}
\end{figure}

\begin{figure}[tbp]
    \centering
    \includegraphics[width=.22\textwidth]{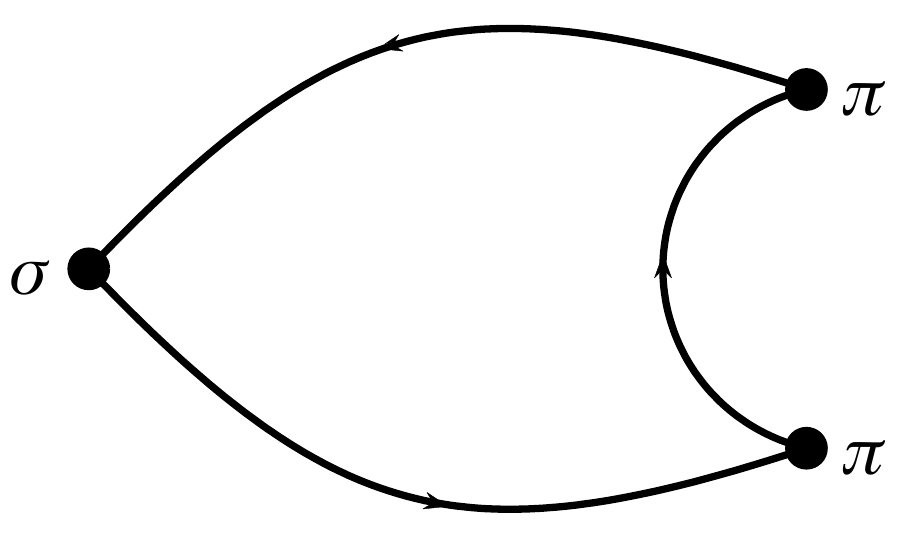}~~~~~~~
    \includegraphics[width=.22\textwidth]{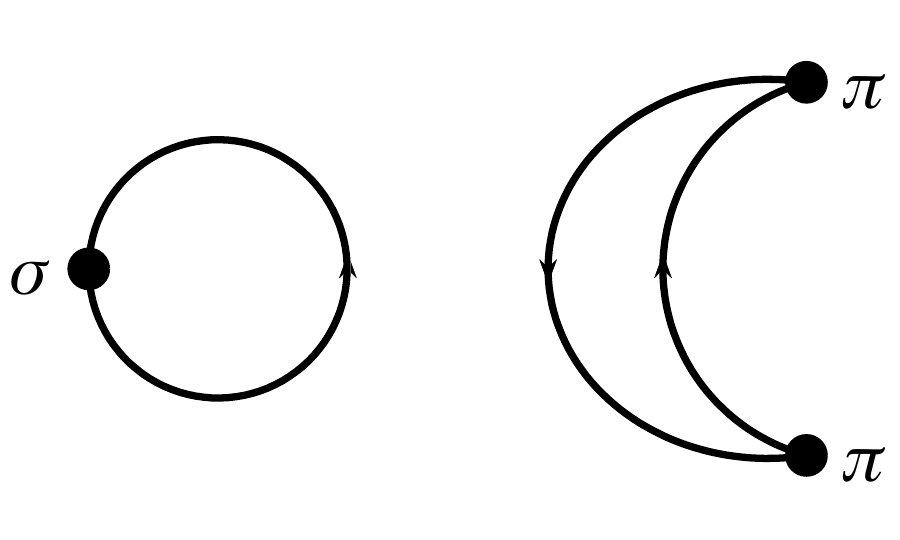}
    \\[6mm]
    \includegraphics[width=.22\textwidth]{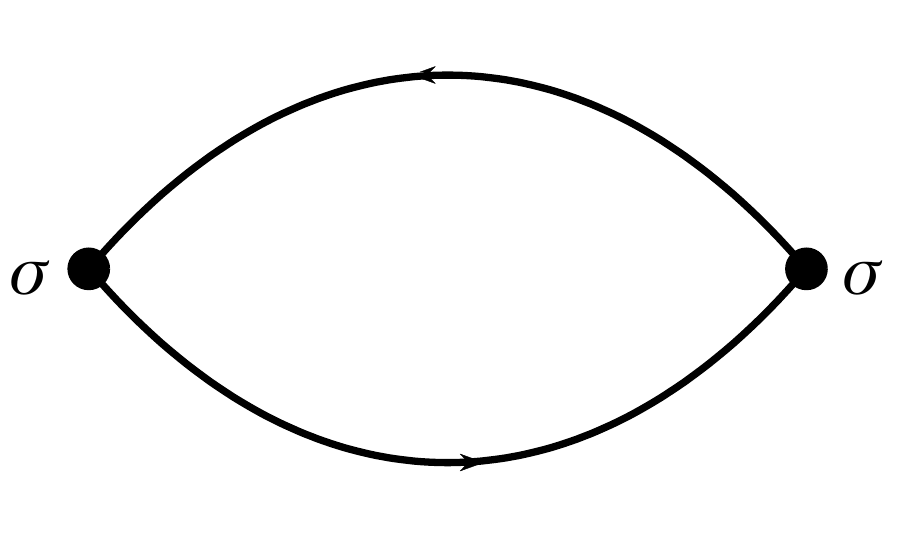}~~~~~~~
    \includegraphics[width=.22\textwidth]{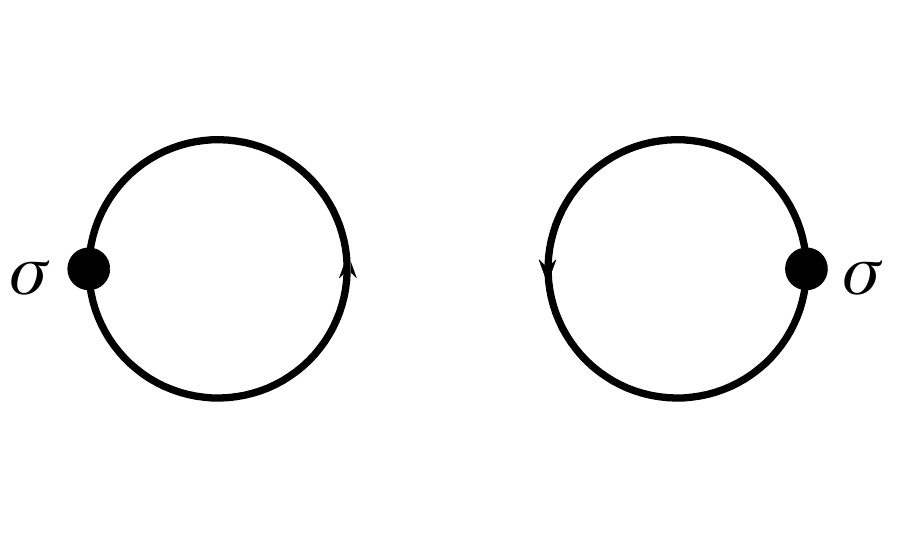}
    \caption{Elemental Wick contractions with one (upper) or two (lower) scalar bilinear operators. These are analoguous to the R (left) and V (right) diagrams in Fig.~\ref{fig:Wick contractions} and relevant for only the $I=0$ channel.}
    \label{fig:Wick contractions sigma}
\end{figure}

The Wick contractions of the two-point functions are shown diagramatically in Figs.~\ref{fig:Wick contractions} and \ref{fig:Wick contractions sigma} .  They are denoted ``direct" (D), ``cross" (C), ``rectangle" (R), and ``vacuum" (V). 
By taking an appropriate linear combination of these elemental contractions, we construct a correlation function of operators carrying definite isospin. 
All diagrams in Figs.~\ref{fig:Wick contractions} and \ref{fig:Wick contractions sigma} contribute to the $I=0$ channel, while only diagrams D and C contribute to the $I=2$ channel.
We compute the disconnected diagrams for every time translation and take the translation average, while the connected diagrams are computed at every several source time slices, which are specified in Section~\ref{sec:lat details}. 
The complete formulae for the $I=2$ and $I=0$ channels are given in Appendix~\ref{sec:contractions}.

For $I=0$ there is an additional complication: 
in the rest frame the ground state in this channel is the vacuum. This contribution dominates the correlation function and must be subtracted,
\begin{align}
    C_{ij}^{0,0}(t)
    &= \left\langle
    O^{0,0}_i(t) {O^{0,0}_j(-\Delta_j)}^\dag
    \right\rangle
    \notag\\
    & - \frac{1}{L_t}\sum_{t_{src}=0}^{L_t-1} \left\langle
    O^{0,0}_i(t+t_{src}) \right\rangle
    \left\langle {O^{0,0}_j(t_{src}-\Delta_j)}^\dag
    \right\rangle,
    \label{eq:vac_subt}
\end{align}
where $L_t$ stands for the time extent of the lattice ensemble in lattice units.
While the second term on the right hand side is independent of $t$ in the limit of infinite statistics, we perform this subtraction time-slice-by-time-slice as we found it provides a minor statistical advantage~\cite{RBC:2020kdj,RBC:2021acc}.

\subsection{Thermal effects}

Due to the finite time size of the lattice ($L_t$) and the pions satisfying periodic boundary conditions in time, unwanted contributions contaminate the correlation function. These so-called around-the-world (ATW), or thermal, effects arise when one of the source pions propagates forward in time while the other goes backwards through the boundary to reach the sink time slice.  They can be seen by inserting a complete set of states in the two-point (``thermal") correlation function and translating the source and sink operators to equal times.
\begin{align}
&\braket{O_{\pi\pi}(t)O_{\pi\pi}(0)^\dagger}
\notag\\
=&
\sum_m\sum_n \braket{m|O_{\pi\pi}(t) \ket{n}\bra{n}O_{\pi\pi}^\dagger(0)|m}\nonumber
\\ =&\sum_n (e^{-E_n^{\pi\pi} t}+e^{-E_n^{\pi\pi} (L_t-t)}) \braket{0|O_{\pi\pi} \ket{n}\bra{n}O_{\pi\pi}^\dagger|0}\nonumber
\\ &+e^{-E_{\pi(\vec p)} t}e^{-E_{\pi(\vec p)}(L_t-t)}
\notag\\&\hspace{5mm}\times
\braket{\pi(\vec p) \left| O_{\pi\pi}\right|\pi(\vec p)}\braket{\pi(\vec p) \left| O^\dagger_{\pi\pi}\right|\pi(\vec p)}+\dots,
\label{eq:atw}
\end{align}
where we omit the isospin superscripts and momentum arguments for simplicity and the sum over $m$ gives the thermal expectation value.  The first term on the right hand side contains the zero temperature expectation value, while the last term is the thermal contribution which vanishes as $L_t\to \infty$.  Notice when the rest frame is employed, the leading ATW contribution is time independent while the ATW effects of an excited-state pion and those in a moving frame are time dependent. All thermal effects are suppressed exponentially with $L_t$.

Since we employ the rest frame, the leading ATW term, which is constant, can be removed simply by a ``matrix" subtraction 
\begin{align}
    C_{ij,subt}^{I,I_z}(t)&\equiv C_{ij}^{I,I_z}(t)-C_{ij}^{I,I_z}(t+\delta_t).
    \label{eq:matdt_subt}
\end{align}
where $\delta_t$ is an arbitrary time shift. 
This subtraction removes all constant contributions to the correlation functions and therefore the vacuum subtraction \eqref{eq:vac_subt} is in principle unnecessary if this subtraction is applied.
In this work, we still apply the vacuum subtraction so that we can investigate the significance of the ATW effects with the absence of vacuum effects by analyzing both matrix-subtracted and unsubtracted correlators.
As seen in Section~\ref{sec:energies}, the ATW effects are significant for two-pion at rest but can well be subtracted by the matrix subtraction.

\subsection{Generalized eigenvalue problem method}
\label{sec:GEVP}

For the rest of the section we omit the superscripts $I$ and $I_z$ and simplify our notation of the correlator matrix $C_{ij,subt}^{I,I_z}(t) \to C_{ij}(t)$, or $C(t)$ when the operator indices $i$ and $j$ can be dropped without confusion.

As is well known, Euclidean space correlation functions are a sum of exponential terms, each term corresponding to one state in a tower of states with fixed quantum numbers and increasing energies, 
\begin{align}
C_{ij}(t)
& = \sum_n A_{n,i}A_{n,j}^*e^{-E_n t},
\label{eq:state_expansion}
\end{align}
where we have neglected the backward propagating contributions proportional to $e^{-E_n(L_t-t)}$ since $t\ll L_t$ in our setup\footnote{This contribution is exponentially suppressed compared to the ATW effect considered earlier.}. $A_{n,i}=\langle 0|O_i|n\rangle (1-e^{-E_n\delta_t})^{1/2}$ is the overlap of the $i$-th operator acting on the $n$-th state and the vacuum multiplied with with a normalization factor due to the matrix subtraction.

To extract the desired excited states in the correlator, we employ the variational method by solving a generalized eigenvalue problem (GEVP)~\cite{Luscher:1990ck,Blossier:2009kd}.  
For an $N\times N$ matrix $C(t)$ we solve the following GEVP:
\begin{eqnarray}
C(t)V_n(t,t_0) &=& \lambda_n(t,t_0) C(t_0)V_n(t,t_0),
\label{eq:GEVP}
\end{eqnarray}
with eigenvalues $\lambda_n(t,t_0)$ and eigenvectors $V_n(t,t_0)$, where in principle we can choose $t_0$ in the range $0 < t_0 < t$.
At asymptotically large time separations the eigenvalue behaves as $\lambda_n(t,t_0) = e^{-E_n (t-t_0)}$ where $E_n$ is $n$-th energy state in the GEVP. In Ref.~\cite{Blossier:2009kd} it was shown that the leading correction behaves like $e^{-(E_{N+1}-E_n)t}$ for $t_0\ge t/2$.  In this work, we use the first $N$ operators of the bases in Eqs.~\eqref{eq:basisI2} and \eqref{eq:basisI0} for the $I=2$ and $I=0$ channels, respectively, and solve GEVP with a $t$-independent value of $t - t_0\equiv\Delta_t$.  While the preferable inequality $t_0\ge t/2$ is violated in the region $t<2\Delta_t$, we do not use data at such short times relative to $\Delta_t$ for our final results.

Effective two-pion energies are defined as~\cite{Luscher:1990ck,Blossier:2009kd}
\begin{equation}
     E^{\rm eff}_n(t,t_0) = \ln{\lambda_n(t,t_0)}- \ln{\lambda_n(t+1,t_0)}.
     \label{eq:def_Eeff}
\end{equation}
The corresponding eigenvector $V_n(t,t_0)$ at asymptotic time separation provides a new operator that couples to the $n$-th state but not with the other states in the GEVP\footnote{The coupling is not perfect: in Ref.~\cite{Blossier:2009kd} it is shown that corrections are $O(e^{-(E_{N+1}-E_n)t_0})$.}~\cite{Blossier:2009kd}
\begin{equation}
    \widetilde O_n = \sum_i V_{n.i} O_i.
    \label{eq:op eigenvector}
\end{equation}
These eigenvectors play a key role in isolating the weak operator matrix elements between an excited two-pion state and the kaon in $K\to\pi\pi$ decays, for example.

In practice, the large statistical error of correlation functions at large time separations may cause misordering of eigenvalues and eigenvectors for specific jackknife samples resulting in incorrectly large errors in two-pion energies and GEVP eigenvectors. 
A brief description of our procedure to ensure the correct order of eigenvectors is given below:
\begin{enumerate}
\item At small time separations where correlators and hence eigenvectors are well resolved, ensure the descending order of eigenvalues.  Then the corresponding effective energies will be obtained in the ascending order.
\label{GEVP1}
\item At large time separations where excited-state contamination is small, solve the GEVP with the correlators that are mostly diagonalized by Eq.~\eqref{eq:op eigenvector} with the eigenvectors obtained at one time slice earlier. 
The eigenvectors from such a GEVP are close to a unit vector for a certain direction and the ordering is fairly trivial.
Then change the basis of these eigenvectors back into the original basis.
\label{GEVP2}
\end{enumerate}
See Appendix~\ref{sec:GEVP details} for more detail.

In this work, we employ procedure~\ref{GEVP1} at $t_0 = 1,2$ and procedure~\ref{GEVP2} at larger time separations $t_0\ge3$.
The reason why we switch the procedure at a certain value of $t_0$ rather than $t$ is because the GEVP eigenvectors $V_n(t,t_0)$ receive the contamination from the higher states by $O(e^{-(E_{N+1}-E_n)t_0})$~\cite{Blossier:2009kd}.

It is mathematically guaranteed that the GEVP eigenvalues and eigenvectors in Eq.~\eqref{eq:GEVP} are real when $C(t_0)$ is a positive-definite real symmetric matrix and some software functions to solve GEVP have this assumption.
While the correlator matrix at large time separations with limited statistics have zero-consistent eigenvalues and it is inevitable that correlator matrix have negative eigenvalues at some time slices, that may not necessarily mean we cannot solve GEVP or obtain any information of two-pion signals at those time separations.
The positivity and real symmetry are sufficient but not necessary to give us real eigenvalues and eigenvectors of GEVP.
As shown in Section~\ref{sec:energies}, we find that GEVP with $C(t_0)$ including a negative eigenvalue can still give us good signals of a few lowest energy states with the ordering strategy explained above and in Appendix~\ref{sec:GEVP details} as long as GEVP eigenvalues and eigenvectors are real.

\subsection{Re-based GEVP}
\label{sec:RGEVP}

The GEVP method provides a decomposition based on the $N$ lightest states.  It assumes sufficiently large time separations, so the contamination of higher excited states ($n\ge N+1$) can be ignored.  On the other hand, the GEVP becomes increasingly difficult with increasing time separation due to the exponentially deteriorating signal-to-noise of the correlator matrix. 
For small $N$, we expect the plateau to start at larger time separations, where the signal-to-noise ratio is already poor.  For large $N$, plateaus for the various energies move to earlier times, but the larger statistical errors on the higher energy states in the GEVP might spoil the signal of the lower states.  Thus for any choice of $N$ (or operator set) there is a chance that the signal loss occurs before a clear plateau is observed.

To address the problem we propose a modified version of the GEVP which we call the {\it re-based} GEVP (RGEVP) whereby we choose a new reduced basis of fewer operators that couple well with a few resolved low energy states.  The idea originates from the fact that the number of resolvable states which dictates the appropriate size of GEVP decreases with increasing time separation  and that the GEVP at short time separations, even before reaching a plateau, provides a set of nearly diagonalized operators.  

The simplest re-basing can be performed by Eq.~\eqref{eq:op eigenvector} with a chosen number $N'(<N)$ of eigenvectors $V_{n(<N')}$ obtained at a chosen time slice $t_0 = t_0'$.  The new basis provides an $N'\times N'$ correlator matrix with which we can perform GEVP without contamination from the states labeled by $N'+1$ to $N$.

The re-basing does not have to be a single step.  When one is interested in the ground (first excited) state, the reduced GEVP size could minimally be one (two).  Doing such a reduction of basis with a single step may not minimize both statistical and systematic errors.  A multi-step re-basing can be made by choosing multiple pairs of re-basing time $t_{0,\alpha}(>t_{0,{\alpha-1}})$ and reduced size $N_\alpha(< N_{\alpha-1})$ with the label $\alpha$ of re-basing steps and repeating the rebasing $N_{\alpha-1}\to N_\alpha$ at $t_0=t_{0,\alpha}$ for each $\alpha$.  See Appendix~\ref{sec:RGEVP details} for more detail.

In this work, we implement the re-basing with the central values eigenvectors for all jackknife samples to maintain the configuration independence of the new operator basis.

\subsection{Phase shifts}
\label{sec:phase shifts theory}

The L\"uscher method~\cite{Luscher:1990ux} allows us to extract the scattering phase shifts from finite-volume energies on the lattice. The interaction region is supposed to be confined to a volume well contained inside a box of larger volume.  Outside this region the solution of the wave equation corresponds to free (non-interacting) particles, and the boundary conditions of the box impose a quantization condition on the complete solution which necessarily relates the phase shifts to finite-volume energies. 

While the extension to moving frames is straightforward \cite{Rummukainen:1995vs}, we limit our discussion to the case of the rest frame, where one obtains the two-pion $s$-wave scattering phase shift $\delta(E_{\pi\pi})$ corresponding to a given two-pion energy as follows:
  \begin{align}
    k&=\sqrt{\frac{E_{\pi\pi}^2}{4}-m_{\pi}^2},
    \label{eq:disp_rel}
    \\q&= \frac{kL}{2\pi},
    \\ \tan~\phi(q)&=\frac{\pi^{3/2}q}{Z_{00}(1;q^2)},
  \\\delta(E_{\pi\pi}) &= -\phi(q)+\pi n,~n\in \mathbb{Z},\label{eq:phase shift}
  \end{align}
where $E_{\pi\pi}$ is the energy of the two-pion state in a finite box of size $L^3$, $m_\pi$ is the pion mass and $Z_{00}(1,q^2)$ is the L\"uscher zeta function, which we compute via an efficient numerical implementation given in Ref.~\cite{Yamazaki:2004qb}.
Eq.~(\ref{eq:phase shift}) is used to compute all $I=0$ and 2 phase shifts. 

The method outlined here for obtaining the phase shifts is strictly valid in the limited region $2m_\pi\le E_{\pi\pi}\le4m_\pi$ and up to neglected higher partial waves. In the present work we apply this method also to energies above the $4\pi$ inelastic threshold, and neglect these sources of systematic errors.  Numerical results presented in Section~\ref{sec:phase shifts res} suggest these systematic errors may not be large within the energy range considered here.

\subsection{The dispersion relation method}
\label{sec:DR}

Up to small finite lattice spacing effects, L\"uscher's method, explained above, gives an accurate prescription for obtaining the phase shift. 
With finite lattice spacing, Eq.~\eqref{eq:disp_rel} needs modification since the appropriate dispersion relation for finite lattice spacing depends on the type of lattice fermion. While simulations at multiple lattice spacings enable the removal of these effects, we remove some of them for each lattice spacing separately~\cite{RBC:2021acc}.

The method is based on the cancellation of artifacts between interacting and non-interacting two-pion energies.
The non-interacting two-pion energies, $E_n^{0}$, are determined from a product of two expectation values of single-pion correlators, $C^{0}(t)$, which is analogous to the D diagram in the interacting case.
Since the correlation functions involving the $\sigma$ operator do not contain the D diagram, they should be treated separately.
We first explain the method for $I=2$, where the $\sigma$ operator is absent, and then explain its generalization to the case including the $\sigma$ operator.

The non-interacting correlator matrix $C^{0}(t)$ is diagonal with analogous two-pion effective energies, $E_1^{0,\rm eff}(t,t_0)<E_2^{0,\rm eff}(t,t_0)<\ldots<E_N^{0,\rm eff}(t,t_0)$. The energy shift of two-pion states due to interactions is
\begin{equation}
    \Delta E^{\rm eff}_n(t,t_0) = E^{\rm eff}_n(t,t_0) - E^{0,\rm eff}_n(t,t_0),
    \label{eq:energy_shift}
\end{equation}
where single-pion discretization errors largely cancel.
Thus we obtain an improved two-pion energy
\begin{equation}
    E_n^{\rm eff\,\prime}(t,t_0)
    = E^{0,\rm disp}_n + \Delta E^{\rm eff}_n(t,t_0),
    \label{eq:newEeff}
\end{equation}
where we define
\begin{equation}
    E^{0,\rm disp}_n = 2\sqrt{m_\pi^2 + |\vec p_n|^2},
    \label{eq:disp_nonintEpipi}
\end{equation}
with a non-interacting pion momentum in the finite box $\vec p_n = (0,0,0), (0,0,2\pi/L), (0,2\pi/L,2\pi/L),\ldots$.

In addition to reducing the scaling violation in the dispersion relation, this method determines two-pion energies and phase shifts with other improvements as well.
The first term on the right hand side of Eq.~\eqref{eq:newEeff}, which is defined in Eq.~\eqref{eq:disp_nonintEpipi}, is as statistically precise as $m_\pi$.
On the other hand, the second term on the right hand side of Eq.~\eqref{eq:newEeff}, which is given in Eq.~\eqref{eq:energy_shift}, is also expected to be more precise than $E_n^{\rm eff}(t,t_0)$ because of the correlation between the first and second terms on the right hand side of Eq.~\eqref{eq:energy_shift}.
Furthermore, the energy difference in Eq.~\eqref{eq:energy_shift} may also remove excited state effects related to single pions and allow effective energies $E_n^{\rm eff\,\prime}$ to plateau sooner.
We find these to be the case, especially for the $I=2$ channel, as discussed in Section~\ref{sec:results}.

For the $I=0$ channel, the statistical errors are dominated by the disconnected diagram which cannot be improved by this method, though we still find some improvement for the combination of diagrams by applying the following procedure with the $\sigma$ operator.
The $I=0$ channel is more complicated not only because of the inclusion of the $\sigma$ operator but also because the interaction between the two pions makes the finite-volume two-pion energies quite unlike the energies of non-interacting two pions, {\it i.e.} $|\Delta E_n^{\rm eff}(t,t_0)|$ for $I=0$ is much larger than that for $I=2$.  Therefore it is less meaningful to identify the one-to-one correspondence between the interacting and non-interacting two-pion energies as in Eq.~\eqref{eq:energy_shift}.
In this work instead of matching the state label $n$ of the interacting and non-interacting two-pion energies, the first and second terms in Eq.~\eqref{eq:energy_shift}, we choose the non-interacting two-pion energy as the one closest to the interacting $n$-th state energy $E_n^{\rm eff}(t,t_0)$ for the procedure explained above.

\section{Ensemble Details and Computational Setup}
\label{sec:lat details}

\begin{table}[tbp]
    \centering
    \begin{tabular}{c|c|c|c|c|c|c}
    \hline
         $m_\pi$& lattice &  & $a^{-1}$ &$L$ & trajectories&  \\
        (MeV) & size & $L_s$ &  (GeV) & (fm) & (MD time units) & configs  \\
        \hline
        142.6(3) & $24^3\times64$ & 24 & 1.023(2) & 4.67 & 250-3860 & 258\\
        143.6(9) & $32^3\times64$ & 12 & 1.378(5) & 4.58 & 200-1320 & 107\\
        \hline
    \end{tabular}
    \caption{Ensemble parameters. 2+1 flavors of M\"obius domain wall fermions, generated by the RBC/UKQCD collaborations~\cite{RBC:2014ntl,Tu:2020vpn}. Trajectories used for measurements are separated by 10 or 20 Monte Carlo time units. The last column refers to the number of configurations in each ensemble used for measurements.}
    \label{tab:ensembles}
\end{table}

Our computations are carried out on two ensembles of 2+1 flavors of M\"obius domain wall fermions (MDWF) with physical masses generated by the RBC/UKQCD collaborations~\cite{RBC:2014ntl,Tu:2020vpn}. Both use the Iwasaki-DSDR gauge action~\cite{Renfrew:2008zfx} and correspond to inverse lattice spacings of about 1.0 and 1.4 GeV, respectively, with similar physical volumes ($L\sim5$ fm). The parameters of each ensemble are listed in Tab.~\ref{tab:ensembles}.

Correlation functions are computed in an all-to-all propagator (A2A)~\cite{Foley:2005ac} framework using 2,000 low-modes of the preconditioned, squared Dirac operator and spin-color-time diluted random source propagators for the high modes. 

We employ all-mode averaging (AMA)~\cite{Blum:2012uh,Shintani:2014vja} to save the computational cost for the conjugate gradient (CG).  While the traditional AMA is to perform fewer exact measurements ($e.g.$ with fewer source locations) for all the configurations with which sloppy measurements are performed, we rather reduce the number of configurations for the exact measurements keeping the A2A procedure as it is.  We first perform both exact and sloppy measurements with $N_{\rm exact}$ configurations and create corresponding jackknife samples of the difference between exact and sloppy correlators, which would correct the bias due to the sloppy CG.  In the case of bin size 1,
\begin{equation}
\Delta C^{(l)}(t) = \frac{1}{N_{\rm exact}-1}\sum_{k\neq l} \left(C_{\rm exact}^{{\cal E}_k}(t)-C_{\rm sloppy}^{{\cal E}_k}(t)\right),
\end{equation}
where $\cal E$ is the list of configurations with which both exact and sloppy measurements are performed and $C_{exact/sloppy}^{{\cal E}_k}(t)$ stands for an exact/sloppy sample of the correlator matrix calculated with a configuration ${\cal E}_k$. 
The average of the difference is defined as
\begin{equation}
\Delta C(t) = \frac{1}{N_{\rm exact}}\sum_{l=1}^{N_{\rm exact}}\Delta C^{(l)}(t).
\end{equation}
We also perform sloppy measurements with $N_{\rm sloppy}$ additional configurations in the list $\cal S$ and create corresponding jackknife samples.  If we set the bin size to 1, they are written as
\begin{equation}
C_{\rm sloppy}^{(l)}(t) = \frac{1}{N_{\rm sloppy}-1}\sum_{k\neq l}C_{\rm sloppy}^{{\cal S}_k}(t),
\end{equation}
and the average is given as
\begin{equation}
C_{\rm sloppy}(t) = \frac{1}{N_{\rm sloppy}}\sum_{l=1}^{N_{\rm sloppy}}C_{\rm sloppy}^{(l)}(t).
\end{equation}
With these jackknife samples and averages, we define super-jackknife samples as
\begin{equation}
C_{\rm AMA}^{(l)}(t) = \Bigg\{\begin{array}{ll}
C_{\rm sloppy}(t) + \Delta C^{(l)}(t) & (1\le l\le N_{\rm exact})
\\[2mm]
C_{\rm sloppy}^{(l-N_{\rm exact})}(t) + \Delta C(t) & (l> N_{\rm exact})
\end{array}.
\end{equation}

In this work, the sloppy (high-mode) propagators are computed with 400 and 330 iterations of CG and the exact propagators are computed to the CG residual of $10^{-8}$ with 14 and 17 configurations for the $24^3$ and $32^3$ lattices, respectively.  For the sloppy part of the measurements on the $24^3$, 1.023~GeV, ensemble we employ the zM\"obius approximation~\cite{Abramczyk:2017oxr} of the M\"obius Dirac operator, reducing the size of the fifth dimension by a factor of two.

As described in Section~\ref{sec:operator def}, the two-pion operators are defined as a product of two single-pion operators separated by a parameter $\Delta$.  We choose $\Delta = 3$ and 4 for the $24^3$ and $32^3$ lattices, respectively, so the separation in physical units is about the same.
The smearing radius $r$ of the single-pion and the sigma operators is set to 1.5 and 2.0 in lattice units for the $24^3$ and $32^3$ lattices, respectively.
We average the correlation functions over time location of the source operator $t_{src}$.
The disconnected diagrams are computed at $t_{src} = 0,1,\ldots,63$, while the connected diagrams are computed at $t_{src} = 0, 8, \dots, 56$ for the $24^3$ lattice and $t_{src} = 0, 10, \dots, 50$ for the $32^3$ lattice.

\section{results}
\label{sec:results}

\subsection{Energies}
\label{sec:energies}

\begin{table}
\centering
\begin{tabular}{ccc}
\hline
momentum & lattice & DR \\
\hline
\multicolumn{3}{c}{$24^3$ lattice}\\
\hline
 (000) & 0.13944(17)[1.2] & 0.13944(17)\\
 (001) & 0.29572(30)[0.9] & 0.296621(80)\\
 (011) & 0.39431(66)[1.2] & 0.395630(60)\\
 (111) & 0.4685(18)[1.0] & 0.474407(50)\\
\hline
\multicolumn{3}{c}{$32^3$ lattice}\\
\hline
 (000) & 0.10422(20)[1.1] & 0.10422(20)\\
 (001) & 0.22190(44)[1.1] & 0.222293(94)\\
 (011) & 0.2954(10)[1.5] & 0.296593(71)\\
 (111) & 0.3559(28)[1.2] & 0.355697(59)\\
\hline
\end{tabular}
\caption{Summary of single-pion energies with the four lowest momenta on the $24^3$ and $32^3$ lattices.  The three-digit number on the first column specifies the spatial momentum of the single pion.  Results from correlated $\chi^2$ fits to single-pion correlators with the cosh function (lattice) and dispersion relation, half of Eq.~\eqref{eq:disp_nonintEpipi}, (DR) are shown in lattice units.  The values of $\chi^2/d.o.f.$ are displayed in the square brackets.}
\label{tab:nonint_energies}
\end{table}

Single-pion energies with the four lowest momenta are summarized in Tab.~\ref{tab:nonint_energies}.  The results from single-pion correlation functions with respective momenta and from the continuum dispersion relation are listed.  The non-interacting two-pion energies can be estimated as the double of these values.  The difference between the values from the two approaches corresponds to the discretization effect on the two-pion energies that can be removed by the dispersion relation (DR) method explained in Section~\ref{sec:DR}.  By definition, the results for zero-momentum energy from the two approaches are identical.  One can recognize the discretization effect in the results for the non-zero momenta on the $24^3$ lattice, while the results on the $32^3$ are not sufficiently resolved to see the difference.

Interacting two-pion energies are tabulated and plotted for several values of Euclidean times $t_0$, $t-t_0$, $\delta_t$, and various types ($N\times N$ or RGEVP) of GEVP, with and without the dispersion relation method, in Appendix~\ref{sec:2pi energies}. The parameters for re-basing are found in captions of respective figures and tables throughout this subsection and tables in Appendix~\ref{sec:2pi energies}.  Some general patterns are apparent. For short times the statistical errors are sub-percent, even down to the per mille level in some cases. The effect of increasing $t-t_0$ is small and likewise for  $\delta_t$. The DR method greatly enhances statistical precision, especially for the $I=2$ ground state but gains are seen generally. It also lessens (single-pion) excited state contamination. In the next two subsections we discuss the results in detail, including comparisons in a range of $t$ and for both lattice spacings to assess residual excited state contamination and lattice artifacts, respectively. 

Our guiding principle throughout this analysis is to stick to as short times as possible, where statistics are better and ATW effects are smaller, taking advantage of the GEVP and the DR method that reduce excited state contamination.
In this section, the state label $n$ begins with 0 so that the ground state is labeled with $n=0$, the first excited state with $n=1$ and so on.

\subsubsection{I=2}

We begin with the ground state for the $24^3$ ensemble. Fig.~\ref{fig:I=2 n=0 energies} shows the ground state energy for several GEVP types and representative values of $\delta_t=2,5,8$, $t-t_0=1,2,3,4$, and $t=4,6,8$. Similar patterns of behavior emerge for the various GEVP types. Without the DR method, increasing either $\delta_t$ or $t-t_0$ tends to decrease the energy, suggesting smaller excited state effects (the statistical errors are relatively large, so we do not draw a strong conclusion). For larger times the effect is smaller. For fixed $t$ we often observe smaller statistical errors as $\delta_t$ increases, though after $\delta_t=5$ the improvement is slight. Increasing $t-t_0$ has little effect. The DR method, on the other hand, shows a dramatic reduction in statistical error but little change after that for the other variables. In either case, there is little dependence on the GEVP type. This is because, in Eq.~\eqref{eq:newEeff}, $E_0^{0,\rm disp} = 2m_\pi$ dominates the statistical error and $\Delta E_0^{\rm eff}(t,t_0)$, which is dependent on GEVP type, is much more precise for the ground state in our measurements.

\begin{figure}
    \centering
    \includegraphics[width=1.\linewidth]{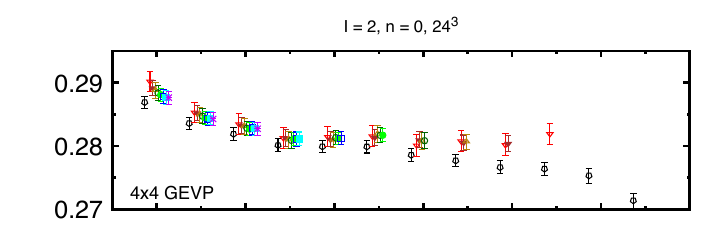}
    \includegraphics[width=1.\linewidth]{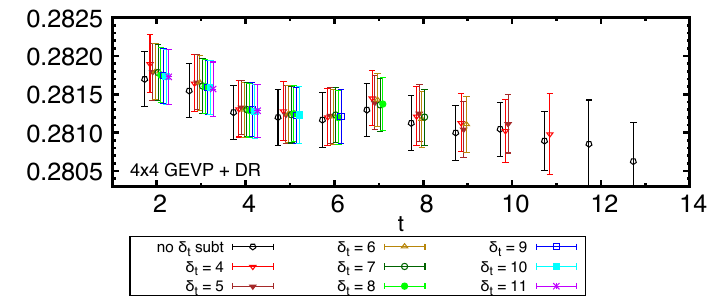}
    \caption{The $I=2$ effective ground state energy with the dispersion relation method (lower) and without (upper). $24^3$, $4\times4$ GEVP, $t-t_0=1$. Without matrix subtraction there is an evident downward shift in the energy, an indication of the ATW effect, for the non-DR result.}
    \label{fig:I=2 4x4 eff energy n=0}
\end{figure}

Fig.~\ref{fig:I=2 4x4 eff energy n=0} shows the effective ground state energy for the $4\times4$ GEVP (our largest basis for $I=2$) with $t-t_0=1$ and matrix subtractions in the range $4\le \delta_t\le11$ with (lower panel) and without (upper panel) the DR method. There are several interesting features. First a clear and stable plateau sets in between $t=4$ and 5 for all non-zero $\delta_t$. Both statistical uncertainties and excited state contamination are significantly reduced by the DR method.
In the upper panel (no matrix subtraction) there is a systematic downward shift for each time slice which grows with increasing $t$. This is a clear indication of the ATW effect, which is eliminated by the matrix subtraction. Perhaps somewhat surprisingly the shift is also eliminated by the DR method. This can be understood as follows. A very similar ATW effect occurs in the single pion case, so the observed cancellation implies that in the interacting case, the two pions do not interact very much. We also observe a small reduction in statistical error as $\delta_t$ increases in the absence of the dispersion method while there is no difference when it is used as the dominant error on the ground state energy after DR is from the error on $m_\pi$. Since contractions were only computed for $t\le16$ to reduce computational cost, the maximum value of $t$ for the effective energy in each case is $t_{max,\delta_t}=15-\delta_t$, as seen in the figure (15 not 16 appears because the effective energy depends on GEVP eigenvalues at $t$ and $t+1$).

\begin{table}[tp]
\centering
\begin{tabular}{|c|ccc|}
\hline
 GEVP & \multicolumn{3}{c|}{fit range} \\
 type & 4--10 & 5--10 & 6--10\\
\hline
$2\times2$ & 0.28132(34)[0.7] & 0.28129(34)[0.8] & 0.28129(34)[1.0]\\
$3\times3$ & 0.28131(34)[0.7] & 0.28129(34)[0.7] & 0.28129(34)[0.9]\\
$4\times4$ & 0.28128(34)[0.6] & 0.28126(34)[0.7] & 0.28126(34)[0.9]\\
RGEVP & 0.28130(34)[0.6] & 0.28128(34)[0.7] & 0.28128(34)[0.9]\\
\hline
\end{tabular}
\caption{Fit results for two-pion energy of the $I=2$ ground state on the $24^3$ lattice with various fit ranges and GEVP methods.  We choose parameters $\delta_t = 5$ and $t-t_0 = 1$.  The $\chi^2/d.o.f.$ is shown in the square brackets.  The re-basing matrix is calculated as: $4\times4\to 3\times3$ at $t_0 = 4$.}
\label{tab:fitres_Epipi_I2_n0_L24}
\end{table}

In Tab.~\ref{tab:fitres_Epipi_I2_n0_L24} we tabulate the fit results for the ground state energy using the dispersion relation method for different fit ranges and GEVP types and display them in Fig.~\ref{fig:I=2 fits 0}. In this work all fits are correlated fit to a constant ($t$-independent) parameter, performed separately for each individual effective energy.  The stability of the fits is quite robust for all values of $t$ and GEVP types considered in the figure.

\begin{figure}
    \centering
    \includegraphics[width=1.\linewidth]{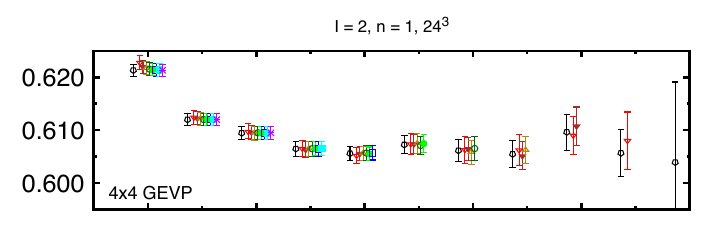}
    \includegraphics[width=1.\linewidth]{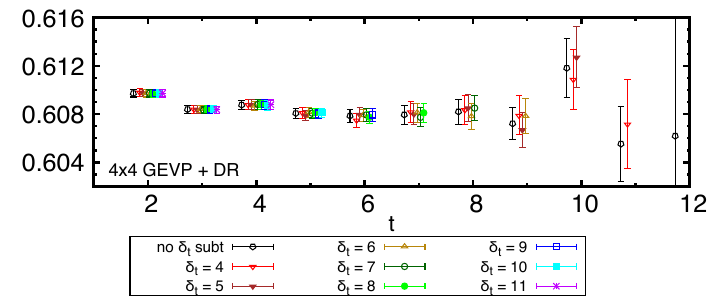}
    \caption{The $I=2$ effective first excited state energy with (lower) and without (upper) the dispersion relation method. $24^3$, $4\times4$ GEVP, $t-t_0=1$. The ATW effect is not observed, $c.f.$ Fig.~\ref{fig:I=2 4x4 eff energy n=0}.}
    \label{fig:I=2 4x4 eff energy n=1}
\end{figure}
\begin{table}[tp]
\centering
\begin{tabular}{|c|ccc|}
\hline
 GEVP & \multicolumn{3}{c|}{fit range} \\
 type & 4--9 & 5--9 & 6--9\\
\hline
$2\times2$ & 0.60859(27)[1.3] & 0.60824(31)[0.3] & 0.60824(43)[0.4]\\
$3\times3$ & 0.60821(27)[0.9] & 0.60793(31)[0.3] & 0.60799(43)[0.4]\\
$4\times4$ & 0.60817(27)[0.8] & 0.60789(31)[0.3] & 0.60797(43)[0.3]\\
RGEVP & 0.60816(27)[0.9] & 0.60789(31)[0.3] & 0.60797(43)[0.3]\\
\hline
\end{tabular}
\caption{Fit results for two-pion energy of the $I=2$ first excited state on the $24^3$ lattice with various fit ranges and GEVP methods.  We choose parameters $\delta_t = 5$ and $t-t_0 = 1$.  The $\chi^2/d.o.f.$ is shown in the square brackets.  The re-basing matrix is calculated as: $4\times4\to 3\times3$ at $t_0 = 4$.}
\label{tab:fitres_Epipi_I2_n1_L24}
\end{table}

The behavior of the first excited state is similar to the ground state. 
In Fig.~\ref{fig:I=2 4x4 eff energy n=1} we again observe a stable plateau beginning with $t=5$; however an ATW effect cannot be seen. But we do see excited state effects, especially without the DR method, and statistical errors are reduced significantly by the DR method. There is little dependence on $\delta_t$.
The small deviation visible at $t_{min}=4$ in the fit results in Tab.~\ref{tab:fitres_Epipi_I2_n1_L24} and Fig.~\ref{fig:I=2 fits 1} may be associated with the slight blip observed in the effective energy at $t=4$ in Fig.~\ref{fig:I=2 4x4 eff energy n=1}. The $2\times2$ GEVP may be systematically high, especially for $t=4$.

\begin{figure}
    \centering
    \includegraphics[width=1.\linewidth]{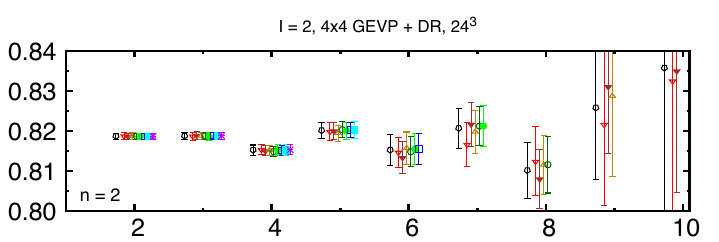}
    \includegraphics[width=1.\linewidth]{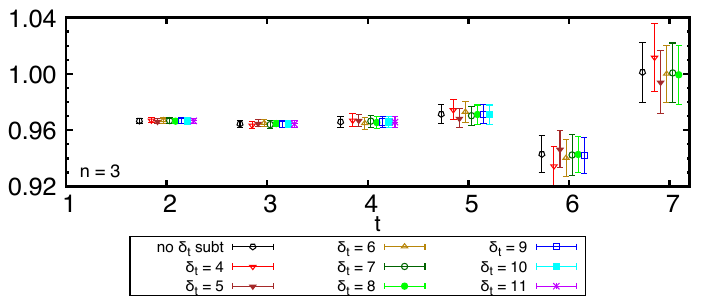}
    \caption{The $I=2$ effective second (upper) and third (lower) excited state energies with the dispersion relation method. $24^3$, $4\times4$ GEVP, $t-t_0=1$. The ATW effect is not observed, $c.f.$ Fig.~\ref{fig:I=2 4x4 eff energy n=0}.}
    \label{fig:I=2 4x4 eff energy n=2,3}
\end{figure}
\begin{table}[tp]
\centering
\begin{tabular}{|c|ccc|}
\hline
 GEVP & \multicolumn{3}{c|}{fit range} \\
 type & 3--9 & 4--9 & 5--9\\
\hline
$3\times3$ & 0.81855(58)[1.5] & 0.81743(92)[1.3] & 0.8183(15)[1.5]\\
$4\times4$ & 0.81743(56)[1.5] & 0.81637(89)[1.3] & 0.8178(15)[1.3]\\
RGEVP & 0.81753(56)[1.4] & 0.81665(87)[1.3] & 0.8185(15)[1.0]\\
\hline
\end{tabular}
\caption{Fit results for two-pion energy of the $I=2$ second excited state on the $24^3$ lattice with various fit ranges and GEVP methods.  We choose parameters $\delta_t = 5$ and $t-t_0 = 1$.  The $\chi^2/d.o.f.$ is shown in the square brackets.  The re-basing matrix is calculated as: $4\times4\to 3\times3$ at $t_0 = 4$.}
\label{tab:fitres_Epipi_I2_n2_L24}
\end{table}

\begin{table}[tp]
\centering
\begin{tabular}{|c|cc|}
\hline
 GEVP & \multicolumn{2}{c|}{fit range} \\
 type & 3--5 & 4--5\\
\hline
$4\times4$ & 0.9658(17)[0.1] & 0.9674(36)[0.0]\\
\hline
\end{tabular}
\caption{Fit results for two-pion energy of the $I=2$ third excited state on the $24^3$ lattice with various fit ranges and GEVP methods.  We choose parameters $\delta_t = 5$ and $t-t_0 = 1$.  The $\chi^2/d.o.f.$ is shown in the square brackets.}
\label{tab:fitres_Epipi_I2_n3_L24}
\end{table}

The second and third excited state energies are shown in Fig.~\ref{fig:I=2 4x4 eff energy n=2,3} and listed for various fit ranges and GEVP types in Tabs.~\ref{tab:fitres_Epipi_I2_n2_L24} and \ref{tab:fitres_Epipi_I2_n3_L24} (see also Fig.~\ref{fig:I=2 fits 2} and Fig.~\ref{fig:I=2 fits 3}). In the upper panel the energy fluctuates down at $t=4$ which leads to an elevated $\chi^2$ in the fit. The same happens at $t=6$ for the third excited state. Again, there is little dependence on $\delta_t$ or GEVP type, except that the $3\times3$ GEVP is a bit high for small $t_{min}$ in the fit (Figs.~\ref{fig:I=2 fits 2} and \ref{fig:I=2 fits 3}).With four interpolating operators for $I=2$, the third excited state is as far as we can go.

\begin{figure}
    \centering
    \includegraphics[width=1.\linewidth]{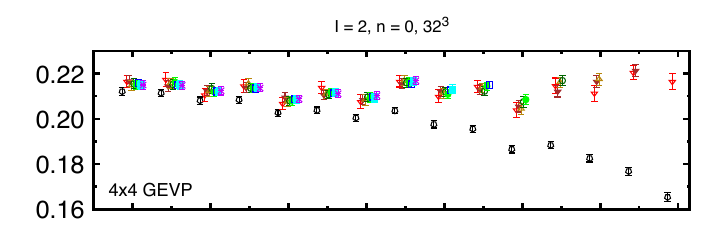}
    \includegraphics[width=1.\linewidth]{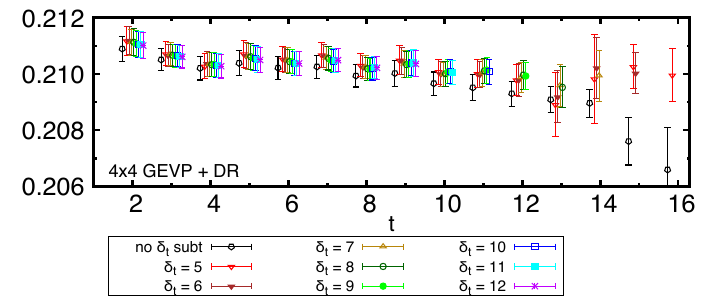}
    \caption{The $I=2$ effective ground state energy with (lower) and without (upper) the dispersion relation method. $32^3$, $4\times4$ GEVP, $t-t_0=1$. Their is a pronounced ATW effect without matrix subtraction (upper panel). The dispersion relation method reduces, but does not entirely eliminate the effect (lower panel).}
    \label{fig:I=2 4x4 eff energy 32c n=0}
\end{figure}
\begin{table}[tp]
\centering
\begin{tabular}{|c|ccc|}
\hline
 GEVP & \multicolumn{3}{c|}{fit range} \\
 type & 4--9 & 5--9 & 6--9\\
\hline
$2\times2$ & 0.21041(38)[0.8] & 0.21041(38)[0.9] & 0.21039(38)[0.7]\\
$3\times3$ & 0.21041(37)[0.8] & 0.21042(37)[1.0] & 0.21039(37)[0.7]\\
$4\times4$ & 0.21039(37)[0.8] & 0.21039(37)[1.0] & 0.21036(37)[0.7]\\
RGEVP & 0.21041(37)[0.8] & 0.21041(37)[1.0] & 0.21038(37)[0.7]\\
\hline
\end{tabular}
\caption{Fit results for two-pion energy of the $I=2$ ground state on the $32^3$ lattice with various fit ranges and GEVP methods.  We choose parameters $\delta_t = 8$ and $t-t_0 = 1$.  The $\chi^2/d.o.f.$ is shown in the square brackets.  The re-basing matrix is calculated as: $4\times4\to 3\times3$ at $t_0 = 5$.}
\label{tab:fitres_Epipi_I2_n0_L32}
\end{table}

The effective ground state energy computed on the $32^3$ lattice is shown in Fig.~\ref{fig:I=2 4x4 eff energy 32c n=0}. The ATW effect is even more pronounced, as expected, since the physical time extent is smaller compared to the $24^3$ lattice, and it is significantly reduced again, but not completely eliminated, by the DR method. The lower panel indicates that the effective energy even with the matrix subtraction might decrease with increasing time from $t=10$ and it could mean there are also higher-order ATW effects that cannot be removed by the matrix subtraction in Eq.~\eqref{eq:matdt_subt}.  Since this tendency is not statistically significant and it is still possible that this is due to statistical fluctuation, we will conclude this point after increasing statistics but do not use data point at $t\ge10$ for our fits.  Again a large reduction in the statistical error and excited state contamination occurs with the dispersion relation method where a plateau emerges beginning at $t=4$. Fig.~\ref{fig:I=2 L32 fits 0} shows fits for several ranges and GEVP types. Results are tabulated in Tab.~\ref{tab:fitres_Epipi_I2_n0_L32}. There is little dependence on $t_{min}$ or GEVP type. 

\begin{figure}
    \centering
    \includegraphics[width=1.\linewidth]{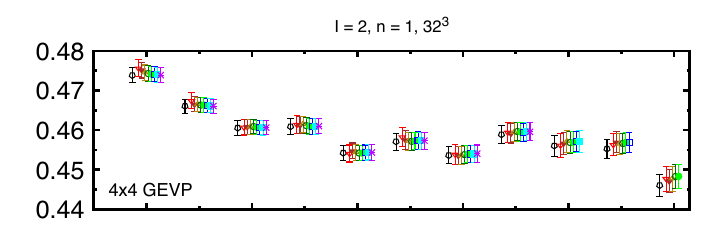}
    \includegraphics[width=1.\linewidth]{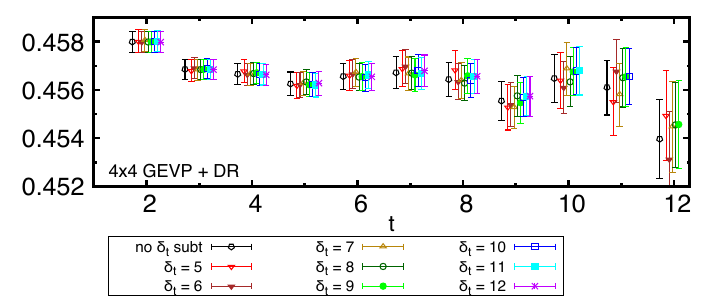}
    \caption{The $I=2$ effective first excited state energy with (lower) and without (upper) the dispersion relation method. $32^3$, $4\times4$ GEVP, $t-t_0=1$.}
    \label{fig:I=2 4x4 eff energy 32c n=1}
\end{figure}
\begin{table}[tp]
\centering
\begin{tabular}{|c|ccc|}
\hline
 GEVP & \multicolumn{3}{c|}{fit range} \\
 type & 4--10 & 5--10 & 6--10\\
\hline
$2\times2$ & 0.45698(37)[0.7] & 0.45669(41)[0.4] & 0.45665(47)[0.5]\\
$3\times3$ & 0.45654(36)[0.3] & 0.45639(41)[0.3] & 0.45638(47)[0.3]\\
$4\times4$ & 0.45648(36)[0.3] & 0.45636(41)[0.2] & 0.45638(47)[0.3]\\
RGEVP & 0.45648(36)[0.3] & 0.45637(41)[0.3] & 0.45639(46)[0.3]\\
\hline
\end{tabular}
\caption{Fit results for two-pion energy of the $I=2$ first excited state on the $32^3$ lattice with various fit ranges and GEVP methods.  We choose parameters $\delta_t = 8$ and $t-t_0 = 1$.  The $\chi^2/d.o.f.$ is shown in the square brackets.  The re-basing matrix is calculated as: $4\times4\to 3\times3$ at $t_0 = 5$.}
\label{tab:fitres_Epipi_I2_n1_L32}
\end{table}

The situation is similar for the first excited state energy (see Fig.~\ref{fig:I=2 4x4 eff energy 32c n=1}) except the ATW effect is not detectable, the same as we saw for the $24^3$ lattice. The plateau begins at $t=4$ or 5 with the DR method. The DR method removes a bit of jitter as well, which is probably due to poor statistics.  Fit results are summarized in Table~\ref{tab:fitres_Epipi_I2_n1_L32} and Fig.~\ref{fig:I=2 L32 fits 1}.

\begin{figure}
    \centering
    \includegraphics[width=1.\linewidth]{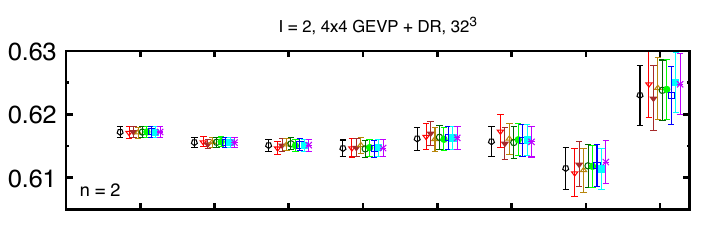}
    \includegraphics[width=1.\linewidth]{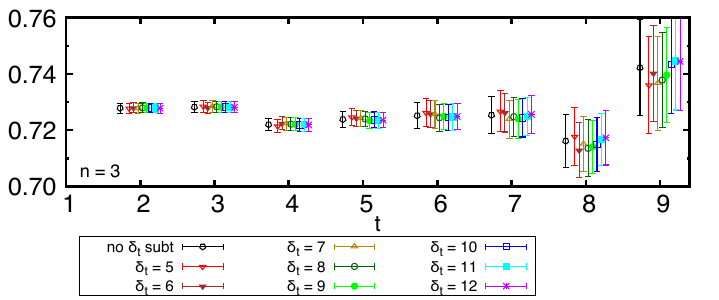}
    \caption{The $I=2$ effective second (upper) and third (lower) excited state energies with the dispersion relation method. $32^3$, $4\times4$ GEVP, $t-t_0=1$.}
    \label{fig:I=2 4x4 eff energy 32c n=2,3}
\end{figure}
\begin{table}[tp]
\centering
\begin{tabular}{|c|ccc|}
\hline
 GEVP & \multicolumn{3}{c|}{fit range} \\
 type & 3--7 & 4--7 & 5--7\\
\hline
$3\times3$ & 0.61671(59)[0.5] & 0.61605(83)[0.3] & 0.6158(12)[0.4]\\
$4\times4$ & 0.61548(59)[0.3] & 0.61523(80)[0.3] & 0.6151(12)[0.4]\\
RGEVP & 0.61552(58)[0.3] & 0.61522(81)[0.3] & 0.6151(12)[0.4]\\
\hline
\end{tabular}
\caption{Fit results for two-pion energy of the $I=2$ second excited state on the $32^3$ lattice with various fit ranges and GEVP methods.  We choose parameters $\delta_t = 8$ and $t-t_0 = 1$.  The $\chi^2/d.o.f.$ is shown in the square brackets.  The re-basing matrix is calculated as: $4\times4\to 3\times3$ at $t_0 = 5$.}
\label{tab:fitres_Epipi_I2_n2_L32}
\end{table}

\begin{table}[tp]
\centering
\begin{tabular}{|c|ccc|}
\hline
 GEVP & \multicolumn{3}{c|}{fit range} \\
 type & 4--7 & 5--7 & 6--7\\
\hline
$4\times4$ & 0.7230(17)[0.1] & 0.7241(25)[0.0] & 0.7245(43)[0.0]\\
\hline
\end{tabular}
\caption{Fit results for two-pion energy of the $I=2$ third excited state on the $32^3$ lattice with various fit ranges and GEVP methods.  We choose parameters $\delta_t = 8$ and $t-t_0 = 1$.  The $\chi^2/d.o.f.$ is shown in the square brackets.}
\label{tab:fitres_Epipi_I2_n3_L32}
\end{table}

The energies for higher excited states are shown in Fig.~\ref{fig:I=2 4x4 eff energy 32c n=2,3}. Fits are summarized in Tabs.~\ref{tab:fitres_Epipi_I2_n2_L32} and~\ref{tab:fitres_Epipi_I2_n3_L32} and shown in Fig.~\ref{fig:I=2 L32 fits 2} and~\ref{fig:I=2 L32 fits 3}.

\subsubsection{I=0}

This case is statistically noisier than $I=2$ due to the disconnected diagrams of the correlation function as well as to the coupling with the vacuum state. In Fig.~\ref{fig:I=0 n=0 energies} the effective energies for the ground state are shown for several GEVP types computed on the $24^3$ ensemble. Values of $t-t_0$ range from 1 to 4 with $3\le t \le 7$ and $\delta_t=2,5,8$. Like $I=2$, there is little dependence on GEVP type. The dispersion relation method reduces excited state contamination and statistical errors as does increasing $\delta_t$. The effect of $t-t_0$ is less clear, though it appears it may also reduce excited state effects. 

\begin{figure}
    \centering
    \includegraphics[width=1.\linewidth]{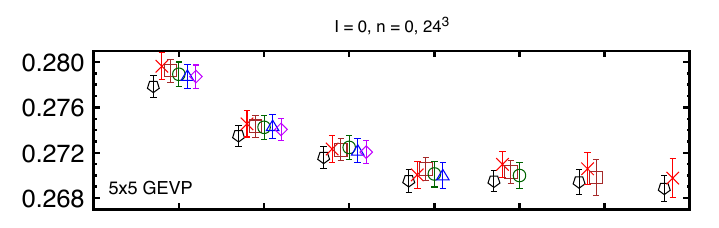}
    \includegraphics[width=1.\linewidth]{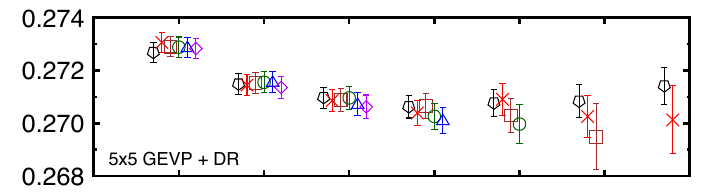}
    \includegraphics[width=1.\linewidth]{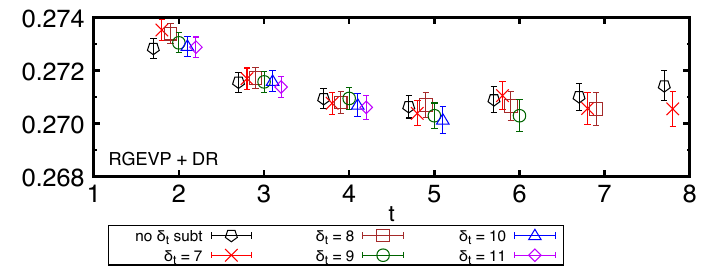}
    \caption{The $I=0$ effective ground state energy. $24^3$, $5\times5$ GEVP, $t-t_0=1$ with (middle, lower) and without (upper) the dispersion relation method. The RGEVP result is shown in the lower panel. The re-basing from $5\times5\to3\times3$ uses GEVP eigenvectors at $t_0=4$.}
    \label{fig:I=0 5x5 eff energy n=0}
\end{figure}

Fig.~\ref{fig:I=0 5x5 eff energy n=0} shows the effective ground state energy for the $5\times5$ GEVP with $t-t_0=1$ and matrix subtractions in the range $7\le \delta_t\le11$, with and without the dispersion relation method and for the RGEVP. A small ATW effect may be visible in the upper panel when no matrix subtraction is performed, and it is largely absent in the middle and lower panels, showing again that the dispersion relation method largely eliminates it. We also observe that the RGEVP makes a moderate improvement on the statistical errors for larger times (lower panel).

\begin{table}[tp]
\centering
\begin{tabular}{|c|ccc|}
\hline
 GEVP & \multicolumn{3}{c|}{fit range} \\
 type & 3--8 & 4--8 & 5--8\\
\hline
$3\times3$ & 0.27116(38)[1.8] & 0.27082(41)[0.9] & 0.27058(47)[0.8]\\
$4\times4$ & 0.27115(38)[1.5] & 0.27087(41)[0.9] & 0.27067(46)[0.9]\\
$5\times5$ & 0.27104(38)[1.6] & 0.27074(41)[1.0] & 0.27054(46)[1.1]\\
RGEVP & 0.27122(38)[3.6] & 0.27069(41)[1.1] & 0.27053(46)[1.3]\\
\hline
\end{tabular}
\caption{Fit results for two-pion energy of the $I=0$ ground state on the $24^3$ lattice with various fit ranges and GEVP methods.  We choose parameters $\delta_t = 7$ and $t-t_0 = 1$.  The $\chi^2/d.o.f.$ is shown in the square brackets.  The re-basing matrix is calculated as: $5\times5\to 3\times3$ at $t_0 = 4$.}
\label{tab:fitres_Epipi_I0_n0_L24}
\end{table}

Fit results are summarized in Tab.~\ref{tab:fitres_Epipi_I0_n0_L24} and Fig.~\ref{fig:I=0 fits 0}. There appears to be a small systematic shift with the minimum time separation in the fit, $t_{\rm min}$, but it is within the statistical uncertainty. 

\begin{figure}
    \centering
    \includegraphics[width=1.\linewidth]{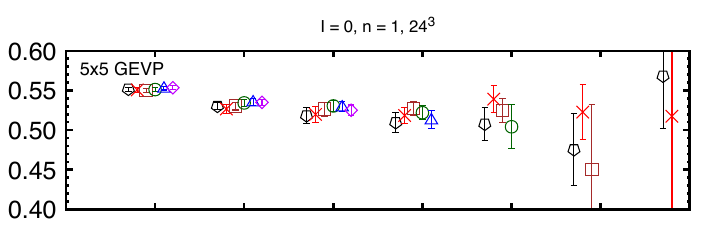}
    \includegraphics[width=1.\linewidth]{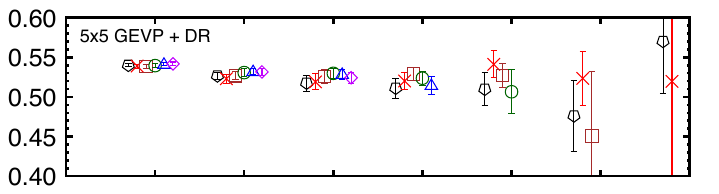}
    \includegraphics[width=1.\linewidth]{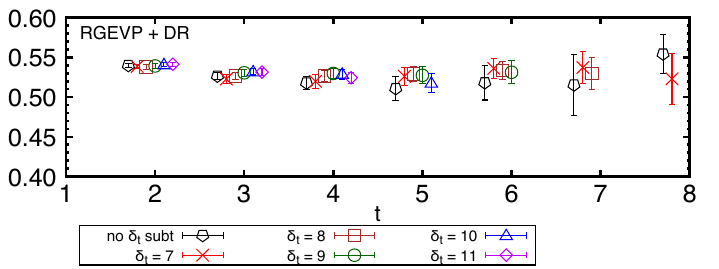}
    \caption{The $I=0$ effective first excited state energy with (middle, lower) and without (upper) the dispersion relation method. $24^3$, $5\times5$ (middle, upper) and $3\times3$ (lower) GEVP, $t-t_0=1$. Note the significant jump from $t=6$ to 7 for $\delta_t=8$ in upper two panels.}
    \label{fig:I=0 5x5 eff energy n=1}
\end{figure}
The first excited state energies are summarized in Tab.~\ref{tab:efm_I0_n1_L24} and Fig.~\ref{fig:I=0 n=1 energies} for a wide range of parameters. In Fig.~\ref{fig:I=0 5x5 eff energy n=1} we show the first excited state energy for the $5\times5$ GEVP and $t-t_0=1$. A plateau begins at $t=3$ or 4, and an interesting systematic begins to emerge as well which is even more pronounced for the second excited state, as we see shortly. A careful inspection of Fig.~\ref{fig:I=0 5x5 eff energy n=1} shows a systematic downward drift of the energy for each $t_{max,\delta_t}$. It is most visible between $t=6$ and 7 for $\delta_t=8$, but a smaller shift appears at smaller times for large values of $\delta_t$. The effect is absent for the RGEVP (lower panel), and leads us to believe it is due to a breakdown of the GEVP at large $t$ due to large statistical errors in the correlation matrix.

\begin{table}
\centering
\begin{tabular}{|c|ccc|}
\hline
 GEVP & \multicolumn{3}{c|}{fit range} \\
 type & 3--6 & 4--6 & 5--6\\
\hline
$3\times3$ & 0.5319(45)[0.1] & 0.5306(59)[0.1] & 0.5298(72)[0.1]\\
$4\times4$ & 0.5302(45)[0.4] & 0.5296(64)[0.6] & 0.5255(94)[0.8]\\
$5\times5$ & 0.5304(45)[0.3] & 0.5296(61)[0.5] & 0.5262(83)[0.6]\\
RGEVP & 0.5308(44)[0.1] & 0.5302(65)[0.1] & 0.528(11)[0.1]\\
\hline
\end{tabular}
\caption{Fit results for two-pion energy of the $I=0$ first excited state on the $24^3$ lattice with various fit ranges and GEVP methods.  We choose parameters $\delta_t = 9$ and $t-t_0 = 1$.  The $\chi^2/d.o.f.$ is shown in the square brackets.  The re-basing matrix is calculated as: $5\times5\to 3\times3$ at $t_0 = 3$.}
\label{tab:fitres_Epipi_I0_n1_L24}
\end{table}

Fit results are given in Tab.~\ref{tab:fitres_Epipi_I0_n1_L24} and displayed in Fig.~\ref{fig:I=0 fits 1}. There is no detectable difference as $t_{min}$ or the GEVP type varies. 

\begin{figure}
    \centering
    \includegraphics[width=1.\linewidth]{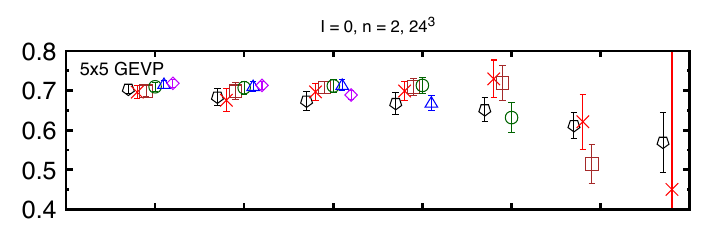}
    \includegraphics[width=1.\linewidth]{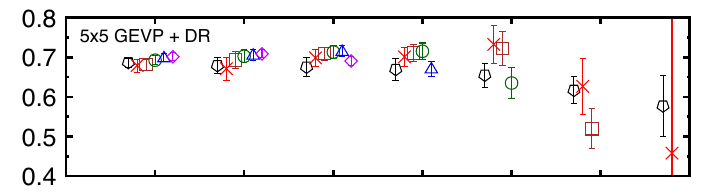}
    \includegraphics[width=1.\linewidth]{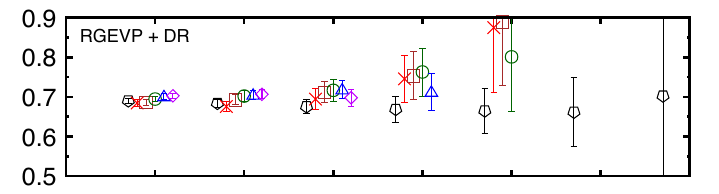}
    \includegraphics[width=1.\linewidth]{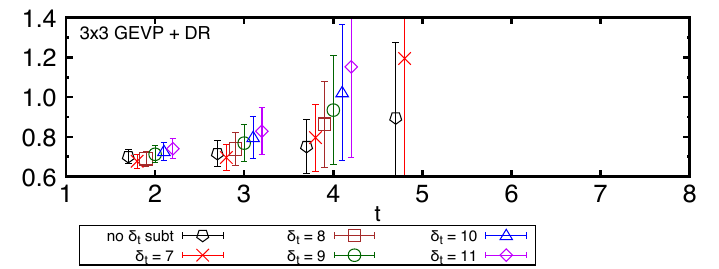}
    \caption{The $I=0$ effective second excited state energy with (2nd, 3rd, 4th) and without (1st) the dispersion relation method. $24^3$, $5\times5$ GEVP (1st, 2nd), $3\times3$ RGEVP (3rd) and $3\times3$ pure GEVP (4th), $t-t_0=1$. Note the significant jump from $t=6$ to 7 for $\delta_t=8$ in upper two panels.  Statistical error on $3\times3$ GEVP is significantly large compared to the others.}
    \label{fig:I=0 5x5 eff energy n=2}
\end{figure}
The second excited state is less well resolved statistically, so we restrict our focus to relatively small times (see Fig.~\ref{fig:I=0 5x5 eff energy n=2}). There is little difference without or with the DR method (upper and middle panels); however the downward trend observed for the first excited state is even more visible here and is also removed by the RGEVP although the statistical errors increase significantly. Fit results are listed in Tab.~\ref{tab:fitres_Epipi_I0_n2_L24} and plotted in Fig.~\ref{fig:I=0 fits 2}.
\begin{table}[tp]
\centering
\begin{tabular}{|c|cc|}
\hline
 GEVP & \multicolumn{2}{c|}{fit range} \\
 type & 3--5 & 4--5\\
\hline
$3\times3$ & 0.713(40)[1.4] & 0.841(89)[0.1]\\
$4\times4$ & 0.709(14)[0.4] & 0.714(15)[0.0]\\
$5\times5$ & 0.710(13)[0.3] & 0.714(14)[0.0]\\
RGEVP & 0.695(12)[0.7] & 0.699(23)[1.3]\\
\hline
\end{tabular}
\caption{Fit results for two-pion energy of the $I=0$ second excited state on the $24^3$ lattice with various fit ranges and GEVP methods.  We choose parameters $\delta_t = 9$ and $t-t_0 = 1$.  The $\chi^2/d.o.f.$ is shown in the square brackets.  The re-basing matrix is calculated as: $5\times5\to 3\times3$ at $t_0 = 1$.}
\label{tab:fitres_Epipi_I0_n2_L24}
\end{table}

Another dramatic difference happens when we reduce the GEVP to $3\times3$ ($4\times4$ is essentially the same as $5\times5$) without re-basing (see lower panel of Fig~\ref{fig:I=0 5x5 eff energy n=2}). The noise increases dramatically because of the significant coupling between the second excited state and the $\pi\pi(011)$ operator which is excluded in the $3\times3$ analysis. This is similar to the effect observed in Refs.~\cite{RBC:2020kdj} and~\cite{RBC:2021acc} where adding a noisy scalar bilinear operator reduced the statistical error of (and excited state contamination in) the ground state, although it is the $\pi\pi(011)$ operator that is not included in the $3\times3$ but in the $4\times4$ and $5\times5$ analyses in this case. 

The GEVP eigenvalues are too noisy even for short times to extract meaningful energies for the higher energy states. They will have to wait for better statistics in the future.

\begin{figure}
    \centering
    \includegraphics[width=1.\linewidth]{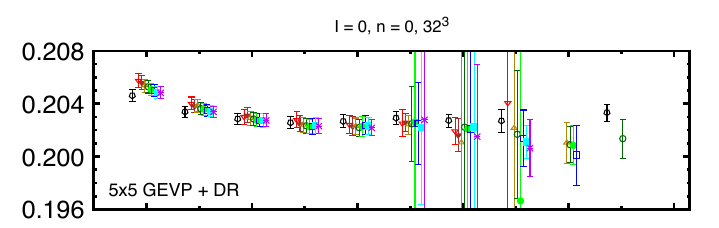}
    \includegraphics[width=1.\linewidth]{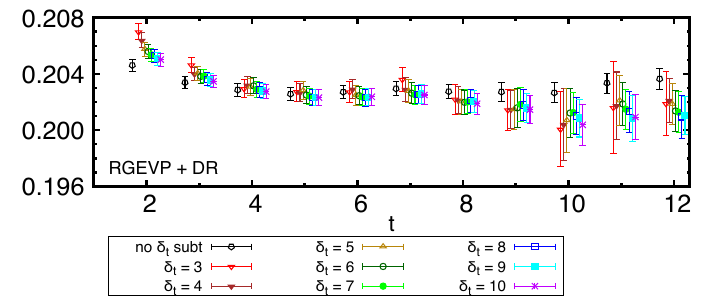}
    \caption{The $I=0$ effective ground state energy with the dispersion relation method. $32^3$, $5\times5$ GEVP (upper) and $2\times2$ RGEVP (lower), $t-t_0=1$. A multi-step RGEVP is performed at $t_0=1$, 2, and 4 to go from $5\times5\to2\times2$.}
    \label{fig:I=0 5x5 eff energy 32c n=0}
\end{figure}
\begin{table}[tp]
\centering
\begin{tabular}{|c|ccc|}
\hline
 GEVP & \multicolumn{3}{c|}{fit range} \\
 type & 4--9 & 5--9 & 6--9\\
\hline
$3\times3$ & 0.20268(44)[0.4] & 0.20252(48)[0.3] & 0.20225(57)[0.2]\\
$4\times4$ & 0.20279(48)[0.4] & 0.20259(52)[0.3] & 0.20234(62)[0.2]\\
$5\times5$ & 0.20288(50)[0.4] & 0.20265(55)[0.2] & 0.20237(67)[0.1]\\
RGEVP & 0.20272(51)[0.5] & 0.20248(56)[0.4] & 0.20235(66)[0.5]\\
\hline
\end{tabular}
\caption{Fit results for two-pion energy of the $I=0$ ground state on the $32^3$ lattice with various fit ranges and GEVP methods.  We choose parameters $\delta_t = 5$ and $t-t_0 = 1$.  The $\chi^2/d.o.f.$ is shown in the square brackets.  The re-basing matrix is calculated as: $5\times5\to 4\times4$ at $t_0 = 1$, $4\times4\to 3\times3$ at $t_0 = 2$ and $3\times3\to 2\times2$ at $t_0 = 4$.}
\label{tab:fitres_Epipi_I0_n0_L32}
\end{table}

On the $32^3$ ensemble, the pattern repeats except the effective energies are even noisier as seen in Fig.~\ref{fig:I=0 5x5 eff energy 32c n=0}.  It is interesting to note that the statistical error in the latter case is smaller at large time when no matrix subtraction is performed. Fits are summarized in Tab.~\ref{tab:fitres_Epipi_I0_n0_L32} and Fig.~\ref{fig:I=0 L32 fits 0}.

\begin{figure}
    \centering
    \includegraphics[width=1.\linewidth]{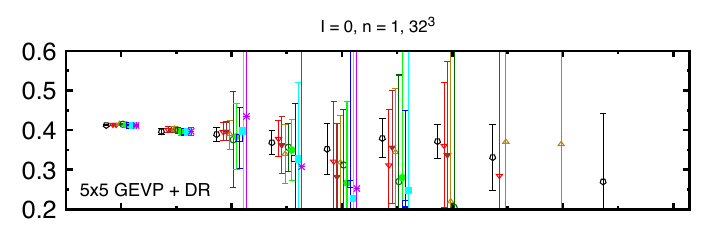}
    \includegraphics[width=1.\linewidth]{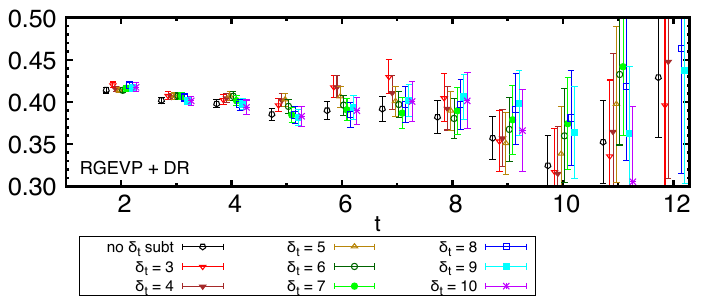}
    \caption{The $I=0$ effective first excited state energy for the $5\times5$ GEVP (upper) and $2\times2$ RGEVP (lower) with the dispersion relation method. $32^3$, $t-t_0=1$.}
    \label{fig:I=0 5x5 eff energy 32c n=1}
\end{figure}
\begin{table}[tp]
\centering
\begin{tabular}{|c|ccc|}
\hline
 GEVP & \multicolumn{3}{c|}{fit range} \\
 type & 3--8 & 4--8 & 5--8\\
\hline
$3\times3$ & 0.3989(60)[0.2] & 0.399(12)[0.3] & 0.394(24)[0.4]\\
$4\times4$ & 0.3998(57)[0.1] & 0.399(14)[0.2] & 0.390(41)[0.2]\\
$5\times5$ & 0.3997(59)[0.2] & 0.398(17)[0.2] & 0.392(28)[0.2]\\
RGEVP & 0.4054(28)[0.5] & 0.4052(41)[0.6] & 0.3989(66)[0.3]\\
\hline
\end{tabular}
\caption{Fit results for two-pion energy of the $I=0$ first excited state on the $32^3$ lattice with various fit ranges and GEVP methods.  We choose parameters $\delta_t = 5$ and $t-t_0 = 1$.  The $\chi^2/d.o.f.$ is shown in the square brackets.  The re-basing matrix is calculated as: $5\times5\to 4\times4$ at $t_0 = 1$, $4\times4\to 3\times3$ at $t_0 = 2$ and $3\times3\to 2\times2$ at $t_0 = 4$.}
\label{tab:fitres_Epipi_I0_n1_L32}
\end{table}

After $t=3$ or 4, the statistical errors get very large for the first excited state even with the DR method (see Fig.~\ref{fig:I=0 5x5 eff energy 32c n=1}).
However, we observe a large reduction in the statistical error using the RGEVP compared to the standard GEVP in this case. Re-basings to go from  $5\times5$ down to $2\times2$ are implemented at $t_0=1$, $t_0=2$ and $t_0=4$. Presumably the improvement occurs because the overlap of the higher states with the lower states is more and more unresolved with increasing $t$, adding only noise to the GEVP. Fit results are summarized in Tab.~\ref{tab:fitres_Epipi_I0_n1_L32} and Fig.~\ref{fig:I=0 L32 fits 1}. 

\begin{figure}
    \centering
    \includegraphics[width=1.\linewidth]{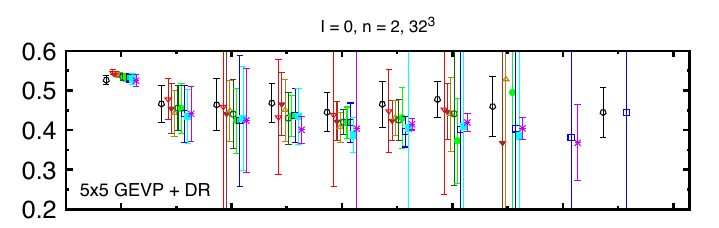}
    \includegraphics[width=1.\linewidth]{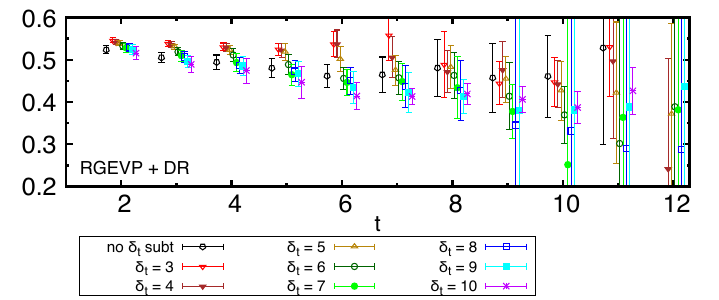}
    \caption{The $I=0$ effective second excited state energy for the $5\times5$ GEVP (upper) and $3\times3$ RGEVP (lower) with the dispersion relation method. $32^3$, $t-t_0=1$.}
    \label{fig:I=0 5x5 eff energy 32c n=2}
\end{figure}
\begin{table}[tp]
\centering
\begin{tabular}{|c|ccc|}
\hline
 GEVP & \multicolumn{3}{c|}{fit range} \\
 type & 3--6 & 4--6 & 5--6\\
\hline
$3\times3$ & 0.428(19)[0.2] & 0.426(26)[0.3] & 0.417(32)[0.4]\\
$4\times4$ & 0.442(44)[0.0] & 0.450(66)[0.0] & 0.445(80)[0.1]\\
$5\times5$ & 0.417(26)[0.1] & 0.414(32)[0.1] & 0.414(35)[0.3]\\
RGEVP & 0.5296(65)[0.5] & 0.524(11)[0.4] & 0.523(19)[0.9]\\
\hline
\end{tabular}
\caption{Fit results for two-pion energy of the $I=0$ second excited state on the $32^3$ lattice with various fit ranges and GEVP methods.  We choose parameters $\delta_t = 5$ and $t-t_0 = 1$.  The $\chi^2/d.o.f.$ is shown in the square brackets.  The re-basing matrix is calculated as: $5\times5\to 4\times4$ at $t_0 = 1$ and $4\times4\to 3\times3$ at $t_0 = 2$.}
\label{tab:fitres_Epipi_I0_n2_L32}
\end{table}

The second excited state is shown in Fig.~\ref{fig:I=0 5x5 eff energy 32c n=2} where again there is a large reduction in statistical error with the RGEVP. A drift downwards with increasing time is observed for larger values of $\delta_t$ for the $3\times3$ RGEVP, though it is not easy to investigate the systematic error at this point because of the accompanied large statistical errors. Fit results are summarized in Tab.~\ref{tab:fitres_Epipi_I0_n2_L32} and Fig.~\ref{fig:I=0 L32 fits 2}.

\begin{table*}[tbp]
\centering
\begin{tabular}{|c|c|c|c|c|c|r|r|r|}
\hline
 $I$& $n$ & GEVP type & fit range & $\delta_t$ & $\chi^2/d.o.f.$& energy & $\delta_0$ (deg) & $m_\pi a_0^I$ \\
\hline
\multicolumn{9}{|c|}{$24^3$ lattice}\\
\hline
2 & 0 & RGEVP & 4--10 & 5[4--8] & 0.6 & 0.28130(34)& $-0.374(13)(4)$ & $-0.0496(11)(5)$ \\
2 & 1 & RGEVP & 5--9 & 5[4--8] & 0.3 & 0.60789(31)& $-12.33(22)(20)$&\\
2 & 2 & RGEVP & 3--9 & 5[4--8] & 1.4 & 0.81753(56)& $-20.18(43)(54)$&\\
2 & 3 & $4\times4$ & 3--5 & 5[4--8] & 0.1 & 0.9658(17)& $-26.5(2.4)(4)$&\\
0 & 0 & RGEVP & 4--8 & 7[5--9] & 1.1 & 0.27069(41)& & 0.2038(70)(160)\\
0 & 1 & RGEVP & 3--6 & 9[6--9] & 0.1 & 0.5308(44)& 45.1(2.9)(2.1)& \\
0 & 2 & RGEVP & 3--5 & 9[7--9] & 0.7& 0.695(12) & 83(11)(15)&\\
\hline
\hline
\multicolumn{9}{|c|}{$32^3$ lattice}\\
\hline
2 & 0 & RGEVP & 4--9 & 8[4--8] & 0.8 & 0.21041(37) & $-0.424(51)(23)$&$-0.0537(42)(22)$\\
2 & 1 & RGEVP & 4--10 & 8[4--8] & 0.3 & 0.45648(36)& $-13.37(32)(11)$&\\
2 & 2 & RGEVP & 3--7 & 8[4--8] & 0.3 & 0.61552(58) & $-22.90(59)(34)$&\\
2 & 3 & $4\times4$ & 4--7 & 8[4--8] & 0.1 & 0.7230(17)& $-24.2(3.2)(2.1)$&\\
0 & 0 & RGEVP & 5--9 & 5[5--8] & 0.4 & 0.20248(56)& & 0.1947(150)(126)\\
0 & 1 & RGEVP & 4--8 & 5[5--8] & 0.6 & 0.4052(41)& 38.6(3.6)(10.1) &\\
0 & 2 & RGEVP & 4--6 & 5[4--6] & 0.4 & 0.5304(91)& 71(12)(16) & \\
\hline
\end{tabular}
\caption{Summary of two-pion energies and phase shifts ($\delta)$. The energies are determined by correlated $\chi^2$ fits to the effective energies with the DR method.  The last column gives the scattering length in units of the pion mass. The first error is statistical, and the second ($\delta$, $m_\pi a_0$) a systematic error, which is estimated by varying fit range as well as $\delta_t$ in the range shown in the square brackets.  See text for more detail.}
\label{tab: energy phase shift summary}
\end{table*}

In preparation for the next subsection, we now summarize in Tab.~\ref{tab: energy phase shift summary} the energies and corresponding time-slice fit ranges and GEVP types that we will use to compute the phase shifts. In all cases except the third excited state for $I=2$, which is only accessible with the $4\times4$ GEVP, we choose the RGEVP since the noise is usually reduced or unchanged from the ordinary GEVP. In addition we believe it is more robust at large times since poorly resolved elements of the correlation function are avoided. As mentioned already, we stick to small time slices where the signal is better resolved, and we use the results of the DR method throughout.  With the DR method we see negligible dependence on $t-t_0$, so we take it to be 1 in all cases. 

For the $I=2$ ground state we choose 4--10 ($24^3$) and 4--9 ($32^3$) for the fit range. From Figs.~\ref{fig:I=2 fits 0}~and~\ref{fig:I=2 L32 fits 0} one can see that the results are very stable with both the range and GEVP type. 

Similarly for the first excited state we choose 5--9 and 4--10 (in the case of $24^3$ a small excited state effect may be visible at $t=4$). In both cases the $2\times2$ GEVP appears to be a little high. For the second excited state we take 3--9 and 3--7. Finally for the third excited state our ranges are 3--5 and 4--7.

For $I=0$ it is not as straightforward to choose central values due to larger statistical errors, especially for the $32^3$ ensemble, and as we have seen already, small times comprise the set of usable time slices.

From Figs.~\ref{fig:I=0 fits 0}($24^3$) and \ref{fig:I=0 L32 fits 0}($32^3$) one sees small variations with $t_{min}$ for the ground states. We choose fit ranges 4--8 and 5--9, respectively. On the $24^3$ lattice we see no significant variation for the first excited state, so we take 3--6. For $32^3$ the errors are large for all but the RGEVP, and we take 4--8. Finally for the second excited state the fit ranges are chosen to be 3--6 and 4--6. In the latter case we use $\delta_t=3$ since we observed a flatter plateau and smaller statistical errors for this choice. Higher excited states cannot be extracted from our data; improved statistics are needed.

\begin{figure}[hbt]
    \includegraphics[width=\linewidth]{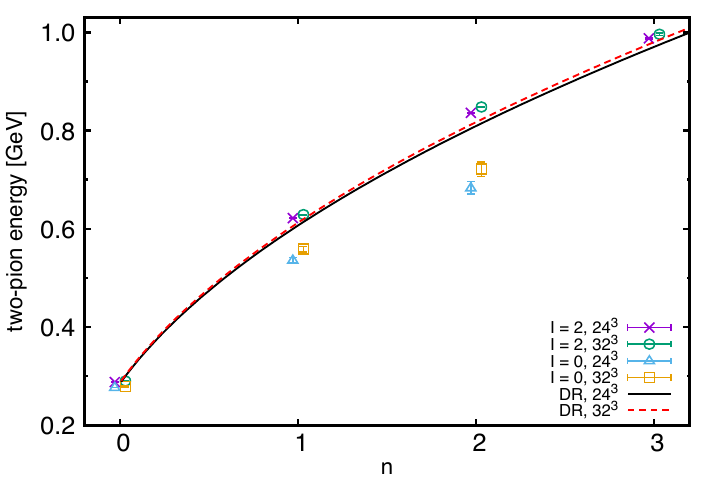}
    \includegraphics[width=\linewidth]{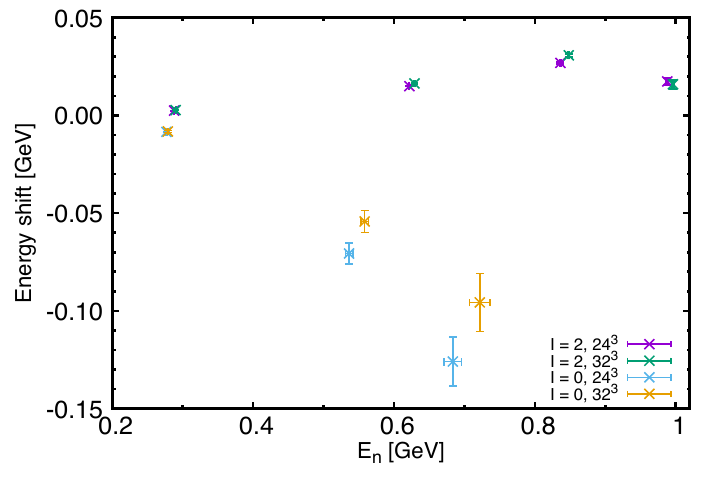}
    \caption{Summary of two-pion energies obtained by correlated fits to DR effective energies (top) and their difference from the corresponding non-interacting energies given by Eq.~\eqref{eq:disp_nonintEpipi} (bottom).  On the top panel the non-interacting energy is also drawn using the continuum dispersion relation~\eqref{eq:disp_nonintEpipi} with $|\vec p_n|^2 = n(2\pi/L)^2$ for both $24^3$ and $32^3$ lattices.}
    \label{fig:energy summary}
\end{figure}

In Fig.~\ref{fig:energy summary} two-pion energies and energy shifts due to pion-pion interactions in finite box are plotted.  The error is statistical only.  The results for the $I=2$ third excited state ($n=3$) show smaller energy shifts than that for the second excited state ($n=2$).  This may indicate there is significant systematic effects.

\subsection{Phase shifts}
\label{sec:phase shifts res}

The phase shifts are computed using the fitted energies described in the previous subsections. Specifically, we take fit values corresponding to Eq.~(\ref{eq:newEeff}) and insert them into Eq.~(\ref{eq:phase shift}). For the $I=2$ ground state with pions at rest, the interacting two-pion energy is above the $2\, m_\pi$ threshold since the interaction is repulsive, and a phase shift is readily calculated using Eq.~(\ref{eq:phase shift}). For $I=0$ the ground state is below the threshold, so the phase shift is purely imaginary.
The phase shifts for each state and various fit ranges for the energies are summarized in Tabs.~\ref{tab:pshift_I2_n0_L24}-\ref{tab:pshift_I0_n2_L32}.

\begin{figure}[tp]
    \includegraphics[width=\linewidth]{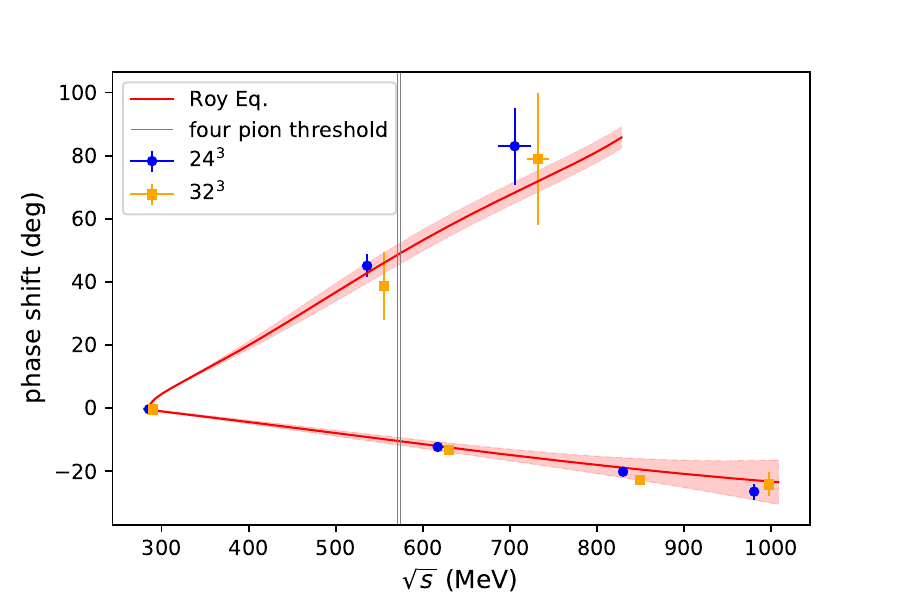}
    \caption{Pion scattering phase shifts for ground and higher excited states. $I=0$ (upper) and $I=2$ (lower). Bars denote statistical and systematic  errors added in quadrature (see Tab.~\ref{tab: energy phase shift summary}. The Roy Equation results~\cite{Ananthanarayan:2000ht} are shown by solid lines and corresponding error bands. Vertical gray lines denote $4 m_\pi$ thresholds for $24^3$ and $32^3$ ensembles, above which the method of determining phase shifts~\cite{Luscher:1990ux} in this work is no longer strictly valid.}
    \label{fig:phase shift}
\end{figure}

Results corresponding to the energies (fit ranges) chosen at the end of the last section are shown in Fig.~\ref{fig:phase shift} and given in Tab.~\ref{tab: energy phase shift summary} for both ensembles. Agreement with the Roy Equation result\footnote{In the figures displaying Roy Equation results we always use the formulae in Ref.~\cite{Ananthanarayan:2000ht}, for technical reasons. In later papers we will switch to the update in ~\cite{Colangelo:2001df}.}~\cite{Ananthanarayan:2000ht,Colangelo:2001df} is observed up to the third (second) excited state for $I=2$(0). Even though, strictly speaking, L\"uscher's method~\cite{Luscher:1990ux} is not valid above the inelastic scattering threshold ($4m_\pi$) which is shown for each ensemble by the vertical gray lines in Fig.~\ref{fig:phase shift}, we see no evidence of a dramatic breakdown above this threshold (or higher pion multiplicities, $K\bar K$, and so on). It will be interesting to see in future works with more precision if a signal of such a breakdown emerges. 

In addition to statistical errors, we also estimate a systematic error due to choice of fit range for the energy and $\delta_t$ in the matrix subtraction to remove the ATW effect. The error is estimated by comparing phase shifts for various fit ranges and $\delta_t$ values and taking half the difference between minimum and maximum central values for a given state. For the comparison of the systematic error, we use the fit ranges shown in Tabs.~\ref{tab:pshift_I2_n0_L24}--\ref{tab:pshift_I0_n2_L32} and a range of $\delta_t$ values shown in the square bracket in Tab.~\ref{tab: energy phase shift summary}.  We remove the fit results that have more than twice as large statistical error as the fit result shown in Tab.~\ref{tab: energy phase shift summary} from the estimation of the systematic error.  In all cases we do not vary the GEVP type and choose the RGEVP except for the $I=2$ third excited state where we need to use the full-size $4\times4$ GEVP. This is because only the RGEVP appears to give reasonable results for the $I=0$ excited states on the $32^3$ lattice, while the other results are less dependent on GEVP type than fit range.  The systematic errors are small for the ground state (less than the statistical error) and become comparable, and even larger than, the statistical error with increasing energy. Since the $I=0$ energies are noisier than their $I=2$ counterparts, so too are their phase shifts.

\begin{figure}[hbt]
    \includegraphics[width=\linewidth]{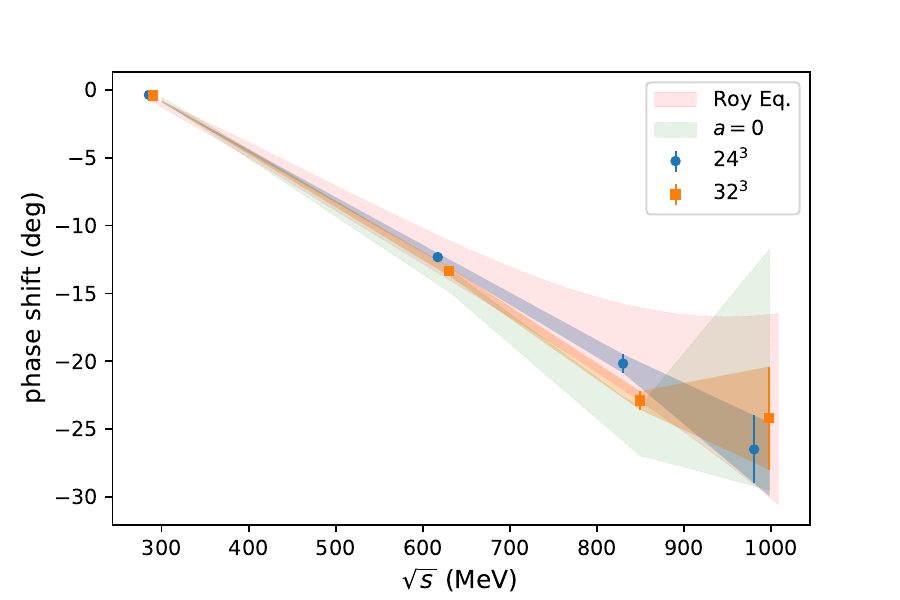}
    \includegraphics[width=\linewidth]{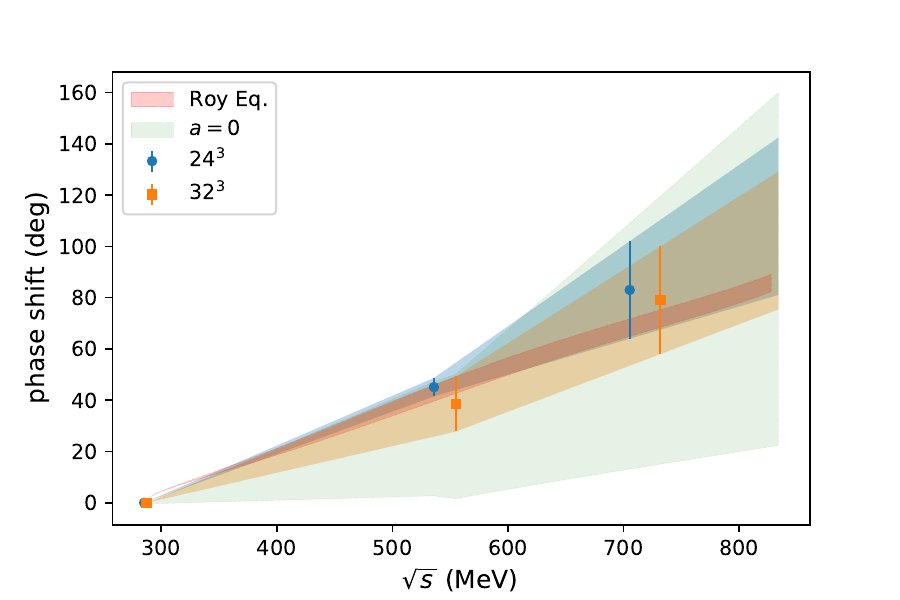}
    \caption{Continuum limit of the pion scattering phase shifts for ground and higher excited states (light green bands) for $I=2$ (upper panel) and $I=0$ (lower).  The continuum extrapolation is performed linearly in $a^2$ after linear interpolation in the energy at fixed lattice spacing. The Roy Equation results~\cite{Ananthanarayan:2000ht} are shown by the light red error bands.}
    \label{fig:phase shift 0 and 2}
\end{figure}

The physical energies corresponding to the various states differ slightly on the two ensembles, so to compare the phase shifts at fixed energy, we do a piece-wise linear interpolation of the phase shifts in discrete energy at fixed lattice spacing and then extrapolate the interpolated phase shifts to the continuum, $a\to0$, at fixed energy. The extrapolation is performed linearly in $a^2$. The results are shown in Fig.~\ref{fig:phase shift 0 and 2}. In each panel the $a\to0$ extrapolation is shown with a green band. The values are compatible with the Roy Equation, albeit within relatively large uncertainties especially for $I=0$. For this study we calculate error bands for the dispersive results using Ref~\cite{Ananthanarayan:2000ht} and we defer a more comprehensive study, including for example the more accurate results of Ref.~\cite{Colangelo:2001df}, to a future publication.

Lastly, we discuss the two-pion scattering length.  Since the relation between the two-pion phase shift $\delta_0$ and scattering length $a_0$ is given as\footnote{The subscript refers to the $s$-wave channel.} 
\begin{equation}
k\cot{\delta_0(k)} = \frac{1}{a_0} + \frac{1}{2}r_0k^2 +O(k^4),
\label{eq:delta_expansion}
\end{equation}
with the effective range parameter $r_0$,
we can calculate the scattering length by
\begin{equation}
a_0 = \frac{\tan{\delta_0(k)}}{k} + O(k^2),
\label{eq:a0}
\end{equation}
with a sufficiently small value of $k$, which can be obtained with the ground state straightforwardly for the $I=2$ channel.
For the $I=0$ channel, $k$ is a pure imaginary number for the ground state and we cannot directly use  Eq.~\eqref{eq:delta_expansion} and the formulae given in Section~\ref{sec:phase shifts theory} for this case.
It is known \cite{Fu:2017apw} that these formulae can be analytically continued with pure imaginary values of $k$. 
With that, we obtain the scattering length as a real number. The values for the scattering lengths for various ranges of fits to the ground state energies are given in Tabs.~\ref{tab:a0_I2_L24}-\ref{tab:a0_I0_L32}. In Tab.~\ref{tab: energy phase shift summary} central values are displayed; systematic errors are computed as before for the phase shifts. A simple linear extrapolation in $a^2$, after combining statistical and systematic errors in quadrature, yields
\begin{eqnarray}
    m_\pi a_0^2 &=& -0.058(11),\\
    m_\pi a_0^0&=& 0.184(47),
\end{eqnarray}
for $I=2$ and $I=0$, respectively.

\begin{figure}[hbt]
    \includegraphics[width=\linewidth]{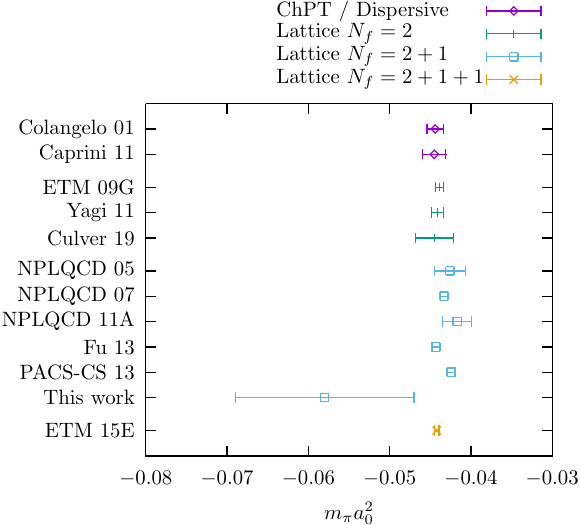}
    \includegraphics[width=\linewidth]{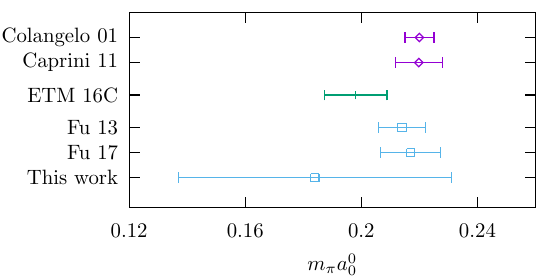}
    \caption{Summary plot of two-pion scattering length $m_\pi a^I_0$ for $I=2$ (upper) and $I=0$ comparing with results from earlier phenomenology and lattice studies summarized by FLAG~\cite{FlavourLatticeAveragingGroupFLAG:2021npn}.  Earlier lattice results quoted with statistical error only are not plotted.  The plotted earlier results are all obtained by a chiral extrapolation (Eqs.~\eqref{eq:I2extrap} and \eqref{eq:I0extrap}) to the physical pion mass using an inputted value of $f_\pi$ and lattice results at unphysical pion masses, while the result from this work is a purely lattice result at the physical pion mass.}
    \label{fig:aI0 with previous}
\end{figure}

Figure~\ref{fig:aI0 with previous} shows comparison of these numbers with results from earlier works including phenomenology and lattice QCD.
The figure indicates our results are consistent with ChPT prediction and earlier lattice calculations \cite{Colangelo:2000jc,FlavourLatticeAveragingGroupFLAG:2021npn} within 1.3$\sigma$ for $I = 2$ and 1$\sigma$ for $I = 0$.  The errors on our results are somewhat larger because the earlier works at unphysical pion masses performed a chiral extrapolation with the assumption
\begin{align}
m_\pi a_0^2 &= -\frac{m_\pi^2}{8\pi f_\pi^2}
\left\{
1+\frac{m_\pi^2}{16\pi^2f_\pi^2}
\left[
3\ln\frac{m_\pi^2}{f_\pi^2}-1-l_{\pi\pi}^2
\right]
\right\},
\label{eq:I2extrap}
\\
m_\pi a_0^0 &= \frac{7m_\pi^2}{16\pi f_\pi^2}
\left\{
1-\frac{m_\pi^2}{16\pi^2f_\pi^2}
\left[
9\ln\frac{m_\pi^2}{f_\pi^2}-5-l_{\pi\pi}^0
\right]
\right\},
\label{eq:I0extrap}
\end{align}
where the leading order with the pion decay constant $f_\pi$ was inputted independently of lattice calculation.
They only used lattice results for determination of the low energy constants $l_{\pi\pi}^I$ which is associated with the sub-leading contribution to the scattering length.  
Our results on the other hand are pure lattice results obtained without assuming the leading contribution. 
Another work \cite{Fischer:2020jzp} carried out at physical pion mass gave $m_\pi a^2_0=-0.0481(86)$, which is omitted from the plot because of the absence of the systematic error but certainly has larger error than others even without the continuum extrapolation.

\section{Cost Comparison with G-Parity boundary conditions}
\label{sec:comparison}

One of the main goals of this study is to extract the signal of an excited state that has the energy near the kaon mass and is useful for calculation of $K\to\pi\pi$ decay matrix elements and $\varepsilon'$, the measure of direct CP violation.
We find from the previous section that the energy of the first excited state is close to the kaon mass and well resolved.
We have carried out the same kind of studies with G-parity boundary conditions (GPBC)~\cite{Bai:2015nea,RBC:2020kdj,RBC:2021acc}, in which pions are anti-periodic in space and therefore must have non-zero spatial momentum, and realized the on-shell $K\to\pi\pi$ kinematics with the corresponding $I=0$ ground two-pion state.
It is valuable to compare the effectiveness of GPBC against the conventional periodic boundary conditions (PBC). 
In this section we carry out an efficiency comparison between GPBC ground state and PBC first-excited state.

In Ref.~\cite{RBC:2021acc} measurements were carried out on an ensemble of 741 gauge field configurations with identical parameters to those used in the $32^3$ ensemble except GPBC were used. Nine hundred low-modes and twenty-four (spin-color-flavor diluted) random source fields on each time slice comprised the A2A~\cite{Foley:2005ac} measurement setup. The high-mode part is double for GPBC, 1,536 $vs$ 768 modes in this work because of no need of flavor dilution with PBC.  The low-mode part also differs, 900 exact eigenvectors (GPBC) $vs.$ 2,000 approximate coarse-grained eigenvectors based on local-coherence (PBC)~\cite{Clark:2017wom}. 
It is not easy to estimate a quantitative difference between this and the GPBC calculation since different setups (low-modes, AMA, solvers etc.) have been used in the two cases. However, assuming similar solver performances, we expect the cost of GPBC being roughly twice as much, for the same quark mass and lattice volume, because of the doubled size of the Dirac operator. Taking the two-pion energy at the kaon mass as the goal of our calculation, this cost has to be contrasted with a possibly easier extraction of energy levels in GPBC compared to PBC.
Another difference is the time-translation interval of connected diagrams.  Most of connected diagrams are calculated with 6 source locations in this work and 8 source locations in the earlier work.
One exception is the connected diagram of $\sigma$--$\sigma$ two-point function, which is computed, again, with 6 source locations in this work but with 64 source locations in the earlier work.  This difference in $\sigma$--$\sigma$ measurement detail would not affect computational cost as it is very cheap, though it might make an impact on the precision.

We will focus on the zero total momentum rest frame. For $I=2$ Tab. X in Ref.~\cite{RBC:2021acc} lists the energy (in lattice units) as 0.41528(46) with $t_0=11$, $t-t_0=1$. In this study the corresponding value is 0.45549(160). Recall that the momentum, in units of $2\pi/L$, of the pions is slightly smaller in the GPBC case, $\sqrt{3\times \left(\frac{1}{2}\right)^2}$ instead of 1 and therefore we do not expect the central values to agree. In fact both the number in our present study and that of Ref.~\cite{RBC:2021acc} agree with the dispersive results at their corresponding momenta. Accounting for the increased error growth due to this slight mismatch in the energy~\cite{Parisi:1983ae} and the difference in number of measurements expects the GPBC result should be more precise the PBC by a factor of $\approx 1.6 \times \sqrt{741/107}=4.2$, which implies the efficiency is about the same for the two methods in this case.

The more salient comparison is for the $I=0$ case since that is the main reason for using GPBC. In Tab.~X in Ref.~\cite{RBC:2021acc} the energy of the two-pion state is given as 0.3479(11)[10], where the first error is statistical and the second is excited state systematic. It is determined from a three-operator, two-state fit with fit range 6--15. A GEVP analysis with $t_0=5$, $t-t_0=1$, without matrix subtraction, gives 0.3489(11), where the error is statistical. The matrix subtraction was not performed for this analysis as the ATW effects with GPBC are due to a long-time propagation of moving pions and the estimates of the size of the dominant ATW contribution gave values ten times smaller than the statistical errors on the data, and separate multi-operator fits with the ATW term included gave results consistent with zero.  The corresponding result for PBC is 0.406(16), where a matrix subtraction was done to remove ATW contamination. While accounting again for the mismatch in energies and number of measurements expects the GPBC result should be more precise by a factor of about 3.5, we do not see significant shift by the matrix subtraction for $I=0$ and it would be interesting to study and understand how significant the ATW effect on $I=0$ is.  In any case the factor 3.5 is not big enough to explain the difference in the statistical error which is about four times bigger in the PBC case for equivalent statistics and energies. Of course the argument about exponential growth of errors is not perfect since the pre-factors may differ between the two cases. 

In the GPBC case, the boundary conditions lead to cubic symmetry breaking at the quark level which is suppressed by averaging pairs of single-pion interpolating operators with different quark and anti-quark momentum assignments but with the same total pion momentum~\cite{Christ:2019sah,RBC:2021acc}. For the ground state eight pairs (averages) of pion interpolating operators are used to construct sixty-four correlation functions, which most strongly overlap with the s-wave G-parity ground state. Here, we use six single-pion operators with momentum $(\pm 1,0,0)$ (plus permutations), or thirty-six correlators, for the corresponding s-wave excited state. While it is difficult to quantify the improvement gained from averaging over more combinations due to correlations, we do expect some benefit and will study this question in future calculations in both setups. 

While not as direct, it is interesting and useful to compare the $24^3$ ensemble to the GPBC case. Here the PBC setup is the same as before, but the statistics are based on 258 configurations. The relevant two-pion energy ($t=4$) is 0.5298(64) in lattice units. Converting the errors to GeV and accounting for the different number of measurements yields a factor $\approx 2.5$, by which the GPBC result is expected to be more precise than the PBC result.  Accounting for the different number of pion momentum combinations as before, this factor is further reduced to somewhere between 1 and 2 (These factors are based on an assumption of statistical independence between the momentum orientations that may not be born out in practice). The exponential factor is roughly one in this case which we ignore.  A more thorough comparison between the GPBC and PBC approaches will be performed when we have measurements performed on an equivalent statistical sample.

\section{Conclusion}
\label{sec:conclusion}

In this work we have carried out a study of pion scattering at the physical point using 2+1 flavor M\"obius domain wall fermion ensembles with inverse lattice spacings of 1.023 and 1.378 GeV using periodic boundary conditions (PBC). The main focus was to extract the first excited state energies in the rest frame, in both $I=0$ and  2 channels, and their corresponding phase shifts, using the finite volume L\"uscher formalism~\cite{Luscher:1990ux}. The first excited state energy (roughly) corresponds to the important case of on-shell $K\to\pi\pi$ decay which is our longer term goal~\cite{Tomii:2022vxb}.

The energies were computed using the GEVP method~\cite{Luscher:1990ck,Blossier:2009kd}. In order to extract the desired first excited state and control excited state contamination, several two-pion interpolating operators were used, including a simple scalar bilinear. The other operators were constructed from pions with equal and opposite momenta. The single pion momentum took values (0,0,0), ($\pm1$,0,0), ($\pm1$,$\pm1$,0), ($\pm1$,$\pm1$,$\pm1$) (and permutations) in units of $2\pi/L$. As found in~\cite{RBC:2020kdj,RBC:2021acc}, the inclusion of the scalar bilinear with the quantum numbers of the vacuum is crucial to disentangle the first and second excited state energies. 

The GEVP, or matrix correlation function, size ranged from $2\times2$ to $5\times5$. We found in most cases the size did not have a large effect after $2\times2$ for $I=2$ and $3\times3$ for $I=0$. However in some cases the noise of the higher states in the correlation matrix at large times adversely impacted the lower energies. In these cases the overlap of higher states with lower states was small, and the extra states only contributed noise to the lower states. This problem led to a modification of the method: as time increases the operator basis of the GEVP is changed (``re-based") using the eigenvectors of the GEVP from earlier times. While this method gives similar results in most cases studied here, it had a dramatic improvement for the $I=0$ first excited state on the $32^3$ ensemble where the statistics were relatively poor (see Figs.~\ref{fig:I=0 L32 fits 0}--\ref{fig:I=0 L32 fits 2}). This RGEVP then allowed a relatively precise phase shift determination (see Fig.~\ref{fig:phase shift}).

We started this series of $K\to\pi\pi$ studies with G-parity boundary conditions (GPBC) \cite{Bai:2015nea,RBC:2020kdj,RBC:2021acc} because we anticipated it would be challenging to extract the signal of an excited two-pion state which is necessary for the $K\to\pi\pi$ study using PBC.
After seeing the successful calculation of two-pion scattering and $K\to\pi\pi$ decay amplitudes with GPBC \cite{RBC:2020kdj,RBC:2021acc}, we launched this PBC project to find a practical alternative, as a further check, and to enable calculations with isospin corrections.
We note once again that we have been successful in extracting signals of multiple states: the four lowest energy states for $I=2$ and the three lowest ones (at least on the $24^3$ lattice with better statistics) for $I=0$, despite the anticipated difficulty of extracting the excited state signals. 

In this first study we have focused on two important systematic effects. First, by computing energies and phase shifts on ensembles with different lattice spacings we find no statistically significant discretization errors for $I=0$ and small, but statistically significant, effects for $I=2$. Second, we studied the time-dependence of the effective energies and observed noticeable excited state contamination for short times. We also saw that much of the effects arising from single-pion excited states could be removed with the dispersion relation method, which also removes leading discretization errors. Moreover, we expect the systematic uncertainties for PBC to be similar to GPBC. In that case statistical and systematic errors on the energies were estimated to be roughly equal~\cite{RBC:2021acc}. 

Another important motivation of employing PBC is that it appears difficult to use GPBC to compute QED and strong-interaction isospin breaking effects since they explicitly enforce isospin symmetry, though isospin symmetry breaking is expected to significantly impact the value of the direct CP violation parameter $\varepsilon'$ (20--30\%)\cite{Cirigliano:2019cpi}.  Controlling such effects precisely is important for the next generation of $\varepsilon^\prime$ calculations.  

We are currently improving the statistics on both ensembles for our companion $K\to\pi\pi$ calculation, and when that is complete, we will have an even better comparison with G-parity and estimates of systematic errors. These results and the ones so far from the kaon decay project leave us optimistic for the PBC method.

\acknowledgments
We thank the U.S. Department of Energy (DOE) for partial support under awards DE-FG02-92ER40716 (TB, DH), DE-SC0010339 (TB, LJ, MT), DE-SC0046548 (ASM), DE-SC0012704 (PAB, TI, CJ, AS) and DE-SC001704 (AS). LJ and MT were also supported in part by the U.S. DOE under Early Career Award No. DE-SC002114.  The work of CK was supported by the U.S. DOE Exascale Computing Project.  PB, TI, CJ acknowledge support under DOE SciDAC-5 LAB 22-2580. 
TI is also supported by the Department of Energy,
Laboratory Directed Research and Development (LDRD No. 23-051) of BNL and RIKEN BNL Research Center.
The research of MB was funded through the MUR program for young researchers, ``Rita Levi Montalcini''.  The work of DH was supported by the Swiss National Science Foundation (SNSF) through grant No. 200020\_208222. The software used for this work is based on \href{https://github.com/RBC-UKQCD/CPS}{CPS},
\href{https://github.com/paboyle/Grid}{Grid}, and \href{https://github.com/aportelli/hadrons}{Hadrons}.  We thank the collaborators in the RBC and UKQCD collaborations, especially N.H. Christ, for fruitful discussions.  Computations were carried out with USQCD resources funded by the US DOE at BNL and JLab.

\appendix


\section{Wick contractions}
\label{sec:contractions}

We write down the explicit forms of two-pion operators and the Wick contractions of two-point functions of these operators to clarify the convention used in this paper.

\begin{align}
    \widetilde O_{\pi\pi}^{2,0}(X_1, X_2)
    &= \frac{1}{\sqrt6}(\pi^+(X_1)\pi^-(X_2) +2\pi^0(X_1)\pi^0(X_2)\notag\\&\hspace{12mm} +\pi^-(X_1)\pi^+(X_2)),
    \\
    \widetilde O_{\pi\pi}^{0,0}(X_1, X_2)
    &= \frac{1}{\sqrt3}(\pi^+(X_1)\pi^-(X_2) -\pi^0(X_1)\pi^0(X_2)\notag\\&\hspace{12mm} +\pi^-(X_1)\pi^+(X_2)),
\end{align}
where $X_i$ denotes the 4D position $x_i$ or the set of the time coordinate and spatial momentum, $(t_i,\vec p_i)$, of the operator labeled by $i$.  
Wick contractions for two-piont functions of these operators yield
\begin{align}
    &\left\langle
    \widetilde O_{\pi\pi}^{2,0}(X_1,X_2) \widetilde O_{\pi\pi}^{2,0}(X_3,X_4)^\dag
    \right\rangle
    \notag\\
    &= 2D(X_1,X_2,X_3,X_4) - 2C(X_1,X_2,X_3,X_4),
    \\
    &\left\langle
    \widetilde O_{\pi\pi}^{0,0}(X_1,X_2) \widetilde O_{\pi\pi}^{0,0}(X_3,X_4)^\dag
    \right\rangle
    \notag\\
    &= 2D(X_1,X_2,X_3,X_4) + C(X_1,X_2,X_3,X_4)
    \notag\\
    &\,\,- 6R(X_1,X_2,X_3,X_4) + 3V(X_1,X_2,X_3,X_4),
\end{align}
where we define the contributions of the diagrams $D,C,R$ and $V$ by
\begin{align}
    D&(x_1,x_2,x_3,x_4)
    \notag\\*
    &= \frac{1}{2}\Big\langle
    \Tr[\gamma_5S_l(x_1,x_3)\gamma_5S_l(x_3,x_1)]
    \notag\\*&\hspace{12mm}
    \cdot \Tr[\gamma_5S_l(x_2,x_4)\gamma_5S_l(x_4,x_2)]
    \notag\\[2mm]
    &\hspace{4mm}
    + \Tr[\gamma_5S_l(x_1,x_4)\gamma_5S_l(x_4,x_1)]
    \notag\\*&\hspace{12mm}
    \cdot \Tr[\gamma_5S_l(x_2,x_3)\gamma_5S_l(x_3,x_2)]
    \Big\rangle,
    \label{eq:diagramD}
    \\
    C&(x_1,x_2,x_3,x_4)
    \notag\\*
    &= \frac{1}{2}\Big\langle
    \Tr[\gamma_5S_l(x_1,x_3)\gamma_5S_l(x_3,x_2)
    \notag\\*&\hspace{12mm}
    \cdot \gamma_5S_l(x_2,x_4)\gamma_5S_l(x_4,x_1)]
    \notag\\[2mm]
    &\hspace{4mm}
    + \Tr[\gamma_5S_l(x_1,x_4)\gamma_5S_l(x_4,x_2)
    \notag\\*&\hspace{12mm}
    \cdot \gamma_5S_l(x_2,x_3)\gamma_5S_l(x_3,x_1)]
    \Big\rangle
    \notag\\
    &= \Big\langle
    \Tr[\gamma_5S_l(x_1,x_3)\gamma_5S_l(x_3,x_2)
    \notag\\*&\hspace{12mm}
    \cdot \gamma_5S_l(x_2,x_4)\gamma_5S_l(x_4,x_1)]
    \Big\rangle,
    \\
    R&(x_1,x_2,x_3,x_4)
    \notag\\*
    &= \frac{1}{4}\Big\langle
    \Tr[\gamma_5S_l(x_1,x_2)\gamma_5S_l(x_2,x_3)
    \notag\\*&\hspace{12mm}
    \cdot \gamma_5S_l(x_3,x_4)\gamma_5S_l(x_4,x_1)]
    \notag\\[2mm]
    &\hspace{4mm}
    + \Tr[\gamma_5S_l(x_1,x_4)\gamma_5S_l(x_4,x_3)
    \notag\\*&\hspace{12mm}
    \cdot \gamma_5S_l(x_3,x_2)\gamma_5S_l(x_2,x_1)]
    \notag\\[2mm]
    &\hspace{4mm}
    + \Tr[\gamma_5S_l(x_1,x_2)\gamma_5S_l(x_2,x_4)
    \notag\\*&\hspace{12mm}
    \cdot \gamma_5S_l(x_4,x_3)\gamma_5S_l(x_3,x_1)]
    \notag\\[2mm]
    &\hspace{4mm}
    + \Tr[\gamma_5S_l(x_1,x_3)\gamma_5S_l(x_3,x_4)
    \notag\\*&\hspace{12mm}
    \cdot \gamma_5S_l(x_4,x_2)\gamma_5S_l(x_2,x_1)]
    \Big\rangle
    \notag\\
    &= \frac{1}{2}\Big\langle
    \Tr[\gamma_5S_l(x_1,x_2)\gamma_5S_l(x_2,x_3)
    \notag\\*&\hspace{12mm}
    \cdot \gamma_5S_l(x_3,x_4)\gamma_5S_l(x_4,x_1)]
    \notag\\[2mm]
    &\hspace{4mm}
    + \Tr[\gamma_5S_l(x_1,x_3)\gamma_5S_l(x_3,x_4)
    \notag\\*&\hspace{12mm}
    \cdot \gamma_5S_l(x_4,x_2)\gamma_5S_l(x_2,x_1)]
    \Big\rangle,
    \\
    V&(x_1,x_2,x_3,x_4)
    \notag\\*
    &= \Big\langle
    \Tr[\gamma_5S_l(x_1,x_2)\gamma_5S_l(x_2,x_1)]
    \notag\\*&\hspace{12mm}
    \cdot \Tr[\gamma_5S_l(x_3,x_4)\gamma_5S_l(x_4,x_3)]
    \Big\rangle.
    \label{eq:diagramV}
\end{align}
for $X_i = x_i$.
For our actual calculation with the A2A quark propagators proposed in Ref.~\cite{Foley:2005ac}, 
we define the pion meson field $\Pi_{ij}^\pi(t,\vec p)$, which is projected to a certain spatial momentum and spin-color singlet but has mode indices that label the low and high modes.
In our calculation, there are 2,000 low modes and $4\times3\times64 = 768$ high modes from the spin-color-time dilution and therefore each mode index runs for 1, 2, \ldots, 2,768.
The expressions of the contractions in Eqs.~\eqref{eq:diagramD}--\eqref{eq:diagramV} in time-momentum space with the pion meson fields are then given as follows:
\begin{align}
    D&(X_1,X_2,X_3,X_4)
    \notag\\*
    &= \frac{1}{2}\bigg\langle\sum_{i,j} \Pi^\pi_{ij}(X_1)\Pi^\pi_{ji}(X_3)
    \cdot\sum_{k,l}\Pi^\pi_{kl}(X_2)\Pi^\pi_{lk}(X_4)
    \notag\\
    &\hspace{6mm}
    + \sum_{i,j} \Pi^\pi_{ij}(X_1)\Pi^\pi_{ji}(X_4)
    \cdot\sum_{k,l}\Pi^\pi_{kl}(X_2)\Pi^\pi_{lk}(X_3)\bigg\rangle,
    \\[2mm]
    C&(X_1,X_2,X_3,X_4)
    \notag\\*
    &= \bigg\langle
    \sum_{i,j,k,l}\Pi^\pi_{ij}(X_1)\Pi^\pi_{jk}(X_3)
    \Pi^\pi_{kl}(X_2)\Pi^\pi_{li}(X_4)
    \bigg\rangle,
    \\[2mm]
    R&(X_1,X_2,X_3,X_4)
    \notag\\*
    &= \frac{1}{2}\bigg\langle
    \sum_{i,j,k,l}\Pi^\pi_{ij}(X_1)\Pi^\pi_{jk}(X_2)
    \Pi^\pi_{kl}(X_3)\Pi^\pi_{li}(X_4)
    \notag\\
    &\hspace{6mm}
    + \sum_{i,j,k,l}\Pi^\pi_{ij}(X_1)\Pi^\pi_{jk}(X_3)
    \Pi^\pi_{kl}(X_4)\Pi^\pi_{li}(X_2)
    \bigg\rangle,
    \\
    V&(X_1,X_2,X_3,X_4)
    \notag\\*
    &= \bigg\langle
    \sum_{i,j} \Pi^\pi_{ij}(X_1)\Pi^\pi_{ji}(X_2)
    \cdot\sum_{k,l}\Pi^\pi_{kl}(X_3)\Pi^\pi_{lk}(X_4)
    \bigg\rangle,
\end{align}
with $X_i = (t_i,\vec p_i)$.

Wick contractions for two-point functions including one or two sigma operators read
\begin{align}
    &\left\langle
    \widetilde O_{\pi\pi}^{2,0}(X_1,X_2) \sigma(X_3)^\dag
    \right\rangle
    \notag\\*
    &= -\sqrt{6}R_\sigma(X_1,X_2,X_3) + \sqrt{6}V_\sigma(X_1,X_2,X_3),
    \\
    &\left\langle\sigma(X_1) \sigma(X_3)^\dag\right\rangle
    \notag\\*
    &= -R_{\sigma\sigma}(X_1,X_3) + 2V_{\sigma\sigma}(X_1,X_3),
\end{align}
Here we define
\begin{align}
    R&_\sigma(x_1,x_2,x_3)
    \notag\\*
    &= \frac{1}{2}\Big\langle
    \Tr[\gamma_5S_l(x_1,x_2)\gamma_5S_l(x_2,x_3)S_l(x_3,x_1)]
    \notag\\*
    &\hspace{5mm}+ \Tr[\gamma_5S_l(x_2,x_1)\gamma_5S_l(x_1,x_3)S_l(x_3,x_2)]
    \Big\rangle
    \notag\\
    &= \Big\langle
    \Tr[\gamma_5S_l(x_1,x_2)\gamma_5S_l(x_2,x_3)S_l(x_3,x_1)]
    \Big\rangle,
    \\[2mm]
    V&_\sigma(x_1,x_2,x_3)
    \notag\\*
    &= \Big\langle
    \Tr[\gamma_5S_l(x_1,x_2)\gamma_5S_l(x_2,x_3)]\cdot\Tr[S_l(x_3,x_1)]
    \Big\rangle,
    \\
    R&_{\sigma\sigma}(x_1,x_3)
    = \Big\langle
    \Tr[S_l(x_1,x_3)S_l(x_3,x_1)]
    \Big\rangle,
    \\[2mm]
    V&_{\sigma\sigma}(x_1,x_3)
    = \Big\langle
    \Tr[S_l(x_1,x_1)S_l(x_3,x_3)]
    \Big\rangle,
\end{align}
which can be expressed in terms of the pion $\Pi_{ij}^\pi$ and sigma $\Pi_{ij}^\sigma$ meson fields as
\begin{align}
    R&_\sigma(X_1,X_2,X_3)
    \notag\\*
    &= \bigg\langle
    \sum_{i,j,k} \Pi_{ij}^\pi(X_1)\Pi_{jk}^\pi(X_2)\Pi_{ki}^\sigma(X_3)
    \bigg\rangle,
    \\
    V&_\sigma(X_1,X_2,X_3)
    \notag\\*
    &= \bigg\langle
    \sum_{i,j} \Pi_{ij}^\pi(X_1)\Pi_{ji}^\pi(X_2)\cdot
    \sum_k\Pi_{kk}^\sigma(X_3)
    \bigg\rangle,
    \\
    R&_{\sigma\sigma}(X_1,X_3)
    = \bigg\langle
    \sum_{i,j} \Pi_{ij}^\sigma(X_1)\Pi_{ji}^\sigma(X_3)
    \bigg\rangle,
    \\
    V&_{\sigma\sigma}(X_1,X_3)
    = \bigg\langle
    \sum_i \Pi_{ii}^\sigma(X_1)\cdot\sum_j\Pi_{jj}^\sigma(X_3)
    \bigg\rangle,
\end{align}
for $X_i = (t_i,\vec p_i)$.

\section{Details of GEVP procedure}

\subsection{Ordering of GEVP eigenvectors}
\label{sec:GEVP details}

While a simple description of how we order GEVP eigenvectors was made in Section~\ref{sec:GEVP}, it is valuable to address the exact procedure using equations.

Since we fix $t-t_0$ to a constant $\Delta_t$, we can drop $t$ or $t_0$ from the arguments of eigenvalues and eigenvectors and it is valuable for the following discussion.  We drop $t$ rather than $t_0$ since the contamination from excited states in eigenvalues and eigenvectors are measured by $t_0$.  The GEVP equation is then rewritten as
\begin{equation}
C(t)V_n(t_0) = \lambda_n(t_0)C(t_0)V_n(t_0).
\label{eq:GEVPt0}
\end{equation}

At short time separations where the statistical errors are small enough, we simply sort eigenvalues into descending order.  This will give us the ascending order of effective energies and ensure.  Since this approach for short times is trivial and has no ambiguity, we spend the rest of the subsection for the ordering at larger time separations where the statistical error is large but the excited-state contamination is expected to be small.

At long distances we employ a recursive approach using the eigenvectors obtained at one time slice earlier.  As explained in Section~\ref{sec:GEVP}, the idea is to use eigenvectors at one time slice earlier to construct a near diagonal correlator matrix with which it is very easy to obtain the correct order of eigenvectors at the current time slice.  The exact procedure is given as follows.

We now suppose the ordering of the eigenvectors $V_n(t_0-1)$ at $t_0-1$ is successful and give a recipe to obtain the correct order of the eigenvectors $V_n(t_0)$ at $t_0$ using $V_n(t_0-1)$.
We define an $N\times N$ matrix
\begin{eqnarray}
    T(t_0-1) = (V_1(t_0-1)\,\,V_2(t_0-1)\,\ldots V_N(t_0-1)),
\end{eqnarray}
using the set of the GEVP eigenvectors $V_n(t_0-1)$ obtained at one time slice earlier.
Then we can calculate approximately diagonal matrices
\begin{align}
    C'(t, t_0-1) &= T(t_0-1)^\dag C(t) T(t_0-1),
    \label{eq:re-basedCt}\\
    C'(t_0, t_0-1) &= T(t_0-1)^\dag C(t_0) T(t_0-1).
    \label{eq:re-basedCt0}
\end{align}
Here, the second argument of $C'$ on the left hand sides corresponds to the argument of $T$ on the right hand sides.  The off-diagonal elements of these matrices are associated only with the statistical fluctuation and systematic effect from excited states.
With these near diagonal matrices, we can express the GEVP \eqref{eq:GEVPt0} as
\begin{align}
C'(t, t_0-1)& V_n'(t_0) = \lambda_n(t_0)C'(t_0,t_0-1)V_n'(t_0),
\label{eq:GEVPprime}
\end{align}
where $V_n'(t_0)$ satisfies
\begin{equation}
V_n(t_0) = T(t_0-1)V_n'(t_0).
\label{eq:Vprime}
\end{equation}
If we can obtain the correct order of $V_n'(t_0)$ when solving the modified GEVP \eqref{eq:GEVPprime}, we can also obtain the corresponding GEVP eigenvectors $V_n(t_0)$ with the original basis through Eq.~\eqref{eq:Vprime}.
In fact eigenvectors $V_n'(t_0)$ are mostly a unit vector for a certain direction and it is easy to recognize their correct order at sufficiently large $t_0$ where the contamination from the $N+1$-th and higher states is small and therefore the correlator matrices in the GEVP~\eqref{eq:GEVPprime} are mostly diagonal.

\subsection{RGEVP}
\label{sec:RGEVP details}

In this work we consider the RGEVP with fixed $t-t_0 \equiv \Delta_t$ and we continue to omit $t$ from the arguments of $V_n$ in this subsection.  In this subsection we give the recipe for finding a new basis of fewer operators.
\begin{enumerate}
    \item Choose the GEVP size $N_{t_0}$ at each time slice
    \begin{eqnarray}
        N_{1} \ge N_{2} \ge \ldots.
    \end{eqnarray}
    Note that $N_{t_0}$ here has a different meaning than $N_\alpha$ used in Section~\ref{sec:RGEVP}.
    \item Solve the $N_{1}\times N_{1}$ GEVP at the beginning time slice $t_0=1$ and order the eigenvectors $V_n(1)$ ensuring the descending order of the corresponding eigenvalues.
    \item For $t_0\ge2$, suppose we have obtained the eigenvectors $V_n(t_0-1)$ at $t_0-1$ for $n=1,2,\ldots,N_{t_0-1}$ and calculate the $N_{t_0}\times N_{t_0}$ re-based correlator matrices by Eqs.~\eqref{eq:re-basedCt} and \eqref{eq:re-basedCt0} with the modified $N_{1}\times N_{t_0}$ re-basing matrix
    \begin{eqnarray}
        T(t_0-1) = (V_1(t_0-1)\,\,V_2(t_0-1)\,\ldots V_{N_{t_0}}(t_0-1)).
    \end{eqnarray}
    \label{step:rebase}
    \item Solve the GEVP \eqref{eq:GEVPprime} with the re-based correlators in Eqs.~\eqref{eq:re-basedCt} and \eqref{eq:re-basedCt0} and obtain the eigenvectors $V_n(t)$ by Eq.~\eqref{eq:Vprime} for $n = 1,2,\ldots,N_t$ using the ordering procedure described in the previous subsection.
    \label{step:solveGEVP}
\end{enumerate}
Note that, in step~\ref{step:rebase}, the number $N_{t_0}$ of columns of $T(t_0-1)$ despite $N_{t_0-1}(\ge N_{t_0})$ eigenvectors $V_n(t_0-1)$ obtained at time slice $t_0-1$ plays a role in reducing the size of the GEVP at time slice $t_0$ when $N_{t_0-1}\neq N_{t_0}$.

While repeating steps \ref{step:rebase} and \ref{step:solveGEVP} and applying the results to Eq.~\eqref{eq:def_Eeff} give us a new series of effective two-pion energies, they are identical to the normal GEVP results at small $t_0$ that satisfies $N_{t_0}=N_1$.
On the other hand, it is interesting to investigate how the GEVP with the reduced new basis behaves at smaller time slices.  Therefore we perform the following additional steps:
\begin{enumerate}
\setcounter{enumi}{4}
\item Repeat steps \ref{step:rebase} and \ref{step:solveGEVP} to obtain $T(\tilde t_0)$ with a chosen time $\tilde t_0$ where the final re-basing is performed, $i.e.$ $N_{\tilde t_0-1} > N_{\tilde t_0} = N_{\tilde t_0+1} = \ldots$.
\item Perform the GEVP analysis with the re-based correlator matrices $T(\tilde t_0)^\dag C(t)T(\tilde t_0)$ for all available time slices.
\end{enumerate}
The $N_1\times N_{\tilde t_0}$ matrix $T(\tilde t_0)$ is the matrix to define the new operator basis.
These steps can be parametrized by the pairs of $t_0$ and $N_{t_0}$ that satisfy $N_{t_0-1}>N_{t_0}$.  A single pair gives us a single-step RGEVP, while multiple pairs correspond to a multi-step RGEVP.
We present the RGEVP results with these parameters in Section~\ref{sec:results}.

\section{Supplemental figures and tables}
\label{sec:2pi energies}

\begin{table}[tpb]
\centering
\begin{tabular}{|c|ccc|}
\hline
 GEVP & \multicolumn{3}{c|}{fit range} \\
 type & 4--10 & 5--10 & 6--10\\
\hline
$2\times2$ & $-0.378(13)^\circ$ & $-0.372(14)^\circ$ & $-0.372(13)^\circ$\\
$3\times3$ & $-0.377(13)^\circ$ & $-0.371(14)^\circ$ & $-0.371(14)^\circ$\\
$4\times4$ & $-0.369(13)^\circ$ & $-0.364(14)^\circ$ & $-0.364(14)^\circ$\\
RGEVP & $-0.374(13)^\circ$ & $-0.369(14)^\circ$ & $-0.369(13)^\circ$\\
\hline
\end{tabular}
\caption{Results for the $I=2$ phase shift on the $24^3$ lattice for two-pion energy of the ground state shown in Table~\ref{tab:fitres_Epipi_I2_n0_L24}.}
\label{tab:pshift_I2_n0_L24}
\end{table}

\begin{table}[tp]
\centering
\begin{tabular}{|c|ccc|}
\hline
 GEVP & \multicolumn{3}{c|}{fit range} \\
 type & 4--9 & 5--9 & 6--9\\
\hline
$2\times2$ & $-12.94(19)^\circ$ & $-12.64(23)^\circ$ & $-12.63(34)^\circ$\\
$3\times3$ & $-12.61(18)^\circ$ & $-12.37(23)^\circ$ & $-12.41(34)^\circ$\\
$4\times4$ & $-12.58(18)^\circ$ & $-12.33(23)^\circ$ & $-12.40(34)^\circ$\\
RGEVP & $-12.57(18)^\circ$ & $-12.33(23)^\circ$ & $-12.40(34)^\circ$\\
\hline
\end{tabular}
\caption{Results for the $I=2$ phase shift on the $24^3$ lattice for two-pion energy of the first excited state shown in Table~\ref{tab:fitres_Epipi_I2_n1_L24}.}
\label{tab:pshift_I2_n1_L24}
\end{table}

\begin{table}[tp]
\centering
\begin{tabular}{|c|ccc|}
\hline
 GEVP & \multicolumn{3}{c|}{fit range} \\
 type & 3--9 & 4--9 & 5--9\\
\hline
$3\times3$ & $-20.97(44)^\circ$ & $-20.10(71)^\circ$ & $-20.8(1.2)^\circ$\\
$4\times4$ & $-20.11(42)^\circ$ & $-19.29(68)^\circ$ & $-20.4(1.2)^\circ$\\
RGEVP & $-20.18(43)^\circ$ & $-19.50(67)^\circ$ & $-20.9(1.1)^\circ$\\
\hline
\end{tabular}
\caption{Results for the $I=2$ phase shift on the $24^3$ lattice for two-pion energy of the second excited state shown in Table~\ref{tab:fitres_Epipi_I2_n2_L24}.}
\label{tab:pshift_I2_n2_L24}
\end{table}

\begin{table}[tp]
\centering
\begin{tabular}{|c|cc|}
\hline
 GEVP & \multicolumn{2}{c|}{fit range} \\
 type & 3--5 & 4--5\\
\hline
$4\times4$ & $-26.5(2.4)^\circ$ & $-28.7(5.0)^\circ$\\
\hline
\end{tabular}
\caption{Results for the $I=2$ phase shift on the $24^3$ lattice for two-pion energy of the third excited state shown in Table~\ref{tab:fitres_Epipi_I2_n3_L24}.}
\label{tab:pshift_I2_n3_L24}
\end{table}

\begin{table}[tp]
\centering
\begin{tabular}{|c|ccc|}
\hline
 GEVP & \multicolumn{3}{c|}{fit range} \\
 type & 4--9 & 5--9 & 6--9\\
\hline
$2\times2$ & $-0.427(50)^\circ$ & $-0.427(51)^\circ$ & $-0.419(52)^\circ$\\
$3\times3$ & $-0.426(51)^\circ$ & $-0.427(51)^\circ$ & $-0.420(52)^\circ$\\
$4\times4$ & $-0.418(53)^\circ$ & $-0.418(53)^\circ$ & $-0.411(54)^\circ$\\
RGEVP & $-0.424(51)^\circ$ & $-0.424(52)^\circ$ & $-0.417(53)^\circ$\\
\hline
\end{tabular}
\caption{Results for the $I=2$ phase shift on the $32^3$ lattice for two-pion energy of the ground state shown in Table~\ref{tab:fitres_Epipi_I2_n0_L32}.}
\label{tab:pshift_I2_n0_L32}
\end{table}

\begin{table}[tp]
\centering
\begin{tabular}{|c|ccc|}
\hline
 GEVP & \multicolumn{3}{c|}{fit range} \\
 type & 4--10 & 5--10 & 6--10\\
\hline
$2\times2$ & $-13.96(32)^\circ$ & $-13.61(40)^\circ$ & $-13.57(47)^\circ$\\
$3\times3$ & $-13.45(32)^\circ$ & $-13.26(39)^\circ$ & $-13.26(47)^\circ$\\
$4\times4$ & $-13.37(32)^\circ$ & $-13.23(39)^\circ$ & $-13.25(47)^\circ$\\
RGEVP & $-13.37(32)^\circ$ & $-13.24(39)^\circ$ & $-13.26(46)^\circ$\\
\hline
\end{tabular}
\caption{Results for the $I=2$ phase shift on the $32^3$ lattice for two-pion energy of the first excited state shown in Table~\ref{tab:fitres_Epipi_I2_n1_L32}.}
\label{tab:pshift_I2_n1_L32}
\end{table}

\begin{table}[tp]
\centering
\begin{tabular}{|c|ccc|}
\hline
 GEVP & \multicolumn{3}{c|}{fit range} \\
 type & 3--7 & 4--7 & 5--7\\
\hline
$3\times3$ & $-24.13(59)^\circ$ & $-23.44(83)^\circ$ & $-23.1(1.2)^\circ$\\
$4\times4$ & $-22.85(59)^\circ$ & $-22.59(81)^\circ$ & $-22.5(1.2)^\circ$\\
RGEVP & $-22.90(59)^\circ$ & $-22.59(81)^\circ$ & $-22.4(1.2)^\circ$\\
\hline
\end{tabular}
\caption{Results for the $I=2$ phase shift on the $32^3$ lattice for two-pion energy of the second excited state shown in Table~\ref{tab:fitres_Epipi_I2_n2_L32}.}
\label{tab:pshift_I2_n2_L32}
\end{table}

\begin{table}[tp]
\centering
\begin{tabular}{|c|ccc|}
\hline
 GEVP & \multicolumn{3}{c|}{fit range} \\
 type & 4--7 & 5--7 & 6--7\\
\hline
$4\times4$ & $-24.2(3.2)^\circ$ & $-26.4(4.7)^\circ$ & $-27.1(8.1)^\circ$\\
\hline
\end{tabular}
\caption{Results for the $I=2$ phase shift on the $32^3$ lattice for two-pion energy of the third excited state shown in Table~\ref{tab:fitres_Epipi_I2_n3_L32}.}
\label{tab:pshift_I2_n3_L32}
\end{table}

\begin{table}[tp]
\centering
\begin{tabular}{|c|ccc|}
\hline
 GEVP & \multicolumn{3}{c|}{fit range} \\
 type & 3--6 & 4--6 & 5--6\\
\hline
$3\times3$ & $44.4(3.0)^\circ$ & $45.3(3.8)^\circ$ & $45.8(4.7)^\circ$\\
$4\times4$ & $45.5(3.0)^\circ$ & $45.9(4.1)^\circ$ & $48.5(6.0)^\circ$\\
$5\times5$ & $45.4(2.9)^\circ$ & $45.9(3.9)^\circ$ & $48.1(5.3)^\circ$\\
RGEVP & $45.1(2.9)^\circ$ & $45.5(4.2)^\circ$ & $46.9(6.8)^\circ$\\
\hline
\end{tabular}
\caption{Results for the $I=0$ phase shift on the $24^3$ lattice for two-pion energy of the first excited state shown in Table~\ref{tab:fitres_Epipi_I0_n1_L24}.}
\label{tab:pshift_I0_n1_L24}
\end{table}

\begin{table}[tp]
\centering
\begin{tabular}{|c|cc|}
\hline
 GEVP & \multicolumn{2}{c|}{fit range} \\
 type & 3--5 & 4--5\\
\hline
$3\times3$ & $ 65(38)^\circ$ & $ 141(72)^\circ$\\
$4\times4$ & $ 69(13)^\circ$ & $ 64(14)^\circ$\\
$5\times5$ & $ 68(13)^\circ$ & $ 65(13)^\circ$\\
RGEVP & $ 83(11)^\circ$ & $ 79(22)^\circ$\\
\hline
\end{tabular}
\caption{Results for the $I=0$ phase shift on the $24^3$ lattice for two-pion energy of the second excited state shown in Table~\ref{tab:fitres_Epipi_I0_n2_L24}.}
\label{tab:pshift_I0_n2_L24}
\end{table}

\begin{table}[tp]
\centering
\begin{tabular}{|c|ccc|}
\hline
 GEVP & \multicolumn{3}{c|}{fit range} \\
 type & 3--8 & 4--8 & 5--8\\
\hline
$3\times3$ & $44.1(5.2)^\circ$ & $ 44(11)^\circ$ & $ 49(20)^\circ$\\
$4\times4$ & $43.4(5.0)^\circ$ & $ 44(12)^\circ$ & $ 52(34)^\circ$\\
$5\times5$ & $43.4(5.2)^\circ$ & $ 45(15)^\circ$ & $ 50(24)^\circ$\\
RGEVP & $38.4(2.5)^\circ$ & $38.6(3.6)^\circ$ & $44.1(5.7)^\circ$\\
\hline
\end{tabular}
\caption{Results for the $I=0$ phase shift on the $32^3$ lattice for two-pion energy of the first excited state shown in Table~\ref{tab:fitres_Epipi_I0_n1_L32}.}
\label{tab:pshift_I0_n1_L32}
\end{table}

\begin{table}[tp]
\centering
\begin{tabular}{|c|ccc|}
\hline
 GEVP & \multicolumn{3}{c|}{fit range} \\
 type & 3--6 & 4--6 & 5--6\\
\hline
$3\times3$ & $ 17(19)^\circ$ & $ 20(25)^\circ$ & $ 28(30)^\circ$\\
$4\times4$ & $40(680)^\circ$ & $140(690)^\circ$ & $80(860)^\circ$\\
$5\times5$ & $ 28(24)^\circ$ & $ 30(30)^\circ$ & $ 31(33)^\circ$\\
RGEVP & $71.6(8.4)^\circ$ & $ 79(14)^\circ$ & $ 80(24)^\circ$\\
\hline
\end{tabular}
\caption{Results for the $I=0$ phase shift on the $32^3$ lattice for two-pion energy of the second excited state shown in Table~\ref{tab:fitres_Epipi_I0_n2_L32}.}
\label{tab:pshift_I0_n2_L32}
\end{table}

\begin{table}[tp]
\centering
\begin{tabular}{|c|ccc|}
\hline
 GEVP & \multicolumn{3}{c|}{fit range} \\
 type & 4--10 & 5--10 & 6--10\\
\hline
$2\times2$ & $-0.0499(11)$ & $-0.0494(12)$ & $-0.0494(12)$\\
$3\times3$ & $-0.0499(12)$ & $-0.0493(12)$ & $-0.0493(12)$\\
$4\times4$ & $-0.0492(11)$ & $-0.0487(13)$ & $-0.0487(12)$\\
RGEVP & $-0.0496(11)$ & $-0.0492(12)$ & $-0.0492(12)$\\
\hline
\end{tabular}
\caption{Results for the $I=2$ scattering length times pion mass $m_\pi a_0$ calculated on the $24^3$ lattice using the phase shift of the ground state shown in Table~\ref{tab:pshift_I2_n0_L24} and Eq.~\eqref{eq:a0}.}
\label{tab:a0_I2_L24}
\end{table}

\begin{table}[tp]
\centering
\begin{tabular}{|c|ccc|}
\hline
 GEVP & \multicolumn{3}{c|}{fit range} \\
 type & 4--9 & 5--9 & 6--9\\
\hline
$2\times2$ & $-0.0538(41)$ & $-0.0539(42)$ & $-0.0532(43)$\\
$3\times3$ & $-0.0538(42)$ & $-0.0539(42)$ & $-0.0533(43)$\\
$4\times4$ & $-0.0531(44)$ & $-0.0532(44)$ & $-0.0525(45)$\\
RGEVP & $-0.0537(42)$ & $-0.0537(42)$ & $-0.0531(44)$\\
\hline
\end{tabular}
\caption{Results for the $I=2$ scattering length times pion mass $m_\pi a_0$ calculated on the $32^3$ lattice using the phase shift of the ground state shown in Table~\ref{tab:pshift_I2_n0_L32} and Eq.~\eqref{eq:a0}.}
\label{tab:a0_I2_L32}
\end{table}

\begin{table}[tp]
\centering
\begin{tabular}{|c|ccc|}
\hline
 GEVP & \multicolumn{3}{c|}{fit range} \\
 type & 3--8 & 4--8 & 5--8\\
\hline
$3\times3$ & $0.1902(52)$ & $0.2001(72)$ & $0.2069(98)$\\
$4\times4$ & $0.1904(53)$ & $0.1986(72)$ & $0.2044(95)$\\
$5\times5$ & $0.1937(53)$ & $0.2023(73)$ & $0.2080(95)$\\
RGEVP & $0.1885(48)$ & $0.2038(70)$ & $0.2083(91)$\\
\hline
\end{tabular}
\caption{Results for the $I=0$ scattering length times pion mass $m_\pi a_0$ calculated on the $24^3$ lattice using Eq.~\eqref{eq:a0}.}
\label{tab:a0_I0_L24}
\end{table}

\begin{table}[tp]
\centering
\begin{tabular}{|c|ccc|}
\hline
 GEVP & \multicolumn{3}{c|}{fit range} \\
 type & 4--9 & 5--9 & 6--9\\
\hline
$3\times3$ & $0.1871(91)$ & $0.193(11)$ & $0.203(17)$\\
$4\times4$ & $0.183(12)$ & $0.190(14)$ & $0.200(21)$\\
$5\times5$ & $0.180(13)$ & $0.188(15)$ & $0.199(22)$\\
RGEVP & $0.185(13)$ & $0.195(15)$ & $0.200(21)$\\
\hline
\end{tabular}
\caption{Results for the $I=0$ scattering length times pion mass $m_\pi a_0$ calculated on the $32^3$ lattice using Eq.~\eqref{eq:a0}.}
\label{tab:a0_I0_L32}
\end{table}

\begin{figure}[tp]
    \includegraphics[width=\linewidth]{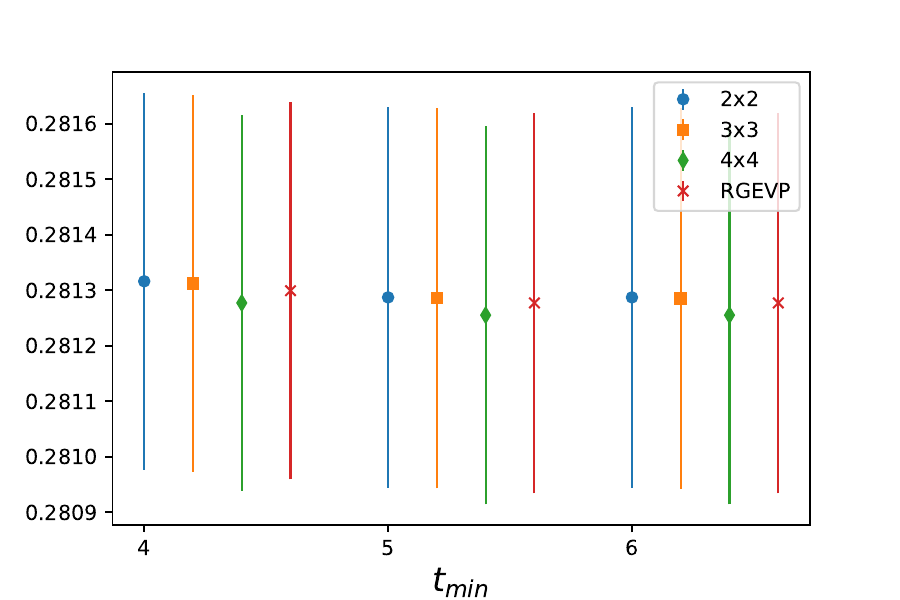}
    \caption{$I=2$ $\pi\pi$ ground state energy on the $24^3$ ensemble obtained from fits to a constant for various fit ranges and GEVP types plotted in lattice units.}
    \label{fig:I=2 fits 0}
\end{figure}

\begin{figure}[tp]
    \includegraphics[width=\linewidth]{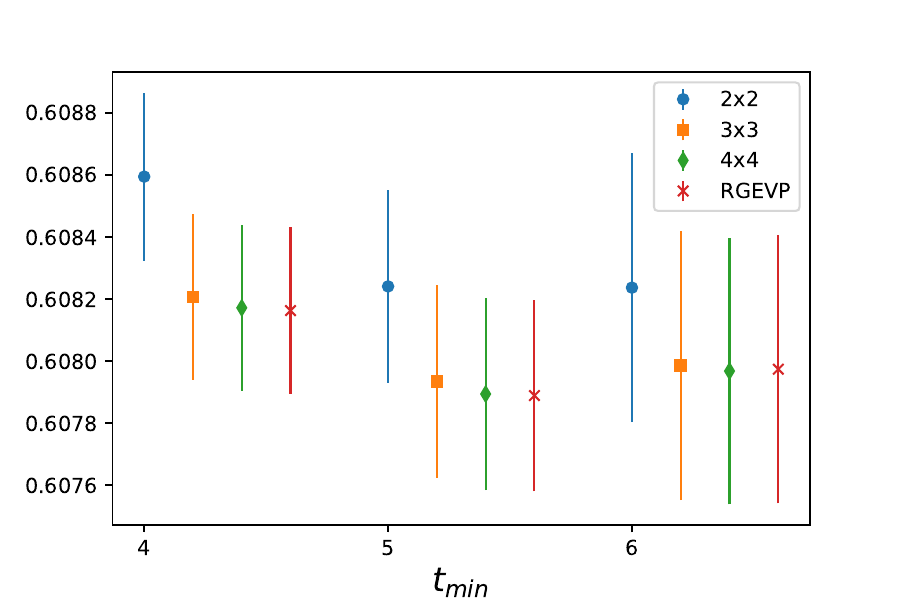}
    \caption{Same as Fig.~\ref{fig:I=2 fits 0} but result for the $I=2$ first-excited state on the $24^3$ ensemble.}
    \label{fig:I=2 fits 1}
\end{figure}

\begin{figure}[tp]
    \includegraphics[width=\linewidth]{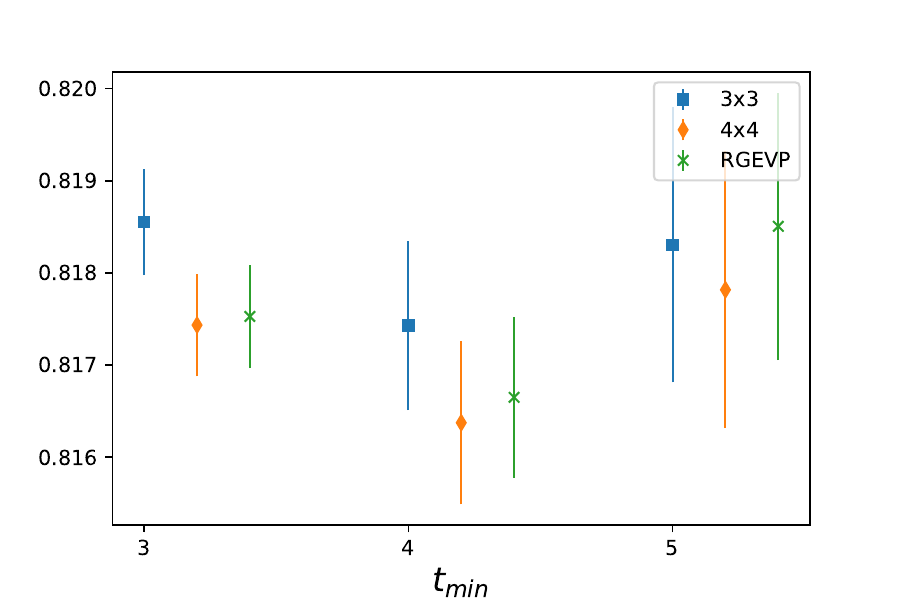}
    \caption{Same as Fig.~\ref{fig:I=2 fits 0} but result for the $I=2$ second-excited state on the $24^3$ ensemble.}
    \label{fig:I=2 fits 2}
\end{figure}

\begin{figure}[tp]
    \includegraphics[width=\linewidth]{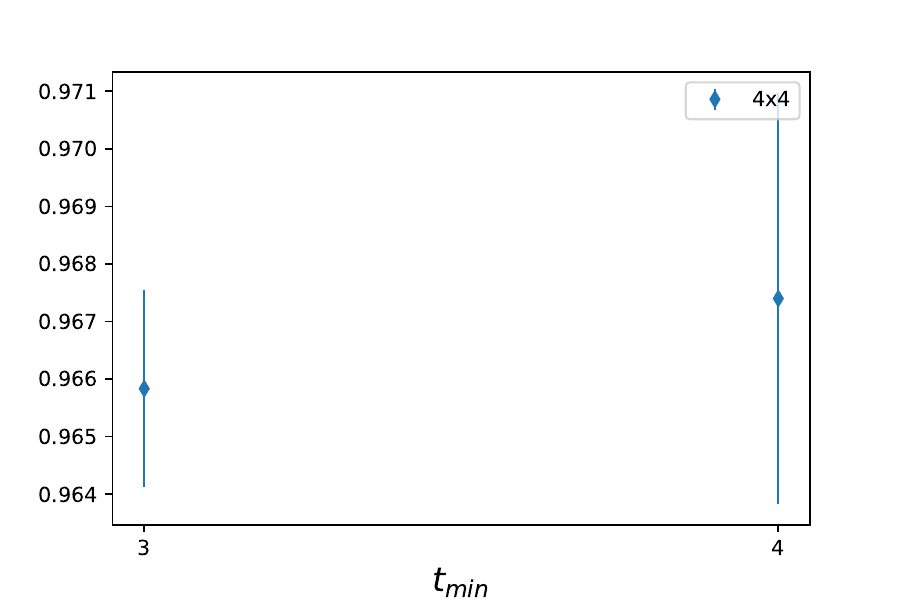}
    \caption{Same as Fig.~\ref{fig:I=2 fits 0} but result for the $I=2$ third-excited state on the $24^3$ ensemble.}
    \label{fig:I=2 fits 3}
\end{figure}

\begin{figure}[tp]
    \includegraphics[width=\linewidth]{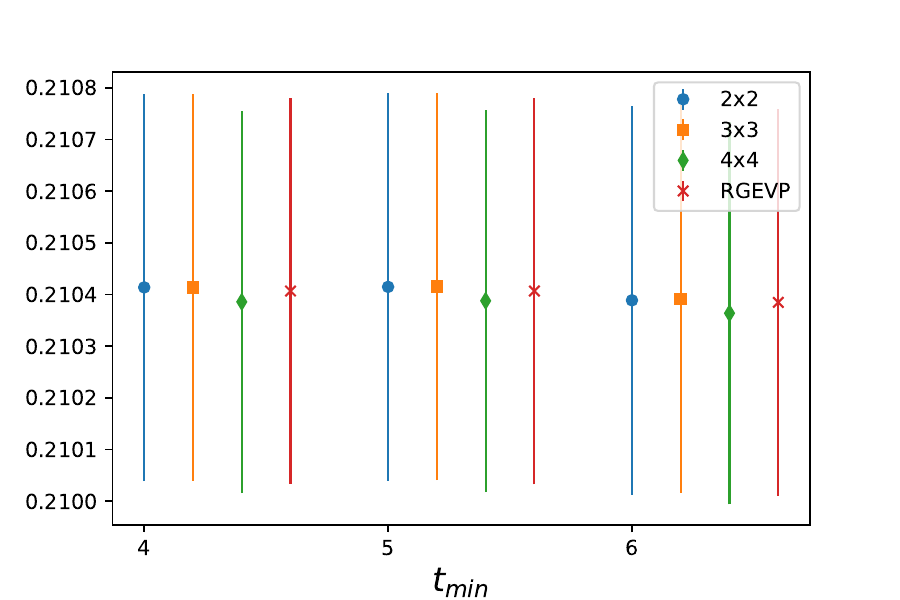}
    \caption{Same as Fig.~\ref{fig:I=2 fits 0} but result for the $I=2$ ground state on the $32^3$ ensemble.}
    \label{fig:I=2 L32 fits 0}
\end{figure}

\begin{figure}[tp]
    \includegraphics[width=\linewidth]{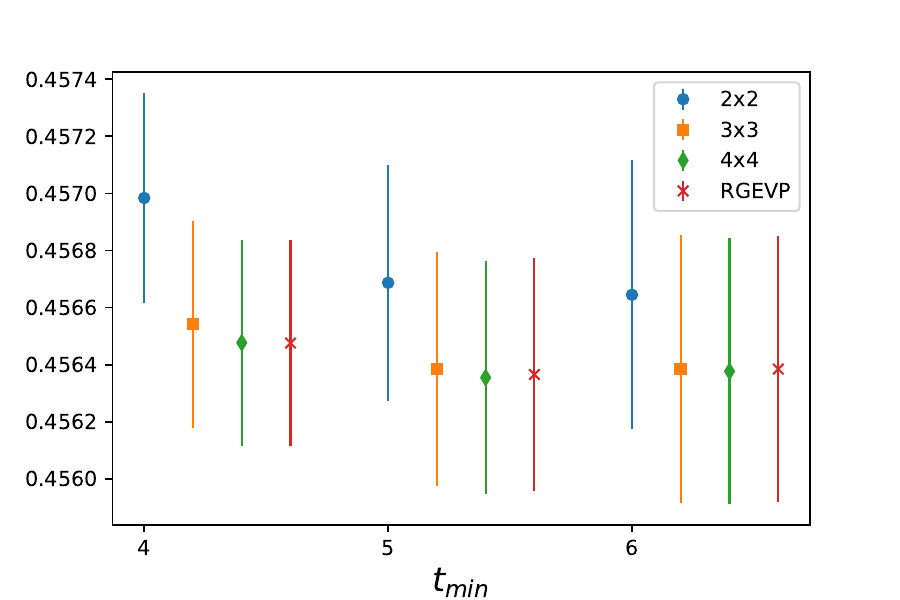}
    \caption{Same as Fig.~\ref{fig:I=2 fits 0} but result for the $I=2$ first-excited state on the $32^3$ ensemble.}
    \label{fig:I=2 L32 fits 1}
\end{figure}

\begin{figure}[tp]
    \includegraphics[width=\linewidth]{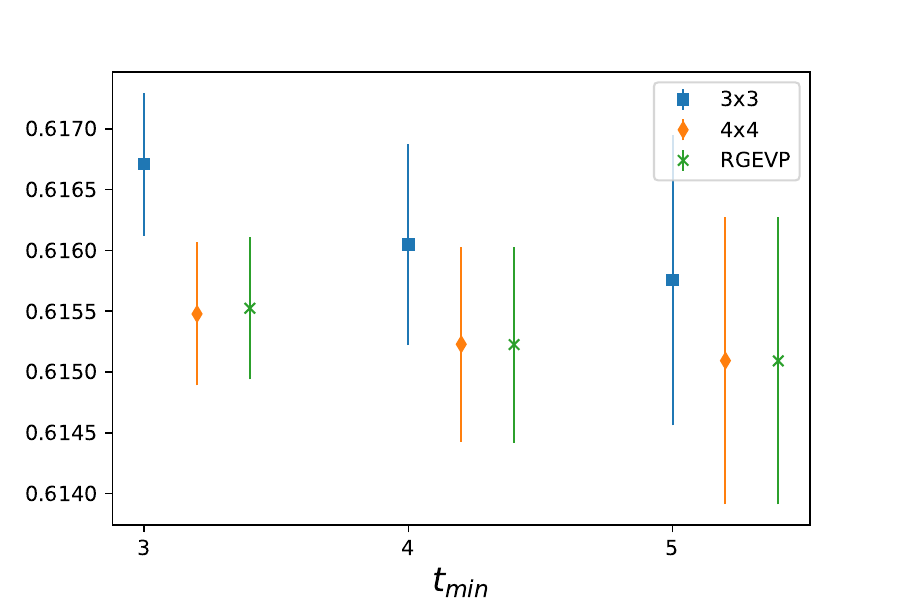}
    \caption{Same as Fig.~\ref{fig:I=2 fits 0} but result for the $I=2$ second-excited state on the $32^3$ ensemble.}
    \label{fig:I=2 L32 fits 2}
\end{figure}

\begin{figure}[tp]
    \includegraphics[width=\linewidth]{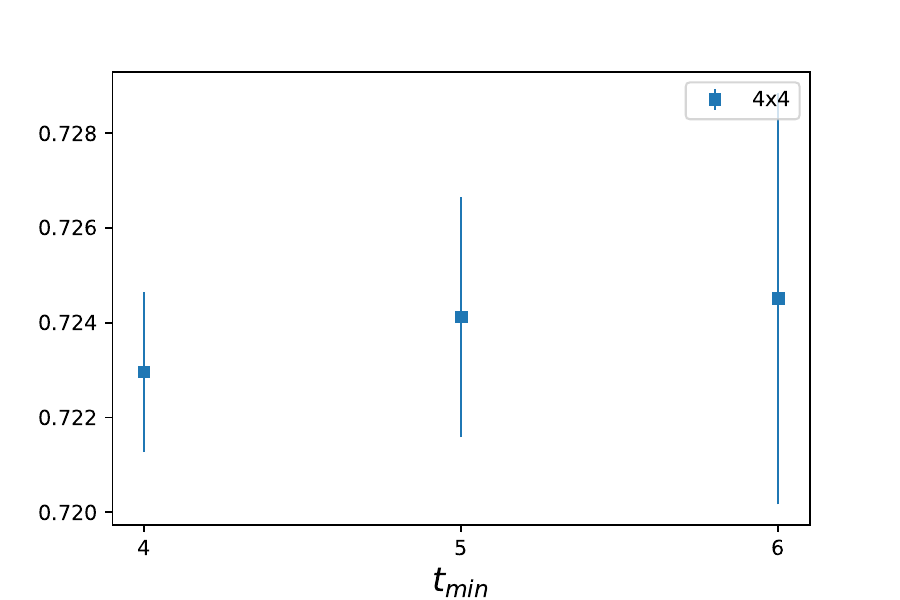}
    \caption{Same as Fig.~\ref{fig:I=2 fits 0} but result for the $I=2$ third-excited state on the $32^3$ ensemble.}
    \label{fig:I=2 L32 fits 3}
\end{figure}

\begin{figure}[tp]
    \includegraphics[width=\linewidth]{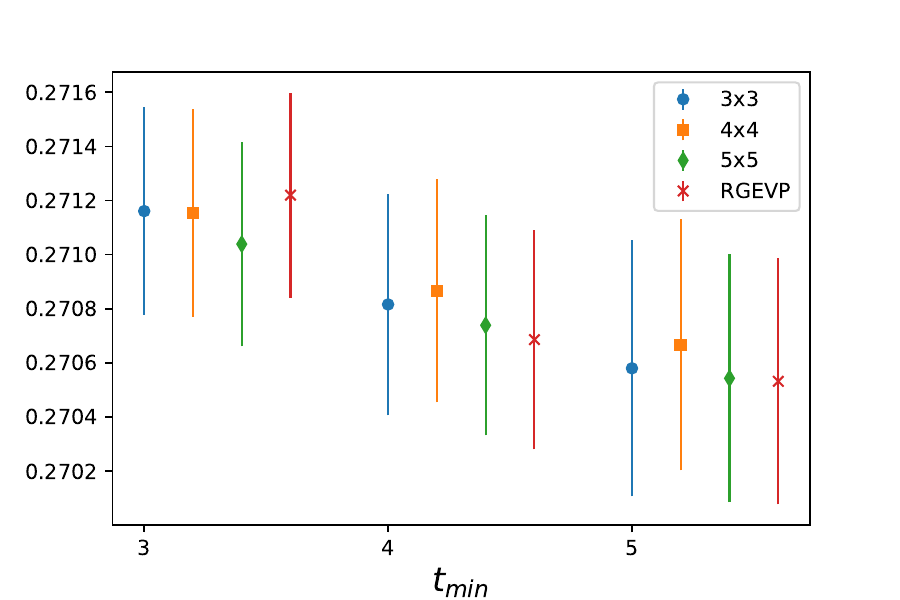}
    \caption{Same as Fig.~\ref{fig:I=2 fits 0} but result for the $I=0$ ground state on the $24^3$ ensemble.}
    \label{fig:I=0 fits 0}
\end{figure}

\begin{figure}[tp]
    \includegraphics[width=\linewidth]{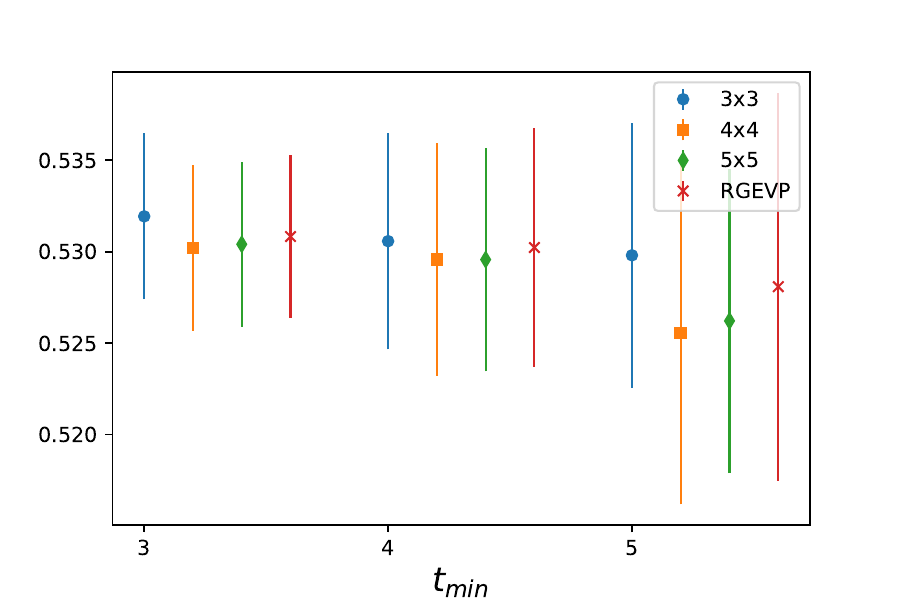}
    \caption{Same as Fig.~\ref{fig:I=2 fits 0} but result for the $I=0$ first-excited state on the $24^3$ ensemble.}
    \label{fig:I=0 fits 1}
\end{figure}

\begin{figure}[tp]
    \includegraphics[width=\linewidth]{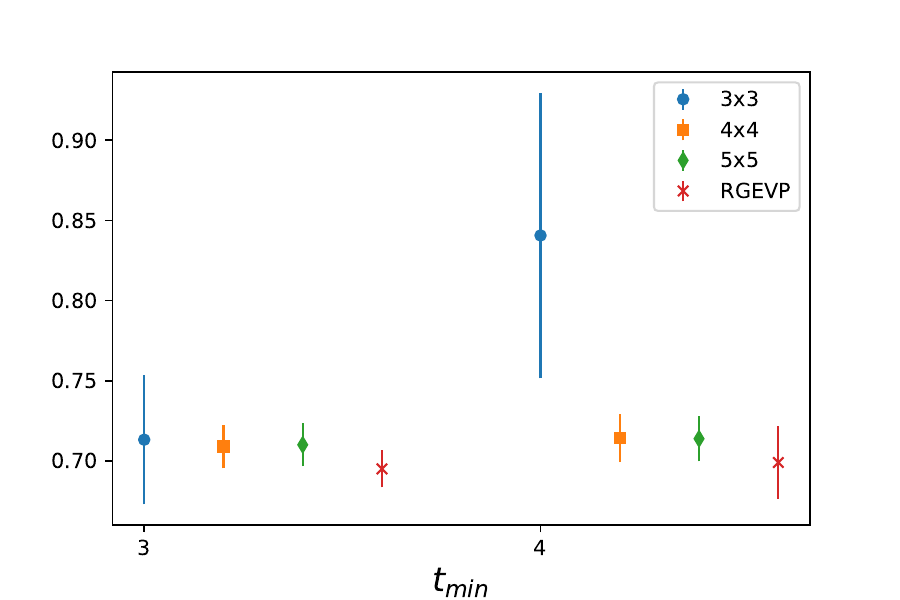}
    \caption{Same as Fig.~\ref{fig:I=2 fits 0} but result for the $I=0$ second-excited state on the $24^3$ ensemble.}
    \label{fig:I=0 fits 2}
\end{figure}

\begin{figure}[tp]
    \includegraphics[width=\linewidth]{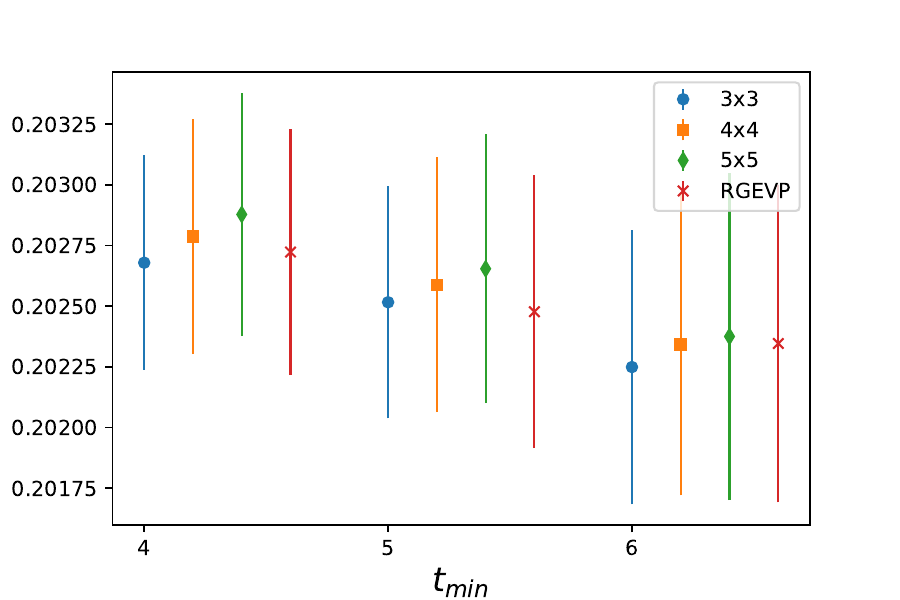}
    \caption{Same as Fig.~\ref{fig:I=2 fits 0} but result for the $I=0$ ground state on the $32^3$ ensemble.}
    \label{fig:I=0 L32 fits 0}
\end{figure}

\begin{figure}[tp]
    \includegraphics[width=\linewidth]{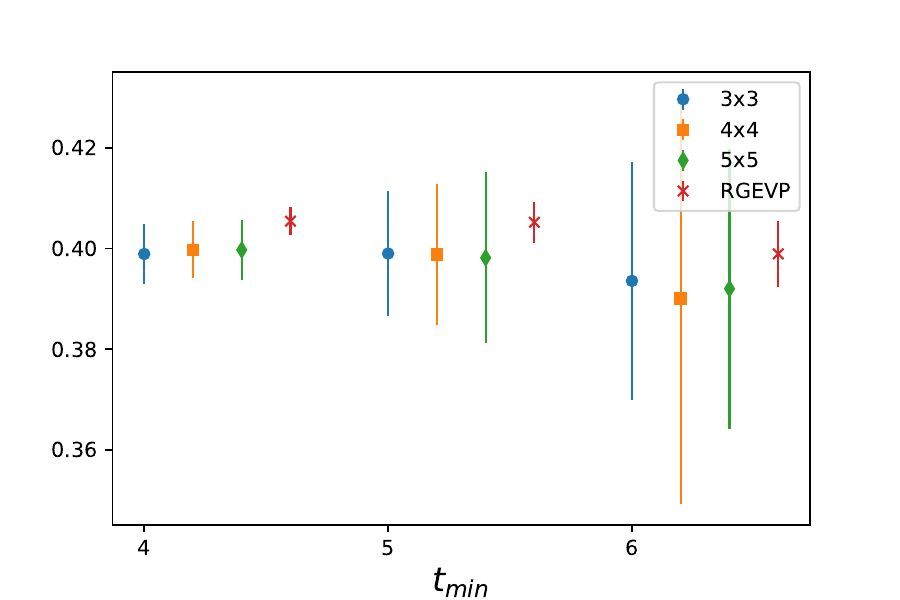}
    \caption{Same as Fig.~\ref{fig:I=2 fits 0} but result for the $I=0$ first-excited state on the $32^3$ ensemble.}
    \label{fig:I=0 L32 fits 1}
\end{figure}

\begin{figure}[tp]
    \includegraphics[width=\linewidth]{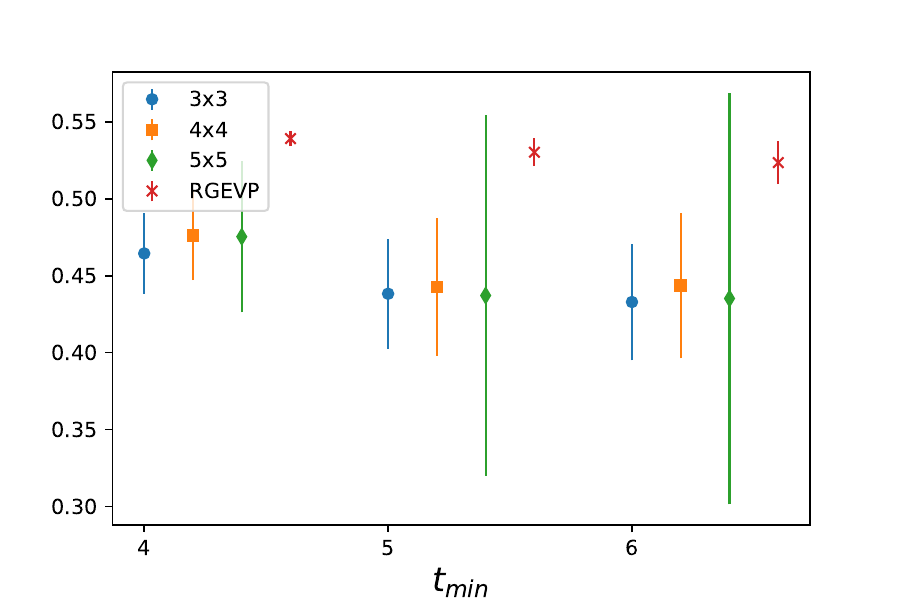}
    \caption{Same as Fig.~\ref{fig:I=2 fits 0} but result for the $I=0$ second-excited state on the $32^3$ ensemble.}
    \label{fig:I=0 L32 fits 2}
\end{figure}

\begin{figure*}
    \includegraphics[width=\linewidth]{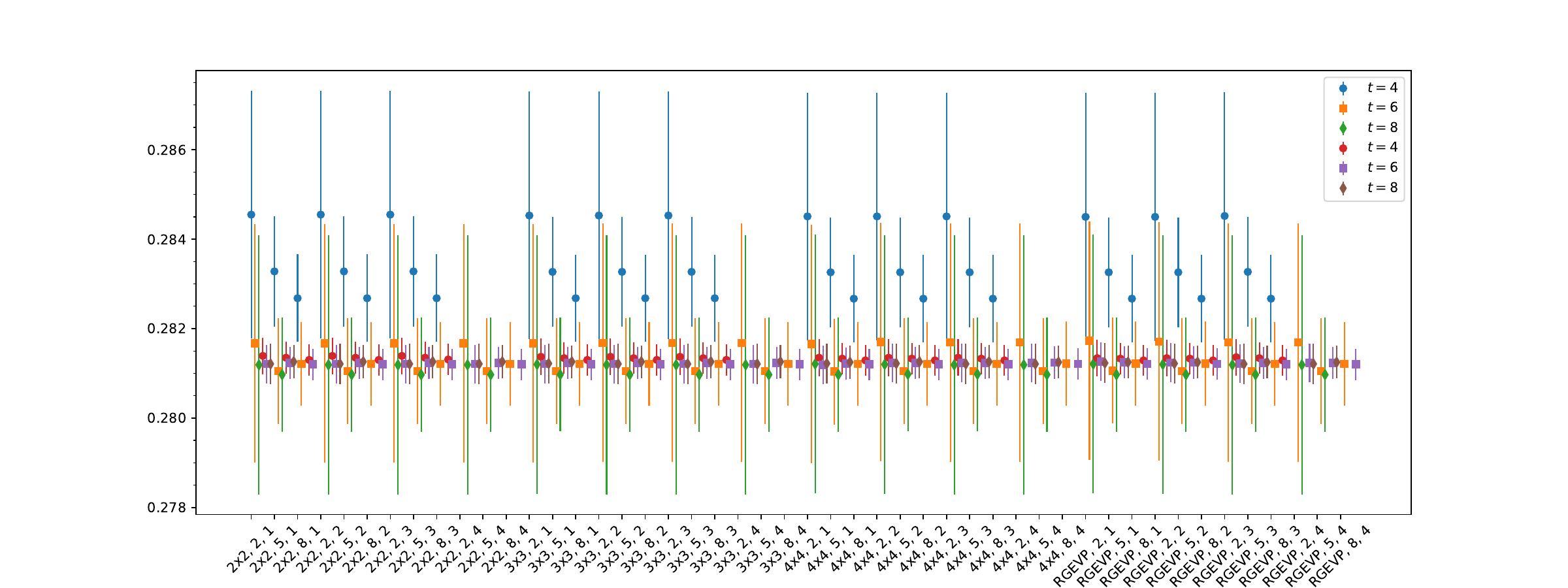}
    \caption{$I=2$ effective ground state $\pi\pi$ energies on the $24^3$ lattice plotted in lattice units. X-axis labels correspond to GEVP type, $\delta_t$ (matrix subtraction), and $t-t_0$. For each label there are up to six values, corresponding to two sets of $t=4$ (circle), 6 (square) and 8 (diamond), one set each for non-dispersion relation method and dispersion relation method, respectively.}
    \label{fig:I=2 n=0 energies}
\end{figure*}
\begin{figure*}
    \includegraphics[width=\linewidth]{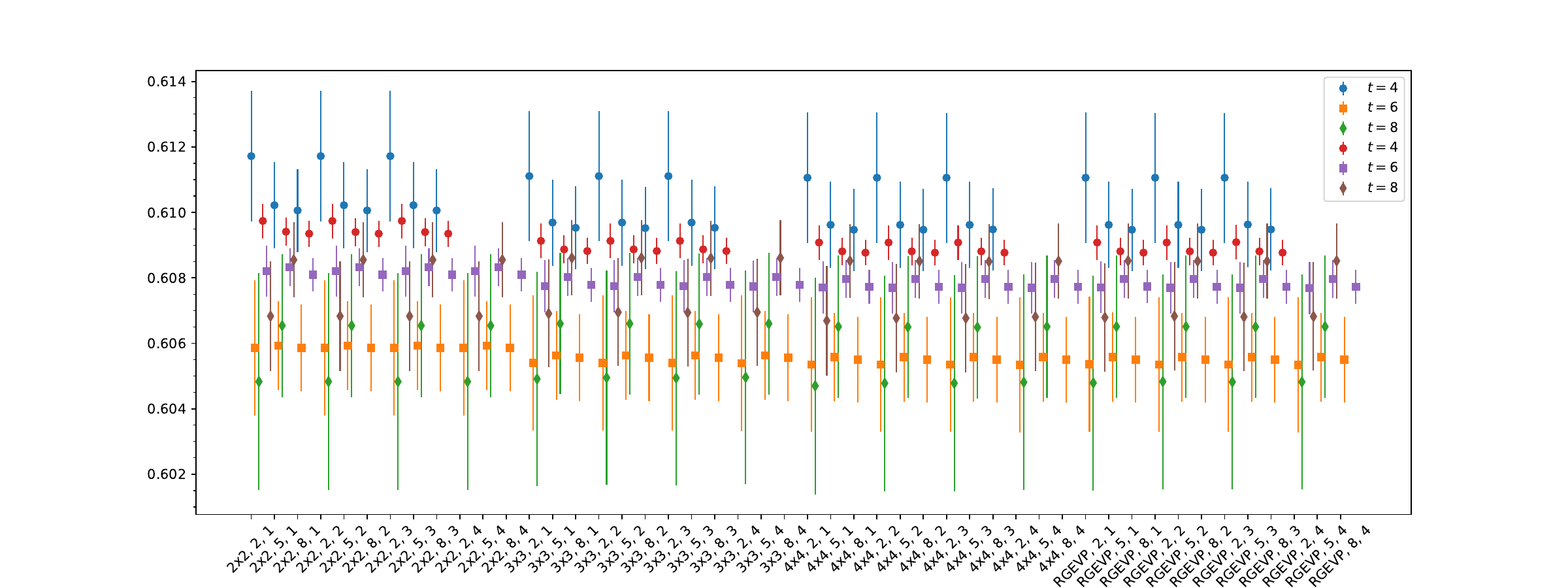}
    \caption{$I=2$ effective first excited state $\pi\pi$ energies on the $24^3$ lattice plotted in lattice units. X-axis labels correspond to GEVP type, $\delta_t$ (matrix subtraction), and $t-t_0$. For each label there are up to six values, corresponding to two sets of $t=4$ (circle), 6 (square) and 8 (diamond), one set each for non-dispersion relation method and dispersion relation method, respectively.}
    \label{fig:I=2 n=1 energies}
\end{figure*}
\begin{figure*}
    \includegraphics[width=\linewidth]{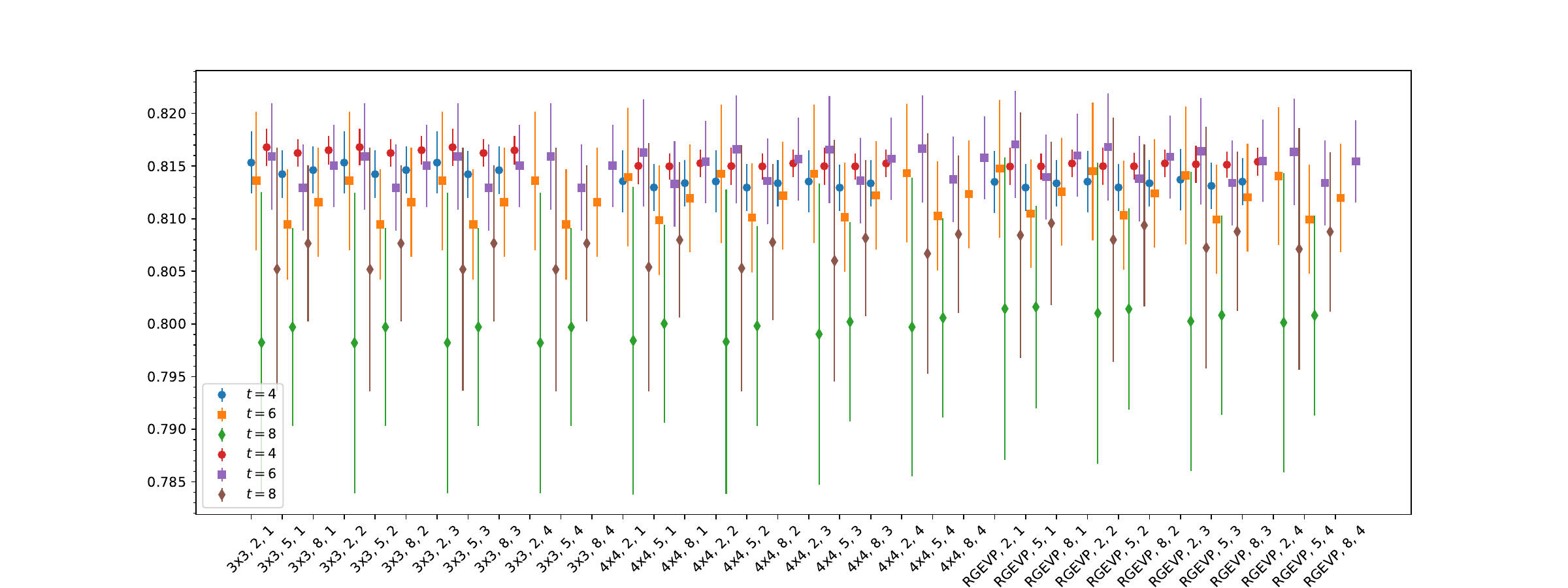}
    \caption{$I=2$ effective second excited state $\pi\pi$ energies on the $24^3$ lattice plotted in lattice units. X-axis labels correspond to GEVP type, $\delta_t$ (matrix subtraction), and $t-t_0$. For each label there are up to six values, corresponding to two sets of $t=4$ (circle), 6 (square) and 8 (diamond), one set each for non-dispersion relation method and dispersion relation method, respectively.}
    \label{fig:I=2 n=2 energies}
\end{figure*}
\begin{figure*}
    \includegraphics[width=\linewidth]{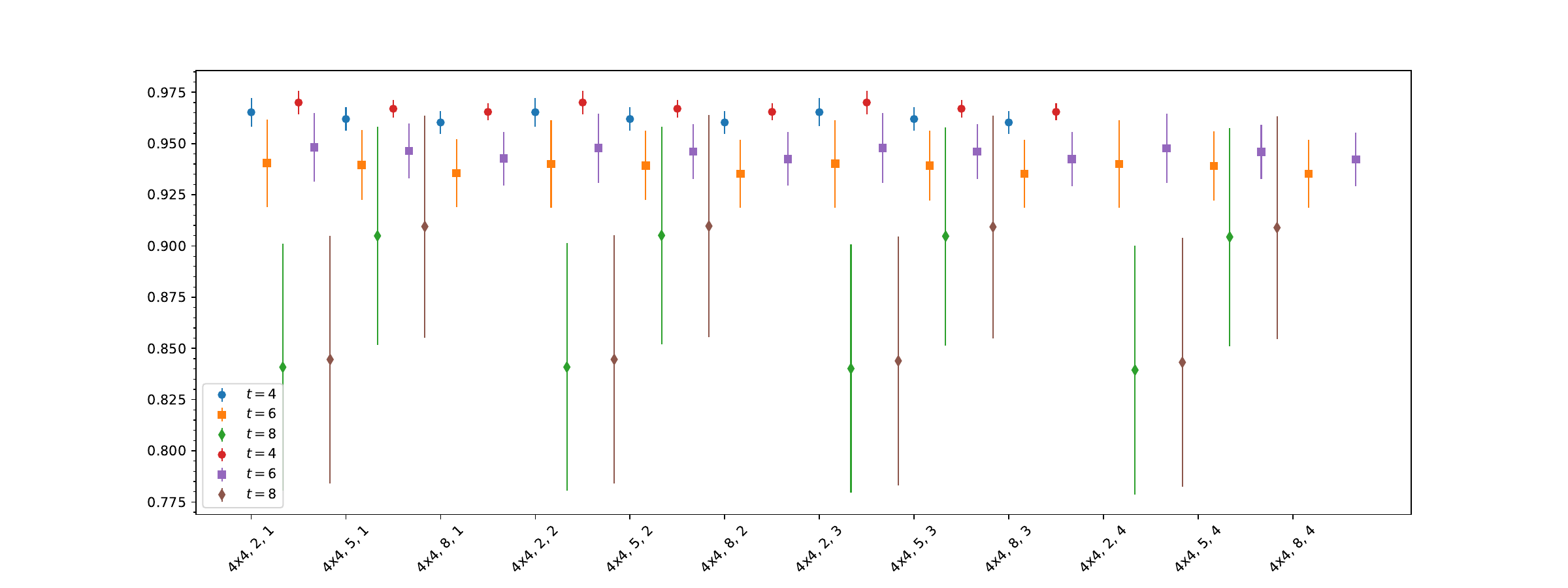}
    \caption{$I=2$ effective third excited state $\pi\pi$ energies on the $24^3$ lattice plotted in lattice units. X-axis labels correspond to GEVP type, $\delta_t$ (matrix subtraction), and $t-t_0$. For each label there are up to six values, corresponding to two sets of $t=4$ (circle), 6 (square) and 8 (diamond), one set each for non-dispersion relation method and dispersion relation method, respectively.}
    \label{fig:I=2 n=3 energies}
\end{figure*}
\begin{figure*}
    \includegraphics[width=\linewidth]{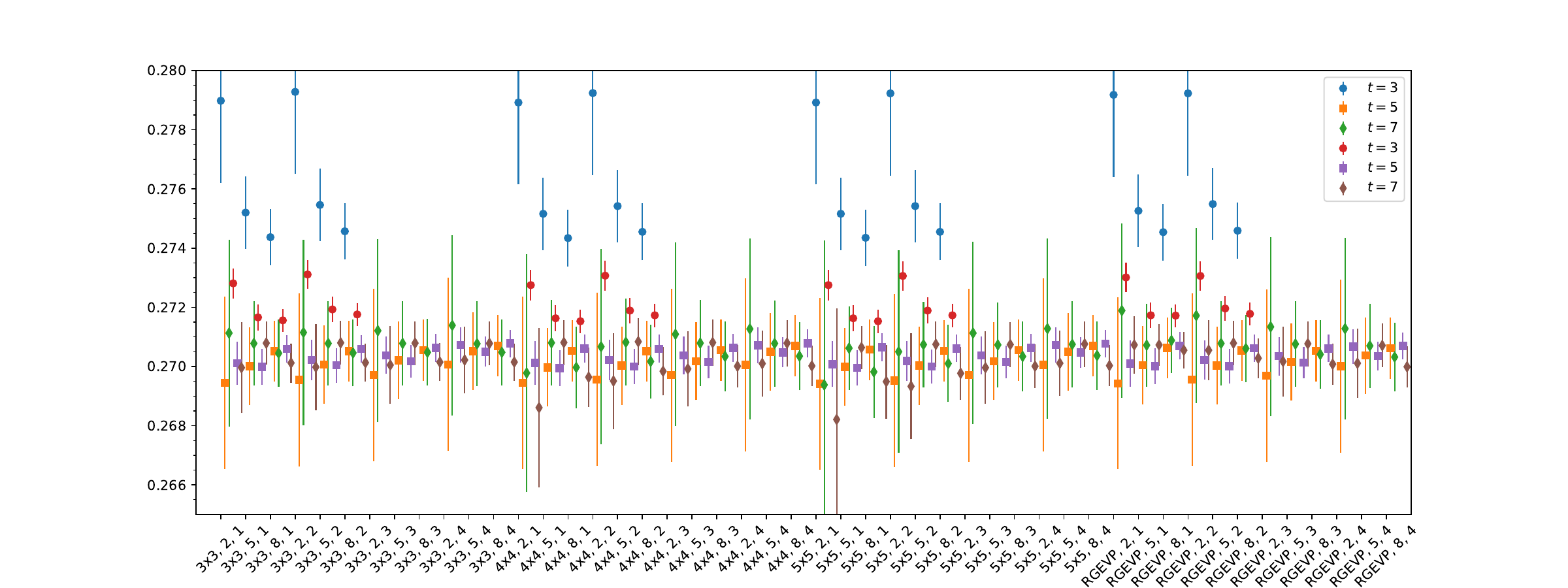}
    \caption{$I=0$ effective ground state $\pi\pi$ energies on the $24^3$ lattice plotted in lattice units. X-axis labels correspond to GEVP type, $\delta_t$ (matrix subtraction), and $t-t_0$. For each label there are up to six values, corresponding to two sets of $t=4$ (circle), 6 (square) and 8 (diamond), one set each for non-dispersion relation method and dispersion relation method, respectively.}
    \label{fig:I=0 n=0 energies}
\end{figure*}
\begin{figure*}
    \includegraphics[width=\linewidth]{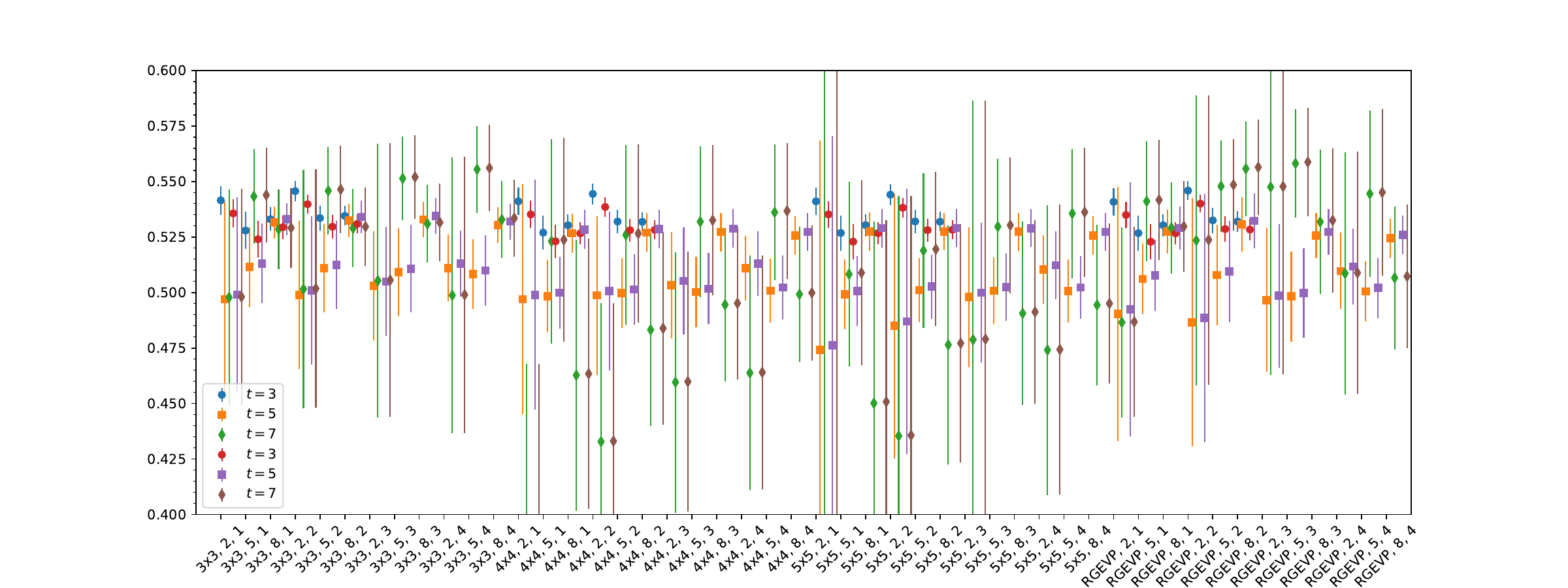}
    \caption{$I=0$ effective first excited state $\pi\pi$ energies on the $24^3$ lattice plotted in lattice units. X-axis labels correspond to GEVP type, $\delta_t$ (matrix subtraction), and $t-t_0$. For each label there are up to six values, corresponding to two sets of $t=3$ (circle), 5 (square) and 7 (diamond), one set each for non-dispersion relation method and dispersion relation method, respectively.}
    \label{fig:I=0 n=1 energies}
\end{figure*}
\begin{figure*}
    \includegraphics[width=\linewidth]{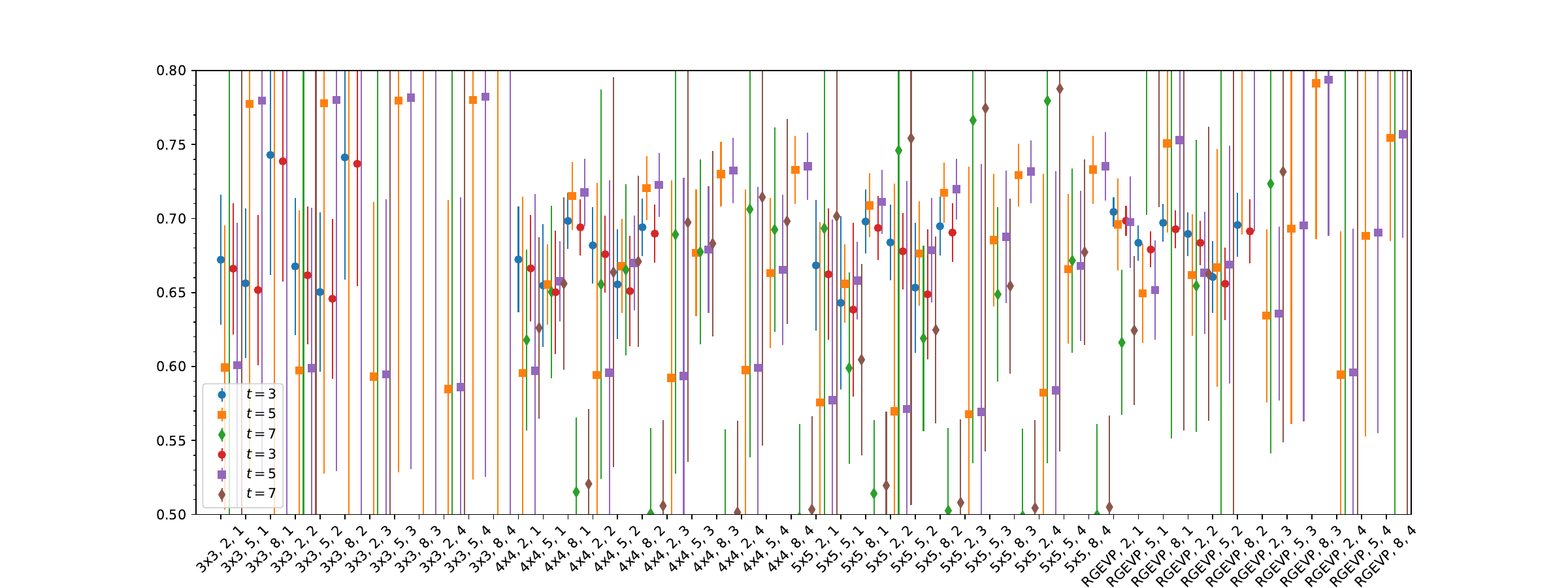}
    \caption{$I=0$ effective second excited state $\pi\pi$ energies on the $24^3$ lattice plotted in lattice units. X-axis labels correspond to GEVP type, $\delta_t$ (matrix subtraction), and $t-t_0$. For each label there are up to six values, corresponding to two sets of $t=3$ (circle), 5 (square) and 7 (diamond), one set each for non-dispersion relation method and dispersion relation method, respectively.}
    \label{fig:I=0 n=2 energies}
\end{figure*}
\begin{figure*}
    \includegraphics[width=\linewidth]{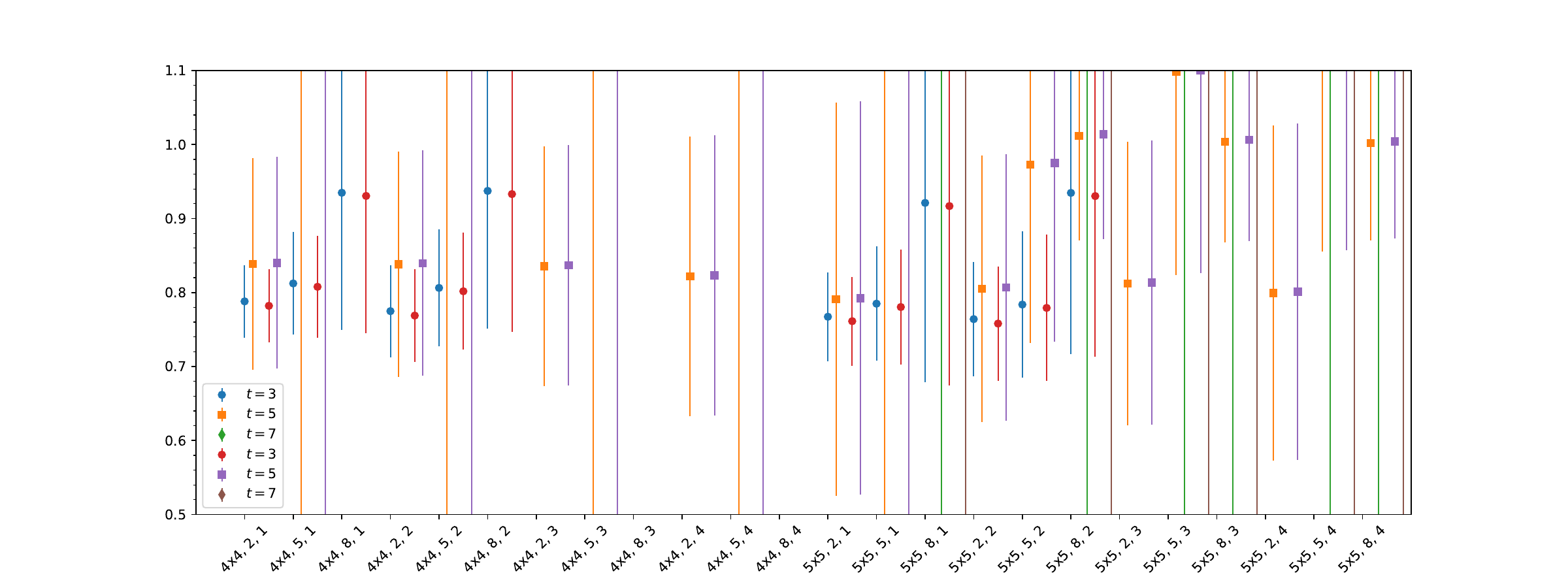}
    \caption{$I=0$ effective third excited state $\pi\pi$ energies on the $24^3$ lattice plotted in lattice units. X-axis labels correspond to GEVP type, $\delta_t$ (matrix subtraction), and $t-t_0$. For each label there are up to six values, corresponding to two sets of $t=3$ (circle), 5 (square) and 7 (diamond), one set each for non-dispersion relation method and dispersion relation method, respectively.}
    \label{fig:I=0 n=3 energies}
\end{figure*}
\begin{figure*}
    \includegraphics[width=\linewidth]{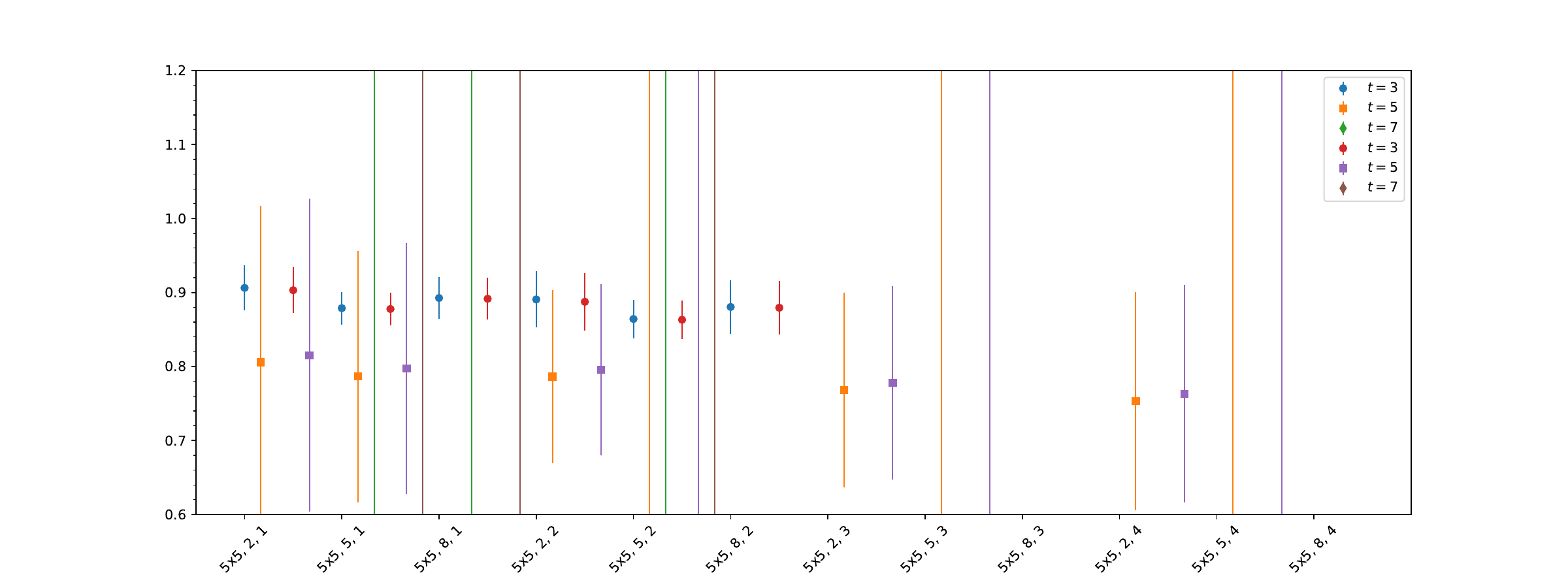}
    \caption{$I=0$ effective fourth excited state $\pi\pi$ energies on the $24^3$ lattice plotted in lattice units. X-axis labels correspond to GEVP type, $\delta_t$ (matrix subtraction), and $t-t_0$. For each label there are up to six values, corresponding to two sets of $t=3$ (circle), 5 (square) and 7 (diamond), one set each for non-dispersion relation method and dispersion relation method, respectively.}
    \label{fig:I=0 n=4 energies}
\end{figure*}
\begin{figure*}
    \includegraphics[width=\linewidth]{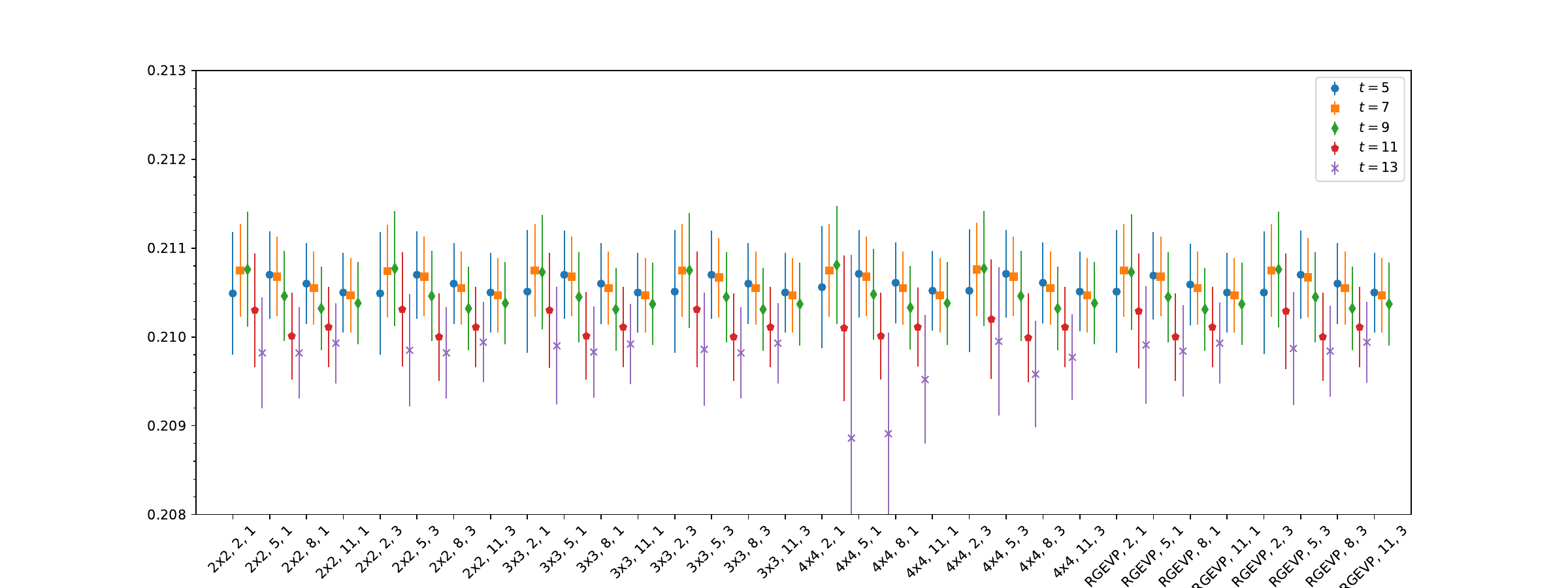}
    \caption{$I=2$ effective ground state $\pi\pi$ energies on the $32^3$ lattice plotted in lattice units. X-axis labels correspond to GEVP type, $\delta_t$ (matrix subtraction), and $t-t_0$. For each label there are results for five values of of $t=5$ (circle), 7 (square), 9 (diamond), 11 (pentagon) and 13 (cross).}
    \label{fig:I=2 n=0 L32 energies}
\end{figure*}
\begin{figure*}
    \includegraphics[width=\linewidth]{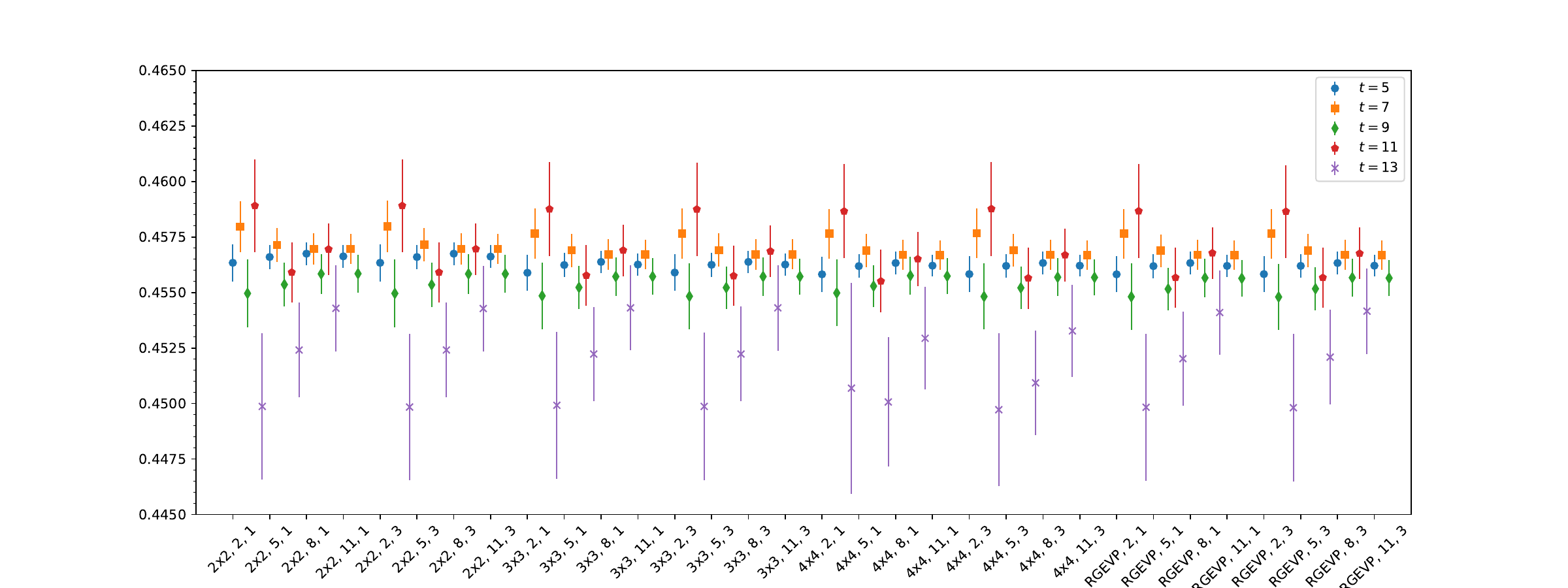}
    \caption{$I=2$ effective first excited state $\pi\pi$ energies on the $32^3$ lattice plotted in lattice units. X-axis labels correspond to GEVP type, $\delta_t$ (matrix subtraction), and $t-t_0$. For each label there are results for five values of of $t=5$ (circle), 7 (square), 9 (diamond), 11 (pentagon) and 13 (cross).}
    \label{fig:I=2 n=1 L32 energies}
\end{figure*}
\begin{figure*}
    \includegraphics[width=\linewidth]{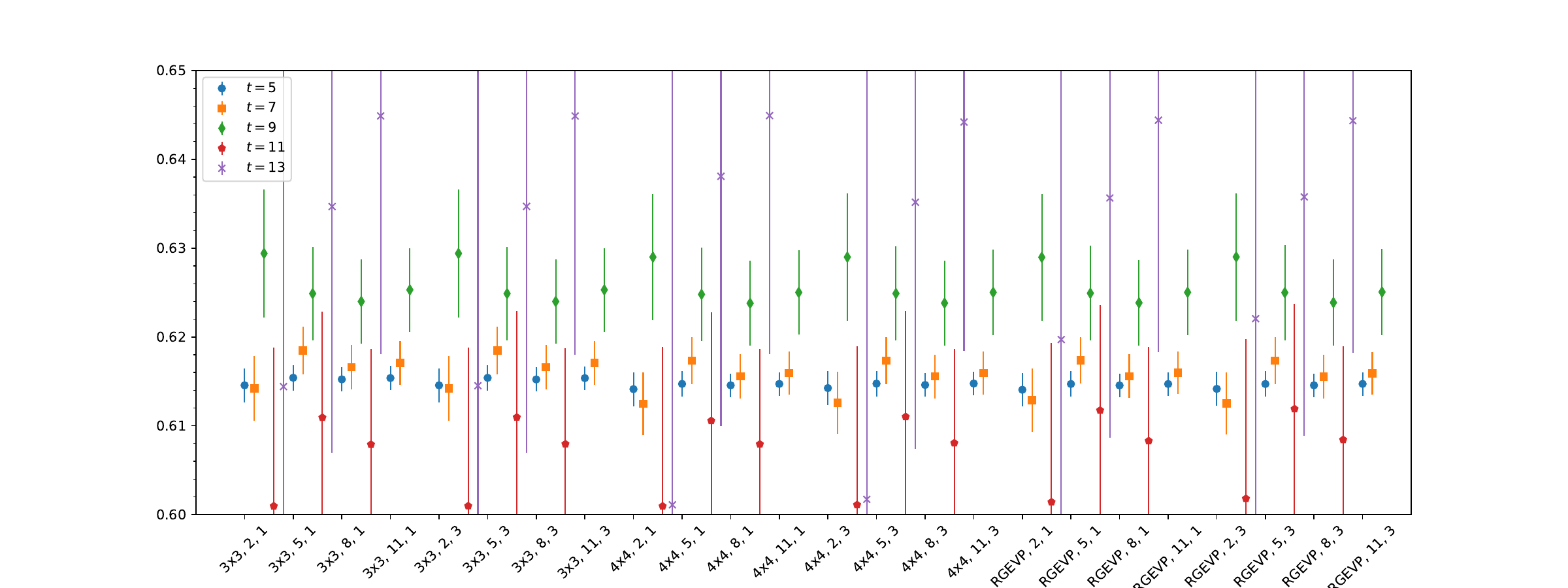}
    \caption{$I=2$ effective second excited state $\pi\pi$ energies on the $32^3$ lattice plotted in lattice units. X-axis labels correspond to GEVP type, $\delta_t$ (matrix subtraction), and $t-t_0$. For each label there are results for five values of of $t=5$ (circle), 7 (square), 9 (diamond), 11 (pentagon) and 13 (cross).}
    \label{fig:I=2 n=2 L32 energies}
\end{figure*}
\begin{figure*}
    \includegraphics[width=\linewidth]{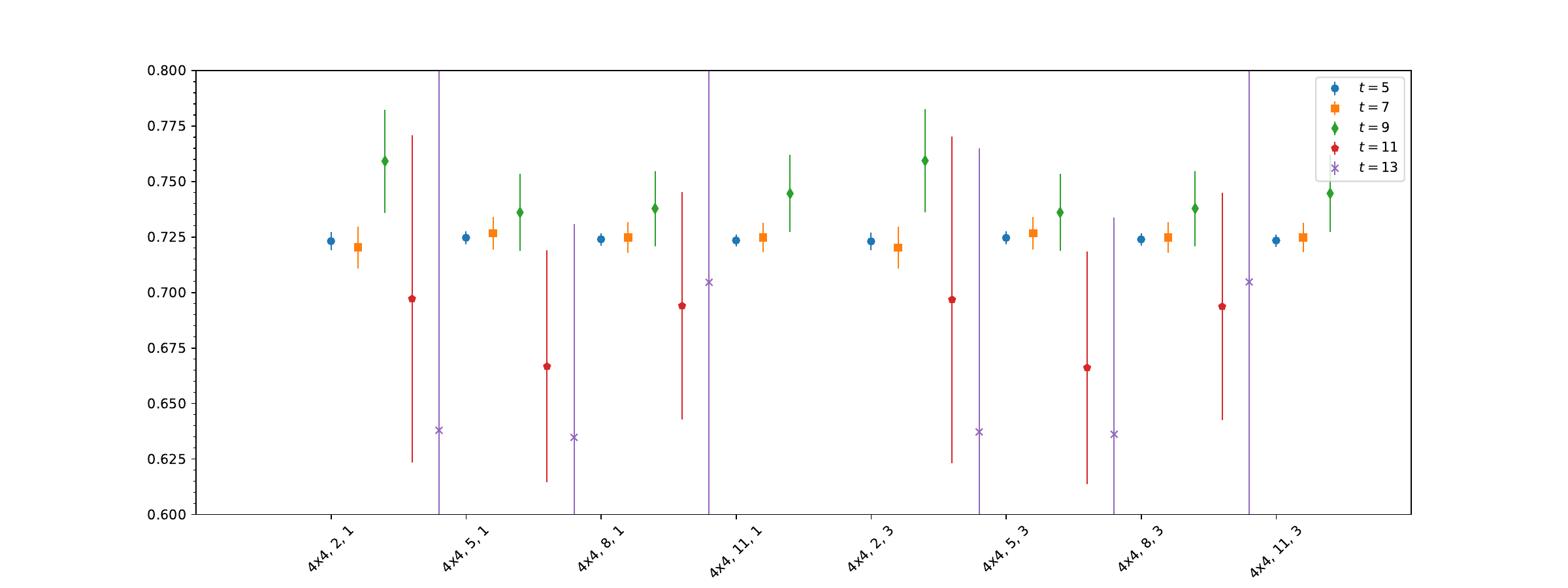}
    \caption{$I=2$ effective third excited state $\pi\pi$ energies on the $32^3$ lattice plotted in lattice units. X-axis labels correspond to GEVP type, $\delta_t$ (matrix subtraction), and $t-t_0$. For each label there are results for five values of of $t=5$ (circle), 7 (square), 9 (diamond), 11 (pentagon) and 13 (cross).}
    \label{fig:I=2 n=3 L32 energies}
\end{figure*}

\begin{figure*}
    \includegraphics[width=\linewidth]{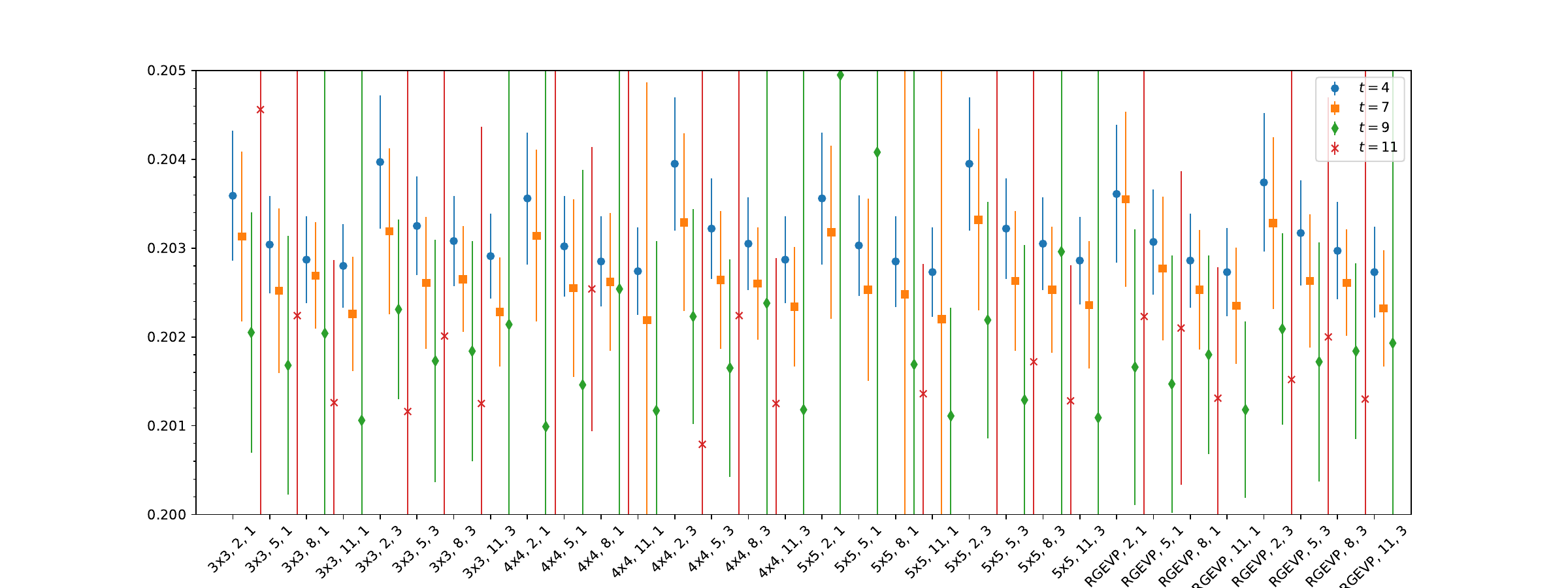}
    \caption{$I=0$ effective ground state $\pi\pi$ energies on the $32^3$ lattice plotted in lattice units. X-axis labels correspond to GEVP type, $\delta_t$ (matrix subtraction), and $t-t_0$. For each label there are results for four values of of $t=4$ (circle), 7 (square), 9 (diamond) and 11 (cross).}
    \label{fig:I=0 n=0 L32 energies}
\end{figure*}
\begin{figure*}
    \includegraphics[width=\linewidth]{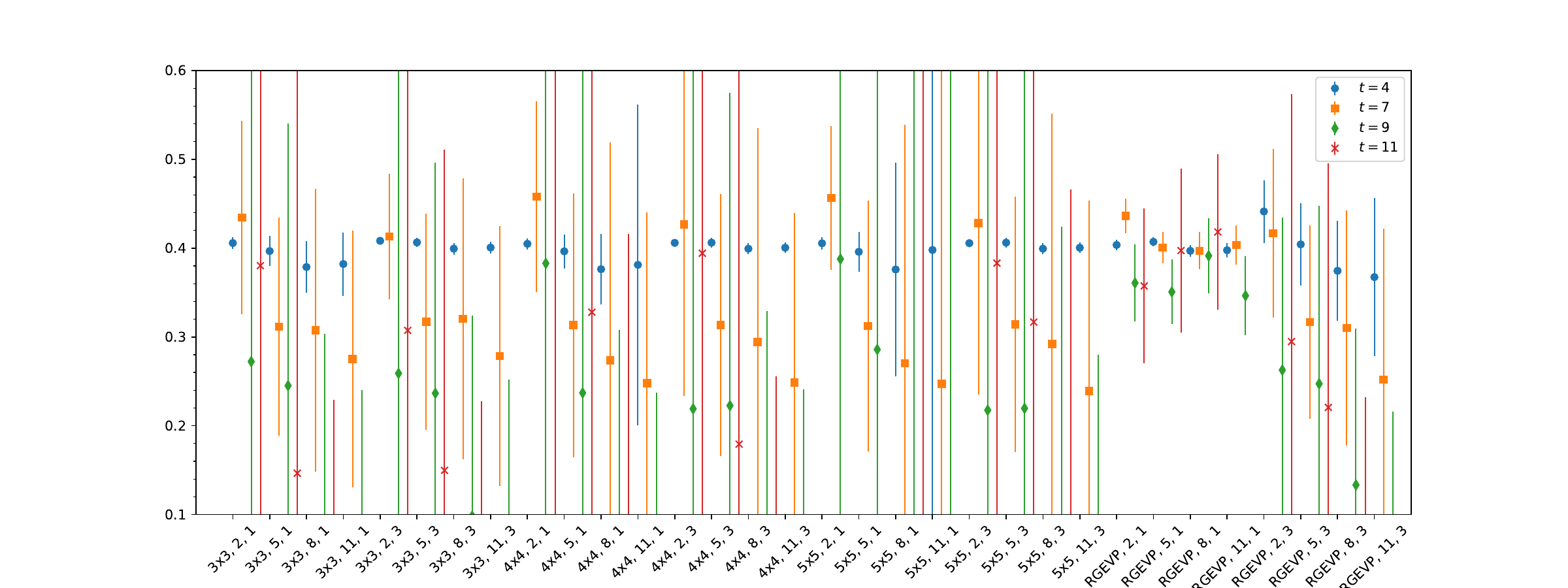}
    \caption{$I=0$ effective first excited state $\pi\pi$ energies on the $32^3$ lattice plotted in lattice units. X-axis labels correspond to GEVP type, $\delta_t$ (matrix subtraction), and $t-t_0$. For each label there are results for four values of of $t=4$ (circle), 7 (square), 9 (diamond) and 11 (cross).}
    \label{fig:I=0 n=1 L32 energies}
\end{figure*}
\begin{figure*}
    \includegraphics[width=\linewidth]{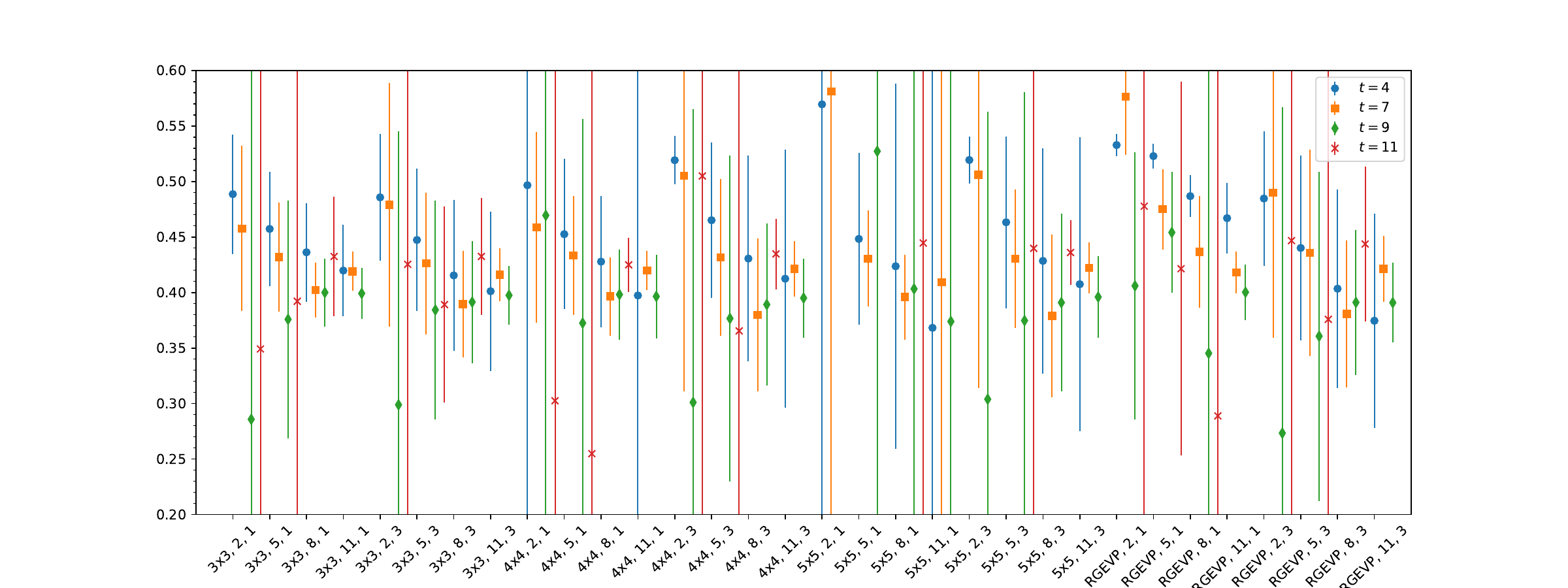}
    \caption{$I=0$ effective second excited state $\pi\pi$ energies on the $32^3$ lattice plotted in lattice units. X-axis labels correspond to GEVP type, $\delta_t$ (matrix subtraction), and $t-t_0$. For each label there are results for four values of of $t=4$ (circle), 7 (square), 9 (diamond) and 11 (cross).}
    \label{fig:I=0 n=2 L32 energies}
\end{figure*}
\begin{figure*}
    \includegraphics[width=\linewidth]{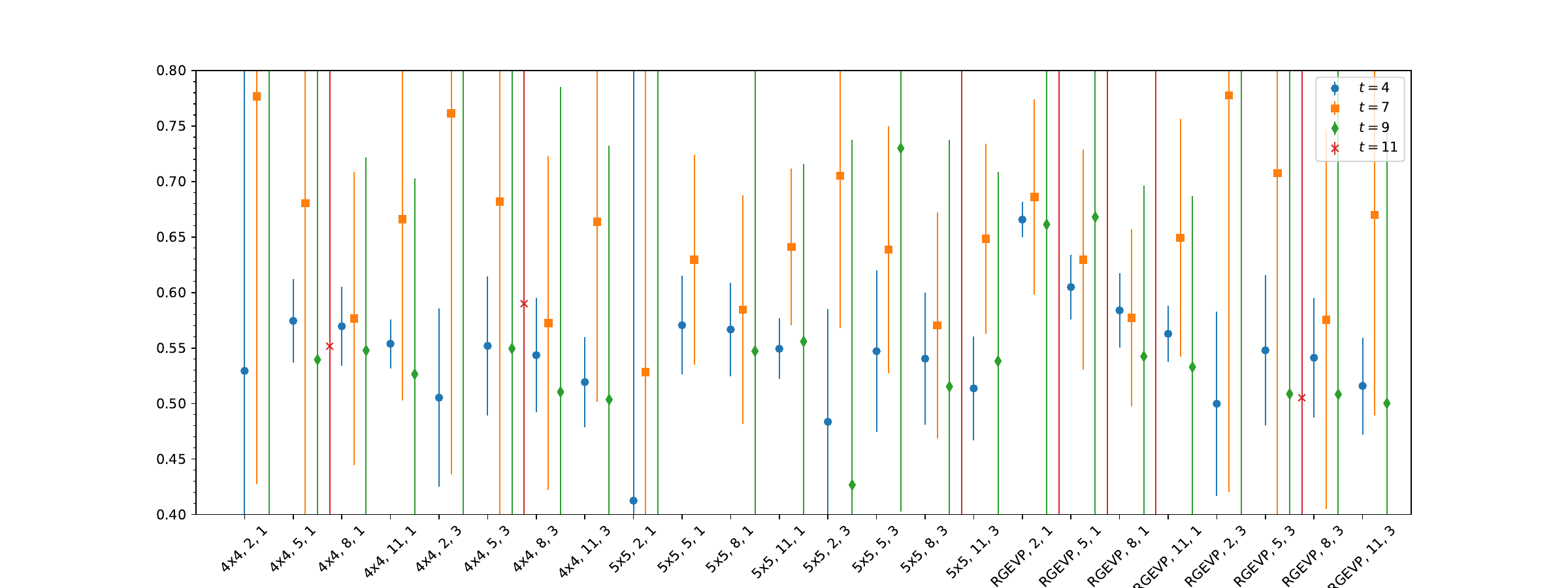}
    \caption{$I=0$ effective third excited state $\pi\pi$ energies on the $32^3$ lattice plotted in lattice units. X-axis labels correspond to GEVP type, $\delta_t$ (matrix subtraction), and $t-t_0$. For each label there are results for four values of of $t=4$ (circle), 7 (square), 9 (diamond) and 11 (cross).}
    \label{fig:I=0 n=3 L32 energies}
\end{figure*}
\begin{figure*}
    \includegraphics[width=\linewidth]{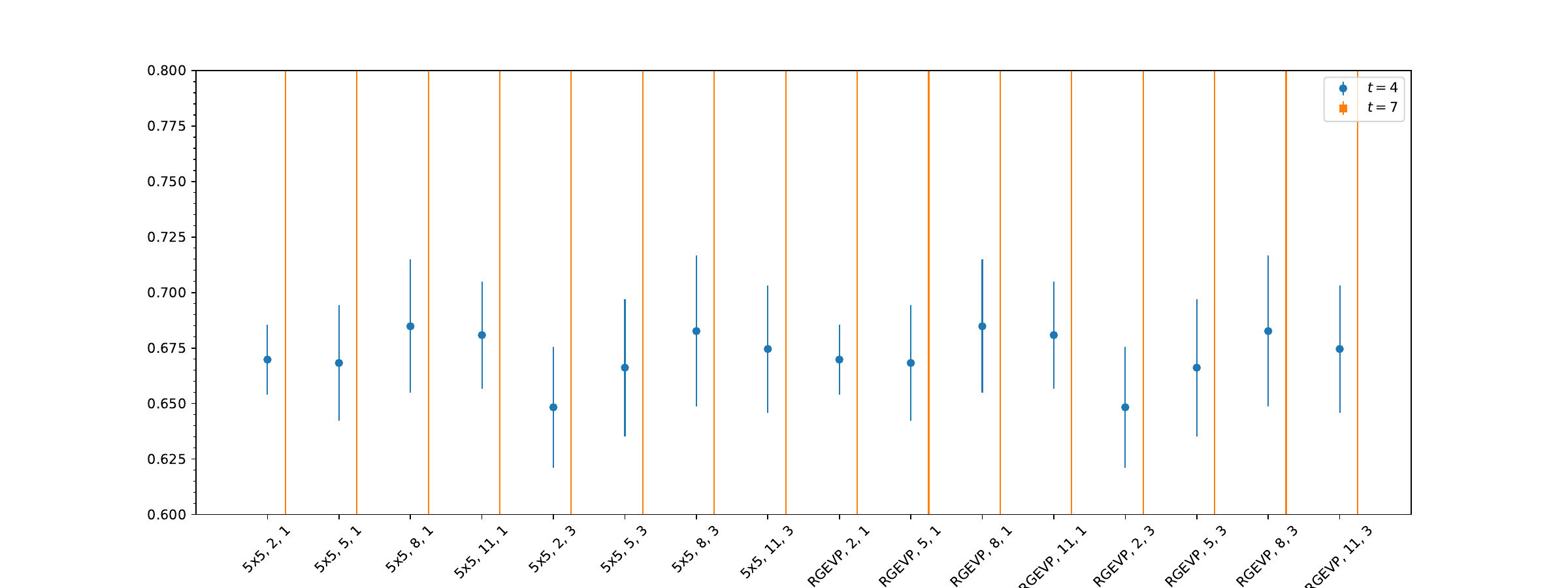}
    \caption{$I=0$ effective fourth excited state $\pi\pi$ energies on the $32^3$ lattice plotted in lattice units. X-axis labels correspond to GEVP type, $\delta_t$ (matrix subtraction), and $t-t_0$. For each label there are results for two values of of $t=4$ (circle) and 7 (square).}
    \label{fig:I=0 n=4 L32 energies}
\end{figure*}

\bibliographystyle{apsrev4-1}
\bibliography{main}

\begin{table*}[tp]
\centering
\begin{tabular}{|ccc|cccccc|}
\hline
 GEVP type & $\delta_t$ & $t-t_0$ & $t = 4$ & $t = 6$ & $t = 8$ & $t = 4$ w DR & $t = 6$ w DR & $t = 8$ w DR\\
\hline
$2\times2$ & 2 & 1 & 0.2845(28)  & 0.2817(27)  & 0.2812(29)  & 0.28139(41)  & 0.28121(43)  & 0.28121(45) \\
$2\times2$ & 5 & 1 & 0.2833(12)  & 0.2810(12)  & 0.2810(13)  & 0.28135(35)  & 0.28123(36)  & 0.28126(37) \\
$2\times2$ & 8 & 1 & 0.28268(98)  & 0.28121(94) & --- & 0.28130(35)  & 0.28120(35) & ---\\
$2\times2$ & 2 & 2 & 0.2845(28)  & 0.2817(27)  & 0.2812(29)  & 0.28139(41)  & 0.28121(43)  & 0.28121(45) \\
$2\times2$ & 5 & 2 & 0.2833(12)  & 0.2810(12)  & 0.2810(13)  & 0.28135(35)  & 0.28123(36)  & 0.28126(37) \\
$2\times2$ & 8 & 2 & 0.28268(98)  & 0.28121(94) & --- & 0.28130(35)  & 0.28120(35) & ---\\
$2\times2$ & 2 & 3 & 0.2846(28)  & 0.2817(27)  & 0.2812(29)  & 0.28139(41)  & 0.28121(43)  & 0.28121(45) \\
$2\times2$ & 5 & 3 & 0.2833(12)  & 0.2810(12)  & 0.2810(13)  & 0.28135(35)  & 0.28123(36)  & 0.28126(37) \\
$2\times2$ & 8 & 3 & 0.28268(98)  & 0.28121(94) & --- & 0.28131(35)  & 0.28120(35) & ---\\
$2\times2$ & 2 & 4& --- & 0.2817(27)  & 0.2812(29) & --- & 0.28121(43)  & 0.28121(45) \\
$2\times2$ & 5 & 4& --- & 0.2810(12)  & 0.2810(13) & --- & 0.28123(36)  & 0.28126(37) \\
$2\times2$ & 8 & 4& --- & 0.28121(94) & ---& --- & 0.28120(35) & ---\\
$3\times3$ & 2 & 1 & 0.2845(28)  & 0.2817(27)  & 0.2812(29)  & 0.28137(41)  & 0.28121(43)  & 0.28122(45) \\
$3\times3$ & 5 & 1 & 0.2833(12)  & 0.2810(12)  & 0.2810(13)  & 0.28134(35)  & 0.28123(36)  & 0.28126(37) \\
$3\times3$ & 8 & 1 & 0.28268(98)  & 0.28121(94) & --- & 0.28130(35)  & 0.28120(35) & ---\\
$3\times3$ & 2 & 2 & 0.2845(28)  & 0.2817(27)  & 0.2812(29)  & 0.28137(41)  & 0.28121(43)  & 0.28121(45) \\
$3\times3$ & 5 & 2 & 0.2833(12)  & 0.2810(12)  & 0.2810(13)  & 0.28134(35)  & 0.28123(36)  & 0.28126(37) \\
$3\times3$ & 8 & 2 & 0.28268(98)  & 0.28121(94) & --- & 0.28130(35)  & 0.28120(35) & ---\\
$3\times3$ & 2 & 3 & 0.2845(28)  & 0.2817(27)  & 0.2812(29)  & 0.28137(41)  & 0.28121(43)  & 0.28121(45) \\
$3\times3$ & 5 & 3 & 0.2833(12)  & 0.2810(12)  & 0.2810(13)  & 0.28134(35)  & 0.28123(36)  & 0.28126(37) \\
$3\times3$ & 8 & 3 & 0.28268(98)  & 0.28121(94) & --- & 0.28130(35)  & 0.28120(35) & ---\\
$3\times3$ & 2 & 4& --- & 0.2817(27)  & 0.2812(29) & --- & 0.28121(43)  & 0.28121(45) \\
$3\times3$ & 5 & 4& --- & 0.2810(12)  & 0.2810(13) & --- & 0.28123(36)  & 0.28126(37) \\
$3\times3$ & 8 & 4& --- & 0.28121(94) & ---& --- & 0.28120(35) & ---\\
$4\times4$ & 2 & 1 & 0.2845(28)  & 0.2817(27)  & 0.2812(29)  & 0.28135(41)  & 0.28119(43)  & 0.28122(45) \\
$4\times4$ & 5 & 1 & 0.2833(12)  & 0.2810(12)  & 0.2810(13)  & 0.28133(35)  & 0.28122(36)  & 0.28125(37) \\
$4\times4$ & 8 & 1 & 0.28267(98)  & 0.28121(94) & --- & 0.28129(35)  & 0.28120(35) & ---\\
$4\times4$ & 2 & 2 & 0.2845(28)  & 0.2817(27)  & 0.2812(29)  & 0.28135(41)  & 0.28123(43)  & 0.28122(45) \\
$4\times4$ & 5 & 2 & 0.2833(12)  & 0.2811(12)  & 0.2810(13)  & 0.28133(35)  & 0.28123(36)  & 0.28126(37) \\
$4\times4$ & 8 & 2 & 0.28267(98)  & 0.28121(94) & --- & 0.28129(35)  & 0.28120(35) & ---\\
$4\times4$ & 2 & 3 & 0.2845(28)  & 0.2817(27)  & 0.2812(29)  & 0.28135(41)  & 0.28123(43)  & 0.28121(45) \\
$4\times4$ & 5 & 3 & 0.2833(12)  & 0.2811(12)  & 0.2810(13)  & 0.28133(35)  & 0.28123(36)  & 0.28126(37) \\
$4\times4$ & 8 & 3 & 0.28267(98)  & 0.28121(94) & --- & 0.28129(35)  & 0.28120(35) & ---\\
$4\times4$ & 2 & 4& --- & 0.2817(27)  & 0.2812(29) & --- & 0.28123(43)  & 0.28121(45) \\
$4\times4$ & 5 & 4& --- & 0.2811(12)  & 0.2810(13) & --- & 0.28124(36)  & 0.28125(37) \\
$4\times4$ & 8 & 4& --- & 0.28122(94) & ---& --- & 0.28121(35) & ---\\
RGEVP & 2 & 1 & 0.2845(28)  & 0.2817(27)  & 0.2812(29)  & 0.28134(41)  & 0.28126(43)  & 0.28123(45) \\
RGEVP & 5 & 1 & 0.2833(12)  & 0.2811(12)  & 0.2810(13)  & 0.28133(35)  & 0.28125(36)  & 0.28125(37) \\
RGEVP & 8 & 1 & 0.28267(98)  & 0.28122(94) & --- & 0.28129(35)  & 0.28121(35) & ---\\
RGEVP & 2 & 2 & 0.2845(28)  & 0.2817(27)  & 0.2812(29)  & 0.28134(41)  & 0.28124(43)  & 0.28122(45) \\
RGEVP & 5 & 2 & 0.2833(12)  & 0.2811(12)  & 0.2810(13)  & 0.28133(35)  & 0.28124(36)  & 0.28125(37) \\
RGEVP & 8 & 2 & 0.28267(98)  & 0.28122(94) & --- & 0.28129(35)  & 0.28121(35) & ---\\
RGEVP & 2 & 3 & 0.2845(28)  & 0.2817(27)  & 0.2812(29)  & 0.28136(41)  & 0.28122(43)  & 0.28121(45) \\
RGEVP & 5 & 3 & 0.2833(12)  & 0.2810(12)  & 0.2810(13)  & 0.28134(35)  & 0.28123(36)  & 0.28125(37) \\
RGEVP & 8 & 3 & 0.28267(98)  & 0.28121(94) & --- & 0.28129(35)  & 0.28120(35) & ---\\
RGEVP & 2 & 4& --- & 0.2817(27)  & 0.2812(29) & --- & 0.28123(43)  & 0.28121(45) \\
RGEVP & 5 & 4& --- & 0.2810(12)  & 0.2810(13) & --- & 0.28123(36)  & 0.28125(37) \\
RGEVP & 8 & 4& --- & 0.28121(94) & ---& --- & 0.28120(35) & ---\\
\hline
\end{tabular}
\caption{Effective energy of the $I=2$ two-pion ground state on the $24^3$ lattice for various GEVP methods and input time parameters: $\delta_t$, $t-t_0$ and $t$.  The values are shown in lattice units.  The re-basing matrix is calculated as: $4\times4\to 3\times3$ at $t_0 = 4$.  The dashes `---' mean that the effective energy could not be evaluated because of one of the following possible reasons: 1.~Correlators at $t+\delta_t+1$ that are needed for calculating effective energy are not computed.  2.~The ratio of corresponding GEVP eigenvalues $\lambda_0(t,t_0)/\lambda_0(t+1,t_0)$ is negative for at least one jackknife sample.}
\label{tab:efm_I2_n0_L24}
\end{table*}

\begin{table*}[tp]
\centering
\begin{tabular}{|ccc|cccccc|}
\hline
 GEVP type & $\delta_t$ & $t-t_0$ & $t = 4$ & $t = 6$ & $t = 8$ & $t = 4$ w DR & $t = 6$ w DR & $t = 8$ w DR\\
\hline
$2\times2$ & 2 & 1 & 0.6117(20)  & 0.6059(21)  & 0.6048(33)  & 0.60974(53)  & 0.60821(78)  & 0.6068(17) \\
$2\times2$ & 5 & 1 & 0.6102(13)  & 0.6059(14)  & 0.6065(22)  & 0.60941(43)  & 0.60833(58)  & 0.6085(11) \\
$2\times2$ & 8 & 1 & 0.6101(13)  & 0.6059(13) & --- & 0.60935(40)  & 0.60810(51) & ---\\
$2\times2$ & 2 & 2 & 0.6117(20)  & 0.6059(21)  & 0.6048(33)  & 0.60974(53)  & 0.60821(78)  & 0.6068(17) \\
$2\times2$ & 5 & 2 & 0.6102(13)  & 0.6059(14)  & 0.6065(22)  & 0.60940(43)  & 0.60833(58)  & 0.6085(11) \\
$2\times2$ & 8 & 2 & 0.6101(13)  & 0.6059(13) & --- & 0.60935(40)  & 0.60810(51) & ---\\
$2\times2$ & 2 & 3 & 0.6117(20)  & 0.6059(21)  & 0.6048(33)  & 0.60974(53)  & 0.60821(78)  & 0.6068(17) \\
$2\times2$ & 5 & 3 & 0.6102(13)  & 0.6059(14)  & 0.6065(22)  & 0.60940(43)  & 0.60833(58)  & 0.6085(11) \\
$2\times2$ & 8 & 3 & 0.6101(13)  & 0.6059(13) & --- & 0.60935(40)  & 0.60810(51) & ---\\
$2\times2$ & 2 & 4& --- & 0.6059(21)  & 0.6048(33) & --- & 0.60821(78)  & 0.6068(17) \\
$2\times2$ & 5 & 4& --- & 0.6059(14)  & 0.6065(22) & --- & 0.60833(58)  & 0.6085(11) \\
$2\times2$ & 8 & 4& --- & 0.6059(13) & ---& --- & 0.60810(51) & ---\\
$3\times3$ & 2 & 1 & 0.6111(20)  & 0.6054(21)  & 0.6049(33)  & 0.60913(53)  & 0.60775(79)  & 0.6069(16) \\
$3\times3$ & 5 & 1 & 0.6097(13)  & 0.6056(14)  & 0.6066(22)  & 0.60887(43)  & 0.60803(58)  & 0.6086(11) \\
$3\times3$ & 8 & 1 & 0.6095(13)  & 0.6056(13) & --- & 0.60882(40)  & 0.60779(52) & ---\\
$3\times3$ & 2 & 2 & 0.6111(20)  & 0.6054(21)  & 0.6050(33)  & 0.60913(53)  & 0.60775(79)  & 0.6070(16) \\
$3\times3$ & 5 & 2 & 0.6097(13)  & 0.6056(14)  & 0.6066(22)  & 0.60887(43)  & 0.60803(58)  & 0.6086(11) \\
$3\times3$ & 8 & 2 & 0.6095(13)  & 0.6056(13) & --- & 0.60882(40)  & 0.60779(52) & ---\\
$3\times3$ & 2 & 3 & 0.6111(20)  & 0.6054(21)  & 0.6049(33)  & 0.60913(53)  & 0.60775(79)  & 0.6069(16) \\
$3\times3$ & 5 & 3 & 0.6097(13)  & 0.6056(14)  & 0.6066(22)  & 0.60887(43)  & 0.60803(58)  & 0.6086(11) \\
$3\times3$ & 8 & 3 & 0.6095(13)  & 0.6056(13) & --- & 0.60882(40)  & 0.60779(52) & ---\\
$3\times3$ & 2 & 4& --- & 0.6054(21)  & 0.6050(33) & --- & 0.60774(79)  & 0.6070(16) \\
$3\times3$ & 5 & 4& --- & 0.6056(14)  & 0.6066(22) & --- & 0.60803(58)  & 0.6086(11) \\
$3\times3$ & 8 & 4& --- & 0.6056(13) & ---& --- & 0.60779(52) & ---\\
$4\times4$ & 2 & 1 & 0.6111(20)  & 0.6054(21)  & 0.6047(33)  & 0.60908(53)  & 0.60771(79)  & 0.6067(17) \\
$4\times4$ & 5 & 1 & 0.6096(13)  & 0.6056(14)  & 0.6065(22)  & 0.60881(42)  & 0.60797(58)  & 0.6085(11) \\
$4\times4$ & 8 & 1 & 0.6095(13)  & 0.6055(13) & --- & 0.60877(39)  & 0.60773(52) & ---\\
$4\times4$ & 2 & 2 & 0.6111(20)  & 0.6053(21)  & 0.6048(33)  & 0.60908(53)  & 0.60770(79)  & 0.6068(17) \\
$4\times4$ & 5 & 2 & 0.6096(13)  & 0.6056(14)  & 0.6065(22)  & 0.60881(42)  & 0.60797(58)  & 0.6085(11) \\
$4\times4$ & 8 & 2 & 0.6095(13)  & 0.6055(13) & --- & 0.60877(39)  & 0.60773(52) & ---\\
$4\times4$ & 2 & 3 & 0.6111(20)  & 0.6053(21)  & 0.6048(33)  & 0.60908(53)  & 0.60770(79)  & 0.6068(17) \\
$4\times4$ & 5 & 3 & 0.6096(13)  & 0.6056(14)  & 0.6065(22)  & 0.60881(42)  & 0.60797(58)  & 0.6085(11) \\
$4\times4$ & 8 & 3 & 0.6095(13)  & 0.6055(13) & --- & 0.60877(39)  & 0.60773(52) & ---\\
$4\times4$ & 2 & 4& --- & 0.6053(21)  & 0.6048(33) & --- & 0.60770(79)  & 0.6068(17) \\
$4\times4$ & 5 & 4& --- & 0.6056(14)  & 0.6065(22) & --- & 0.60797(58)  & 0.6085(11) \\
$4\times4$ & 8 & 4& --- & 0.6055(13) & ---& --- & 0.60773(52) & ---\\
RGEVP & 2 & 1 & 0.6111(20)  & 0.6054(21)  & 0.6048(33)  & 0.60908(53)  & 0.60771(79)  & 0.6068(17) \\
RGEVP & 5 & 1 & 0.6096(13)  & 0.6056(14)  & 0.6065(22)  & 0.60881(42)  & 0.60797(58)  & 0.6085(11) \\
RGEVP & 8 & 1 & 0.6095(13)  & 0.6055(13) & --- & 0.60877(39)  & 0.60774(52) & ---\\
RGEVP & 2 & 2 & 0.6111(20)  & 0.6053(21)  & 0.6048(33)  & 0.60908(53)  & 0.60770(79)  & 0.6068(17) \\
RGEVP & 5 & 2 & 0.6096(13)  & 0.6056(14)  & 0.6065(22)  & 0.60881(42)  & 0.60797(58)  & 0.6085(11) \\
RGEVP & 8 & 2 & 0.6095(13)  & 0.6055(13) & --- & 0.60877(39)  & 0.60773(52) & ---\\
RGEVP & 2 & 3 & 0.6111(20)  & 0.6053(21)  & 0.6048(33)  & 0.60909(53)  & 0.60770(79)  & 0.6068(17) \\
RGEVP & 5 & 3 & 0.6096(13)  & 0.6056(14)  & 0.6065(22)  & 0.60881(42)  & 0.60797(58)  & 0.6085(11) \\
RGEVP & 8 & 3 & 0.6095(13)  & 0.6055(13) & --- & 0.60877(39)  & 0.60773(52) & ---\\
RGEVP & 2 & 4& --- & 0.6053(21)  & 0.6048(33) & --- & 0.60769(79)  & 0.6068(17) \\
RGEVP & 5 & 4& --- & 0.6056(14)  & 0.6065(22) & --- & 0.60797(58)  & 0.6085(11) \\
RGEVP & 8 & 4& --- & 0.6055(13) & ---& --- & 0.60773(52) & ---\\
\hline
\end{tabular}
\caption{Same as Table~\ref{tab:efm_I2_n0_L24} but for the $I=2$ two-pion first excited state on the $24^3$ lattice.  The re-basing matrix is calculated as: $4\times4\to 3\times3$ at $t_0 = 4$.}
\label{tab:efm_I2_n1_L24}
\end{table*}

\begin{table*}[tp]
\centering
\begin{tabular}{|ccc|cccccc|}
\hline
 GEVP type & $\delta_t$ & $t-t_0$ & $t = 4$ & $t = 6$ & $t = 8$ & $t = 4$ w DR & $t = 6$ w DR & $t = 8$ w DR\\
\hline
$3\times3$ & 2 & 1 & 0.8153(29)  & 0.8136(66)  & 0.798(14)  & 0.8168(18)  & 0.8159(51)  & 0.805(12) \\
$3\times3$ & 5 & 1 & 0.8142(23)  & 0.8095(52)  & 0.7997(94)  & 0.8162(13)  & 0.8129(41)  & 0.8077(74) \\
$3\times3$ & 8 & 1 & 0.8146(22)  & 0.8116(52) & --- & 0.8165(14)  & 0.8150(39) & ---\\
$3\times3$ & 2 & 2 & 0.8153(29)  & 0.8136(66)  & 0.798(14)  & 0.8168(18)  & 0.8159(51)  & 0.805(12) \\
$3\times3$ & 5 & 2 & 0.8142(23)  & 0.8095(52)  & 0.7997(94)  & 0.8162(13)  & 0.8129(41)  & 0.8077(74) \\
$3\times3$ & 8 & 2 & 0.8146(22)  & 0.8116(52) & --- & 0.8165(14)  & 0.8150(39) & ---\\
$3\times3$ & 2 & 3 & 0.8153(29)  & 0.8136(66)  & 0.798(14)  & 0.8168(18)  & 0.8159(51)  & 0.805(12) \\
$3\times3$ & 5 & 3 & 0.8142(22)  & 0.8095(52)  & 0.7997(94)  & 0.8162(13)  & 0.8129(41)  & 0.8077(74) \\
$3\times3$ & 8 & 3 & 0.8146(22)  & 0.8116(52) & --- & 0.8165(14)  & 0.8150(39) & ---\\
$3\times3$ & 2 & 4& --- & 0.8136(66)  & 0.798(14) & --- & 0.8159(51)  & 0.805(12) \\
$3\times3$ & 5 & 4& --- & 0.8095(52)  & 0.7997(94) & --- & 0.8129(41)  & 0.8077(74) \\
$3\times3$ & 8 & 4& --- & 0.8116(52) & ---& --- & 0.8150(39) & ---\\
$4\times4$ & 2 & 1 & 0.8136(29)  & 0.8140(66)  & 0.798(15)  & 0.8150(17)  & 0.8163(51)  & 0.805(12) \\
$4\times4$ & 5 & 1 & 0.8130(22)  & 0.8098(52)  & 0.8000(94)  & 0.8150(12)  & 0.8133(41)  & 0.8080(74) \\
$4\times4$ & 8 & 1 & 0.8134(22)  & 0.8119(51) & --- & 0.8153(13)  & 0.8154(39) & ---\\
$4\times4$ & 2 & 2 & 0.8135(29)  & 0.8143(66)  & 0.798(14)  & 0.8150(18)  & 0.8166(51)  & 0.805(12) \\
$4\times4$ & 5 & 2 & 0.8130(22)  & 0.8101(52)  & 0.7998(95)  & 0.8150(12)  & 0.8136(41)  & 0.8078(74) \\
$4\times4$ & 8 & 2 & 0.8134(22)  & 0.8122(51) & --- & 0.8153(13)  & 0.8157(39) & ---\\
$4\times4$ & 2 & 3 & 0.8135(29)  & 0.8143(66)  & 0.799(14)  & 0.8150(18)  & 0.8166(51)  & 0.806(12) \\
$4\times4$ & 5 & 3 & 0.8130(22)  & 0.8102(52)  & 0.8002(94)  & 0.8150(12)  & 0.8136(40)  & 0.8082(74) \\
$4\times4$ & 8 & 3 & 0.8134(22)  & 0.8122(51) & --- & 0.8153(13)  & 0.8157(39) & ---\\
$4\times4$ & 2 & 4& --- & 0.8143(66)  & 0.800(14) & --- & 0.8167(51)  & 0.807(11) \\
$4\times4$ & 5 & 4& --- & 0.8103(52)  & 0.8006(95) & --- & 0.8137(41)  & 0.8085(75) \\
$4\times4$ & 8 & 4& --- & 0.8123(51) & ---& --- & 0.8158(39) & ---\\
RGEVP & 2 & 1 & 0.8135(29)  & 0.8147(66)  & 0.801(14)  & 0.8150(18)  & 0.8171(51)  & 0.808(12) \\
RGEVP & 5 & 1 & 0.8130(22)  & 0.8105(52)  & 0.8016(96)  & 0.8150(12)  & 0.8139(40)  & 0.8096(78) \\
RGEVP & 8 & 1 & 0.8134(22)  & 0.8126(51) & --- & 0.8153(13)  & 0.8160(39) & ---\\
RGEVP & 2 & 2 & 0.8135(29)  & 0.8145(66)  & 0.801(14)  & 0.8150(18)  & 0.8168(51)  & 0.808(12) \\
RGEVP & 5 & 2 & 0.8130(22)  & 0.8103(52)  & 0.8014(96)  & 0.8150(12)  & 0.8138(40)  & 0.8094(77) \\
RGEVP & 8 & 2 & 0.8134(22)  & 0.8124(51) & --- & 0.8153(13)  & 0.8159(39) & ---\\
RGEVP & 2 & 3 & 0.8137(29)  & 0.8141(66)  & 0.800(14)  & 0.8152(18)  & 0.8164(51)  & 0.807(11) \\
RGEVP & 5 & 3 & 0.8131(22)  & 0.8099(52)  & 0.8008(95)  & 0.8151(13)  & 0.8134(40)  & 0.8088(76) \\
RGEVP & 8 & 3 & 0.8135(22)  & 0.8120(51) & --- & 0.8154(13)  & 0.8155(39) & ---\\
RGEVP & 2 & 4& --- & 0.8141(66)  & 0.800(14) & --- & 0.8164(51)  & 0.807(11) \\
RGEVP & 5 & 4& --- & 0.8099(52)  & 0.8008(95) & --- & 0.8134(40)  & 0.8088(76) \\
RGEVP & 8 & 4& --- & 0.8120(51) & ---& --- & 0.8154(39) & ---\\
\hline
\end{tabular}
\caption{Same as Table~\ref{tab:efm_I2_n0_L24} but for the $I=2$ two-pion second excited state on the $24^3$ lattice.  The re-basing matrix is calculated as: $4\times4\to 3\times3$ at $t_0 = 4$.}
\label{tab:efm_I2_n2_L24}
\end{table*}

\begin{table*}[tp]
\centering
\begin{tabular}{|ccc|cccccc|}
\hline
 GEVP type & $\delta_t$ & $t-t_0$ & $t = 4$ & $t = 6$ & $t = 8$ & $t = 4$ w DR & $t = 6$ w DR & $t = 8$ w DR\\
\hline
$4\times4$ & 2 & 1 & 0.9653(70)  & 0.940(21)  & 0.841(60)  & 0.9700(57)  & 0.948(17)  & 0.845(60) \\
$4\times4$ & 5 & 1 & 0.9620(58)  & 0.940(17)  & 0.905(53)  & 0.9670(43)  & 0.946(13)  & 0.909(54) \\
$4\times4$ & 8 & 1 & 0.9603(57)  & 0.936(17) & --- & 0.9654(41)  & 0.943(13) & ---\\
$4\times4$ & 2 & 2 & 0.9653(70)  & 0.940(21)  & 0.841(60)  & 0.9700(57)  & 0.948(17)  & 0.845(61) \\
$4\times4$ & 5 & 2 & 0.9620(58)  & 0.939(17)  & 0.905(53)  & 0.9670(43)  & 0.946(13)  & 0.910(54) \\
$4\times4$ & 8 & 2 & 0.9603(57)  & 0.935(17) & --- & 0.9655(41)  & 0.942(13) & ---\\
$4\times4$ & 2 & 3 & 0.9653(70)  & 0.940(21)  & 0.840(61)  & 0.9700(57)  & 0.948(17)  & 0.844(61) \\
$4\times4$ & 5 & 3 & 0.9620(58)  & 0.939(17)  & 0.905(53)  & 0.9670(43)  & 0.946(13)  & 0.909(54) \\
$4\times4$ & 8 & 3 & 0.9603(57)  & 0.935(17) & --- & 0.9655(42)  & 0.942(13) & ---\\
$4\times4$ & 2 & 4& --- & 0.940(21)  & 0.839(61) & --- & 0.948(17)  & 0.843(61) \\
$4\times4$ & 5 & 4& --- & 0.939(17)  & 0.904(53) & --- & 0.946(13)  & 0.909(54) \\
$4\times4$ & 8 & 4& --- & 0.935(17) & ---& --- & 0.942(13) & ---\\
\hline
\end{tabular}
\caption{Same as Table~\ref{tab:efm_I2_n0_L24} but for the $I=2$ two-pion third excited state on the $24^3$ lattice.  }
\label{tab:efm_I2_n3_L24}
\end{table*}

\begin{table*}[tp]
\centering
\begin{tabular}{|ccc|cccccc|}
\hline
 GEVP type & $\delta_t$ & $t-t_0$ & $t = 3$ & $t = 5$ & $t = 7$ & $t = 3$ w DR & $t = 5$ w DR & $t = 7$ w DR\\
\hline
$3\times3$ & 2 & 1 & 0.2790(28)  & 0.2694(29)  & 0.2711(32)  & 0.27281(51)  & 0.27011(74)  & 0.2700(15) \\
$3\times3$ & 5 & 1 & 0.2752(12)  & 0.2700(13)  & 0.2708(14)  & 0.27166(44)  & 0.26999(60)  & 0.27079(73) \\
$3\times3$ & 8 & 1 & 0.27437(96)  & 0.2705(10)  & 0.2705(11)  & 0.27156(39)  & 0.27060(47)  & 0.27012(66) \\
$3\times3$ & 2 & 2 & 0.2793(28)  & 0.2696(29)  & 0.2711(31)  & 0.27311(49)  & 0.27022(68)  & 0.2700(15) \\
$3\times3$ & 5 & 2 & 0.2755(12)  & 0.2701(13)  & 0.2708(14)  & 0.27193(43)  & 0.27004(58)  & 0.27080(74) \\
$3\times3$ & 8 & 2 & 0.27457(96)  & 0.2705(10)  & 0.2705(11)  & 0.27176(39)  & 0.27060(47)  & 0.27013(65) \\
$3\times3$ & 2 & 3& --- & 0.2697(29)  & 0.2712(31) & --- & 0.27038(63)  & 0.2700(13) \\
$3\times3$ & 5 & 3& --- & 0.2702(13)  & 0.2708(14) & --- & 0.27018(55)  & 0.27079(74) \\
$3\times3$ & 8 & 3& --- & 0.2705(10)  & 0.2705(11) & --- & 0.27063(46)  & 0.27015(64) \\
$3\times3$ & 2 & 4& --- & 0.2701(29)  & 0.2714(31) & --- & 0.27073(59)  & 0.2702(11) \\
$3\times3$ & 5 & 4& --- & 0.2705(13)  & 0.2708(14) & --- & 0.27050(51)  & 0.27078(74) \\
$3\times3$ & 8 & 4& --- & 0.2707(10)  & 0.2705(11) & --- & 0.27078(46)  & 0.27015(63) \\
$4\times4$ & 2 & 1 & 0.2789(28)  & 0.2695(29)  & 0.2698(40)  & 0.27275(52)  & 0.27012(74)  & 0.2686(27) \\
$4\times4$ & 5 & 1 & 0.2752(12)  & 0.2700(13)  & 0.2708(14)  & 0.27163(44)  & 0.26995(61)  & 0.27081(76) \\
$4\times4$ & 8 & 1 & 0.27434(96)  & 0.2705(10)  & 0.2700(14)  & 0.27153(39)  & 0.27061(47)  & 0.2696(10) \\
$4\times4$ & 2 & 2 & 0.2792(28)  & 0.2696(29)  & 0.2707(33)  & 0.27307(50)  & 0.27022(68)  & 0.2695(16) \\
$4\times4$ & 5 & 2 & 0.2754(12)  & 0.2700(13)  & 0.2708(15)  & 0.27189(44)  & 0.27000(59)  & 0.27084(80) \\
$4\times4$ & 8 & 2 & 0.27455(96)  & 0.2705(10)  & 0.2702(12)  & 0.27173(40)  & 0.27060(47)  & 0.26984(82) \\
$4\times4$ & 2 & 3& --- & 0.2697(29)  & 0.2711(31) & --- & 0.27038(64)  & 0.2699(13) \\
$4\times4$ & 5 & 3& --- & 0.2702(13)  & 0.2708(15) & --- & 0.27015(55)  & 0.27081(78) \\
$4\times4$ & 8 & 3& --- & 0.2705(10)  & 0.2703(12) & --- & 0.27063(47)  & 0.27001(71) \\
$4\times4$ & 2 & 4& --- & 0.2701(29)  & 0.2713(31) & --- & 0.27072(60)  & 0.2701(11) \\
$4\times4$ & 5 & 4& --- & 0.2705(13)  & 0.2708(15) & --- & 0.27048(51)  & 0.27079(78) \\
$4\times4$ & 8 & 4& --- & 0.2707(10)  & 0.2704(12) & --- & 0.27078(47)  & 0.27002(68) \\
$5\times5$ & 2 & 1 & 0.2789(28)  & 0.2694(29)  & 0.2694(49)  & 0.27275(52)  & 0.27008(77)  & 0.2682(38) \\
$5\times5$ & 5 & 1 & 0.2752(12)  & 0.2700(13)  & 0.2706(14)  & 0.27163(44)  & 0.26996(60)  & 0.27064(72) \\
$5\times5$ & 8 & 1 & 0.27435(96)  & 0.2706(10)  & 0.2698(16)  & 0.27153(39)  & 0.27065(47)  & 0.2695(12) \\
$5\times5$ & 2 & 2 & 0.2792(28)  & 0.2695(29)  & 0.2705(34)  & 0.27306(50)  & 0.27019(68)  & 0.2693(18) \\
$5\times5$ & 5 & 2 & 0.2754(12)  & 0.2700(13)  & 0.2707(14)  & 0.27189(44)  & 0.27000(58)  & 0.27075(77) \\
$5\times5$ & 8 & 2 & 0.27455(96)  & 0.2705(10)  & 0.2701(13)  & 0.27173(40)  & 0.27061(46)  & 0.26977(90) \\
$5\times5$ & 2 & 3& --- & 0.2697(29)  & 0.2711(31) & --- & 0.27038(64)  & 0.2700(12) \\
$5\times5$ & 5 & 3& --- & 0.2702(13)  & 0.2707(14) & --- & 0.27015(55)  & 0.27074(76) \\
$5\times5$ & 8 & 3& --- & 0.2705(10)  & 0.2703(12) & --- & 0.27063(47)  & 0.27001(73) \\
$5\times5$ & 2 & 4& --- & 0.2701(29)  & 0.2713(31) & --- & 0.27073(60)  & 0.2701(11) \\
$5\times5$ & 5 & 4& --- & 0.2705(13)  & 0.2707(14) & --- & 0.27048(51)  & 0.27076(77) \\
$5\times5$ & 8 & 4& --- & 0.2707(10)  & 0.2704(12) & --- & 0.27077(47)  & 0.27003(69) \\
RGEVP & 2 & 1 & 0.2792(28)  & 0.2694(29)  & 0.2719(29)  & 0.27301(49)  & 0.27010(78)  & 0.27073(97) \\
RGEVP & 5 & 1 & 0.2753(12)  & 0.2700(13)  & 0.2707(14)  & 0.27173(43)  & 0.27001(60)  & 0.27073(71) \\
RGEVP & 8 & 1 & 0.27454(96)  & 0.2706(10)  & 0.2709(11)  & 0.27172(39)  & 0.27070(47)  & 0.27055(62) \\
RGEVP & 2 & 2 & 0.2792(28)  & 0.2696(29)  & 0.2717(30)  & 0.27306(49)  & 0.27022(67)  & 0.2706(10) \\
RGEVP & 5 & 2 & 0.2755(12)  & 0.2700(13)  & 0.2708(14)  & 0.27196(42)  & 0.27001(59)  & 0.27079(75) \\
RGEVP & 8 & 2 & 0.27459(96)  & 0.2705(10)  & 0.2706(11)  & 0.27178(38)  & 0.27062(46)  & 0.27028(68) \\
RGEVP & 2 & 3& --- & 0.2697(29)  & 0.2713(30) & --- & 0.27035(65)  & 0.2702(12) \\
RGEVP & 5 & 3& --- & 0.2702(13)  & 0.2708(14) & --- & 0.27013(53)  & 0.27077(76) \\
RGEVP & 8 & 3& --- & 0.2705(10)  & 0.2704(12) & --- & 0.27061(46)  & 0.27008(70) \\
RGEVP & 2 & 4& --- & 0.2700(29)  & 0.2713(31) & --- & 0.27067(59)  & 0.2701(12) \\
RGEVP & 5 & 4& --- & 0.2704(13)  & 0.2707(14) & --- & 0.27035(49)  & 0.27071(74) \\
RGEVP & 8 & 4& --- & 0.2706(10)  & 0.2703(12) & --- & 0.27070(45)  & 0.26999(70) \\
\hline
\end{tabular}
\caption{Same as Table~\ref{tab:efm_I2_n0_L24} but for the $I=0$ two-pion ground state on the $24^3$ lattice.  The re-basing matrix is calculated as: $5\times5\to 3\times3$ at $t_0 = 4$.}
\label{tab:efm_I0_n0_L24}
\end{table*}

\begin{table*}[tp]
\centering
\begin{tabular}{|ccc|cccccc|}
\hline
 GEVP type & $\delta_t$ & $t-t_0$ & $t = 3$ & $t = 5$ & $t = 7$ & $t = 3$ w DR & $t = 5$ w DR & $t = 7$ w DR\\
\hline
$3\times3$ & 2 & 1 & 0.5415(65)  & 0.497(44)  & 0.498(49)  & 0.5356(63)  & 0.499(44)  & 0.498(49) \\
$3\times3$ & 5 & 1 & 0.5279(84)  & 0.512(18)  & 0.543(21)  & 0.5240(83)  & 0.513(18)  & 0.544(21) \\
$3\times3$ & 8 & 1 & 0.5331(53)  & 0.5315(72)  & 0.528(18)  & 0.5294(53)  & 0.5331(72)  & 0.529(18) \\
$3\times3$ & 2 & 2 & 0.5457(46)  & 0.499(33)  & 0.502(54)  & 0.5398(43)  & 0.501(34)  & 0.502(54) \\
$3\times3$ & 5 & 2 & 0.5336(56)  & 0.511(20)  & 0.546(20)  & 0.5296(54)  & 0.512(20)  & 0.546(20) \\
$3\times3$ & 8 & 2 & 0.5345(44)  & 0.5325(74)  & 0.529(18)  & 0.5309(43)  & 0.5341(75)  & 0.530(18) \\
$3\times3$ & 2 & 3& --- & 0.503(25)  & 0.505(62) & --- & 0.505(25)  & 0.506(62) \\
$3\times3$ & 5 & 3& --- & 0.509(20)  & 0.551(19) & --- & 0.511(20)  & 0.552(19) \\
$3\times3$ & 8 & 3& --- & 0.5329(79)  & 0.531(17) & --- & 0.5345(80)  & 0.532(17) \\
$3\times3$ & 2 & 4& --- & 0.511(15)  & 0.499(62) & --- & 0.513(15)  & 0.499(62) \\
$3\times3$ & 5 & 4& --- & 0.508(16)  & 0.555(19) & --- & 0.510(16)  & 0.556(19) \\
$3\times3$ & 8 & 4& --- & 0.5304(81)  & 0.533(17) & --- & 0.5320(81)  & 0.534(17) \\
$4\times4$ & 2 & 1 & 0.5411(63)  & 0.497(52)  & 0.393(75)  & 0.5352(61)  & 0.499(52)  & 0.393(74) \\
$4\times4$ & 5 & 1 & 0.5270(76)  & 0.498(16)  & 0.523(46)  & 0.5231(75)  & 0.500(16)  & 0.524(46) \\
$4\times4$ & 8 & 1 & 0.5303(50)  & 0.5268(89)  & 0.463(61)  & 0.5267(49)  & 0.5284(89)  & 0.463(61) \\
$4\times4$ & 2 & 2 & 0.5444(48)  & 0.499(36)  & 0.433(62)  & 0.5385(45)  & 0.501(36)  & 0.433(62) \\
$4\times4$ & 5 & 2 & 0.5320(54)  & 0.500(16)  & 0.526(40)  & 0.5281(52)  & 0.501(16)  & 0.527(40) \\
$4\times4$ & 8 & 2 & 0.5319(44)  & 0.5270(87)  & 0.483(43)  & 0.5282(43)  & 0.5286(87)  & 0.484(43) \\
$4\times4$ & 2 & 3& --- & 0.503(24)  & 0.460(59) & --- & 0.505(24)  & 0.460(59) \\
$4\times4$ & 5 & 3& --- & 0.500(16)  & 0.532(34) & --- & 0.502(16)  & 0.533(34) \\
$4\times4$ & 8 & 3& --- & 0.5272(87)  & 0.495(35) & --- & 0.5288(87)  & 0.495(34) \\
$4\times4$ & 2 & 4& --- & 0.511(15)  & 0.464(53) & --- & 0.513(15)  & 0.464(53) \\
$4\times4$ & 5 & 4& --- & 0.501(14)  & 0.536(31) & --- & 0.502(14)  & 0.537(31) \\
$4\times4$ & 8 & 4& --- & 0.5257(86)  & 0.499(31) & --- & 0.5273(86)  & 0.500(30) \\
$5\times5$ & 2 & 1 & 0.5411(62)  & 0.474(94)  & 0.4(1.6)  & 0.5352(60)  & 0.476(94)  & 0.4(1.6) \\
$5\times5$ & 5 & 1 & 0.5268(80)  & 0.499(16)  & 0.508(42)  & 0.5229(79)  & 0.501(16)  & 0.509(42) \\
$5\times5$ & 8 & 1 & 0.5304(50)  & 0.5274(87)  & 0.450(82)  & 0.5267(49)  & 0.5290(87)  & 0.451(82) \\
$5\times5$ & 2 & 2 & 0.5441(48)  & 0.485(60)  & 0.44(11)  & 0.5382(45)  & 0.487(60)  & 0.44(11) \\
$5\times5$ & 5 & 2 & 0.5320(53)  & 0.501(14)  & 0.519(35)  & 0.5281(52)  & 0.503(14)  & 0.520(35) \\
$5\times5$ & 8 & 2 & 0.5319(44)  & 0.5275(84)  & 0.476(54)  & 0.5283(43)  & 0.5290(85)  & 0.477(54) \\
$5\times5$ & 2 & 3& --- & 0.498(32)  & 0.48(11) & --- & 0.500(32)  & 0.48(11) \\
$5\times5$ & 5 & 3& --- & 0.501(15)  & 0.530(31) & --- & 0.502(15)  & 0.530(31) \\
$5\times5$ & 8 & 3& --- & 0.5274(85)  & 0.491(41) & --- & 0.5290(85)  & 0.491(41) \\
$5\times5$ & 2 & 4& --- & 0.510(15)  & 0.474(65) & --- & 0.512(15)  & 0.474(65) \\
$5\times5$ & 5 & 4& --- & 0.501(14)  & 0.536(29) & --- & 0.502(14)  & 0.536(29) \\
$5\times5$ & 8 & 4& --- & 0.5257(86)  & 0.494(36) & --- & 0.5273(86)  & 0.495(36) \\
RGEVP & 2 & 1 & 0.5409(62)  & 0.490(57)  & 0.487(43)  & 0.5350(60)  & 0.492(57)  & 0.487(43) \\
RGEVP & 5 & 1 & 0.5268(79)  & 0.506(16)  & 0.541(27)  & 0.5229(79)  & 0.508(16)  & 0.542(27) \\
RGEVP & 8 & 1 & 0.5303(50)  & 0.5275(98)  & 0.529(21)  & 0.5267(49)  & 0.5291(98)  & 0.530(21) \\
RGEVP & 2 & 2 & 0.5459(43)  & 0.487(56)  & 0.523(65)  & 0.5400(40)  & 0.489(56)  & 0.524(65) \\
RGEVP & 5 & 2 & 0.5325(57)  & 0.508(23)  & 0.548(21)  & 0.5286(56)  & 0.510(23)  & 0.549(21) \\
RGEVP & 8 & 2 & 0.5320(48)  & 0.531(12)  & 0.556(21)  & 0.5284(47)  & 0.532(12)  & 0.556(21) \\
RGEVP & 2 & 3& --- & 0.497(32)  & 0.548(85) & --- & 0.499(32)  & 0.548(85) \\
RGEVP & 5 & 3& --- & 0.498(20)  & 0.558(24) & --- & 0.500(20)  & 0.559(24) \\
RGEVP & 8 & 3& --- & 0.526(10)  & 0.532(32) & --- & 0.527(10)  & 0.532(32) \\
RGEVP & 2 & 4& --- & 0.510(17)  & 0.509(55) & --- & 0.512(17)  & 0.509(55) \\
RGEVP & 5 & 4& --- & 0.501(14)  & 0.544(38) & --- & 0.502(13)  & 0.545(38) \\
RGEVP & 8 & 4& --- & 0.5244(88)  & 0.507(32) & --- & 0.5260(88)  & 0.507(32) \\
\hline
\end{tabular}
\caption{Same as Table~\ref{tab:efm_I2_n0_L24} but for the $I=0$ two-pion first excited state on the $24^3$ lattice.  The re-basing matrix is calculated as: $5\times5\to 3\times3$ at $t_0 = 3$.}
\label{tab:efm_I0_n1_L24}
\end{table*}

\begin{table*}[tp]
\centering
\begin{tabular}{|ccc|cccccc|}
\hline
 GEVP type & $\delta_t$ & $t-t_0$ & $t = 3$ & $t = 5$ & $t = 7$ & $t = 3$ w DR & $t = 5$ w DR & $t = 7$ w DR\\
\hline
$3\times3$ & 2 & 1 & 0.672(44)  & 0.599(96)  & 1.13(94)  & 0.666(44)  & 0.601(96)  & 1.14(94) \\
$3\times3$ & 5 & 1 & 0.656(51)  & 0.78(25)  & ---  & 0.652(51)  & 0.78(25)  & --- \\
$3\times3$ & 8 & 1 & 0.743(81)  & 1.5(1.3)  & ---  & 0.739(81)  & 1.5(1.3)  & --- \\
$3\times3$ & 2 & 2 & 0.668(46)  & 0.60(11)  & 1.13(94)  & 0.662(47)  & 0.60(11)  & 1.14(94) \\
$3\times3$ & 5 & 2 & 0.650(54)  & 0.78(25)  & ---  & 0.646(54)  & 0.78(25)  & --- \\
$3\times3$ & 8 & 2 & 0.741(83)  & 1.5(1.3)  & ---  & 0.737(83)  & 1.5(1.3)  & --- \\
$3\times3$ & 2 & 3& --- & 0.59(12)  & 1.13(93) & --- & 0.59(12)  & 1.14(93) \\
$3\times3$ & 5 & 3& --- & 0.78(25)  & --- & --- & 0.78(25)  & --- \\
$3\times3$ & 8 & 3& --- & 1.5(1.3)  & --- & --- & 1.5(1.3)  & --- \\
$3\times3$ & 2 & 4& --- & 0.58(13)  & 1.13(92) & --- & 0.59(13)  & 1.14(92) \\
$3\times3$ & 5 & 4& --- & 0.78(26)  & --- & --- & 0.78(26)  & --- \\
$3\times3$ & 8 & 4& --- & 1.5(1.3)  & --- & --- & 1.5(1.3)  & --- \\
$4\times4$ & 2 & 1 & 0.672(36)  & 0.60(12)  & 0.618(61)  & 0.666(36)  & 0.60(12)  & 0.626(61) \\
$4\times4$ & 5 & 1 & 0.655(42)  & 0.655(27)  & 0.650(58)  & 0.650(42)  & 0.658(27)  & 0.656(58) \\
$4\times4$ & 8 & 1 & 0.698(19)  & 0.715(23)  & 0.515(50)  & 0.694(19)  & 0.718(23)  & 0.521(50) \\
$4\times4$ & 2 & 2 & 0.682(26)  & 0.59(13)  & 0.66(13)  & 0.676(26)  & 0.60(13)  & 0.66(13) \\
$4\times4$ & 5 & 2 & 0.656(37)  & 0.668(32)  & 0.665(58)  & 0.651(37)  & 0.670(32)  & 0.671(58) \\
$4\times4$ & 8 & 2 & 0.694(19)  & 0.720(22)  & 0.500(58)  & 0.690(20)  & 0.723(22)  & 0.506(58) \\
$4\times4$ & 2 & 3& --- & 0.59(13)  & 0.69(16) & --- & 0.59(13)  & 0.70(16) \\
$4\times4$ & 5 & 3& --- & 0.677(43)  & 0.677(63) & --- & 0.679(43)  & 0.683(63) \\
$4\times4$ & 8 & 3& --- & 0.730(22)  & 0.496(61) & --- & 0.732(22)  & 0.501(62) \\
$4\times4$ & 2 & 4& --- & 0.60(12)  & 0.71(17) & --- & 0.60(12)  & 0.71(17) \\
$4\times4$ & 5 & 4& --- & 0.663(51)  & 0.693(69) & --- & 0.665(51)  & 0.698(69) \\
$4\times4$ & 8 & 4& --- & 0.733(23)  & 0.498(63) & --- & 0.735(23)  & 0.503(63) \\
$5\times5$ & 2 & 1 & 0.668(44)  & 0.58(12)  & 0.69(22)  & 0.662(44)  & 0.58(12)  & 0.70(22) \\
$5\times5$ & 5 & 1 & 0.643(59)  & 0.656(26)  & 0.599(65)  & 0.638(59)  & 0.658(26)  & 0.605(65) \\
$5\times5$ & 8 & 1 & 0.698(21)  & 0.709(22)  & 0.514(50)  & 0.694(22)  & 0.711(22)  & 0.520(50) \\
$5\times5$ & 2 & 2 & 0.684(26)  & 0.57(15)  & 0.75(25)  & 0.678(26)  & 0.57(15)  & 0.75(25) \\
$5\times5$ & 5 & 2 & 0.653(44)  & 0.676(35)  & 0.619(63)  & 0.649(44)  & 0.679(35)  & 0.625(63) \\
$5\times5$ & 8 & 2 & 0.695(20)  & 0.717(21)  & 0.503(56)  & 0.691(20)  & 0.720(20)  & 0.508(56) \\
$5\times5$ & 2 & 3& --- & 0.57(17)  & 0.77(23) & --- & 0.57(17)  & 0.77(23) \\
$5\times5$ & 5 & 3& --- & 0.685(45)  & 0.649(59) & --- & 0.688(45)  & 0.654(59) \\
$5\times5$ & 8 & 3& --- & 0.729(21)  & 0.499(59) & --- & 0.732(21)  & 0.504(60) \\
$5\times5$ & 2 & 4& --- & 0.58(15)  & 0.78(25) & --- & 0.58(15)  & 0.79(25) \\
$5\times5$ & 5 & 4& --- & 0.666(51)  & 0.672(62) & --- & 0.668(51)  & 0.677(63) \\
$5\times5$ & 8 & 4& --- & 0.733(23)  & 0.500(62) & --- & 0.735(23)  & 0.505(62) \\
RGEVP & 2 & 1 & 0.7045(98)  & 0.696(31)  & 0.616(49)  & 0.6985(100)  & 0.698(31)  & 0.624(50) \\
RGEVP & 5 & 1 & 0.684(12)  & 0.650(34)  & 0.96(26)  & 0.679(12)  & 0.652(34)  & 0.97(26) \\
RGEVP & 8 & 1 & 0.697(13)  & 0.751(60)  & 1.01(46)  & 0.693(13)  & 0.753(60)  & 1.02(46) \\
RGEVP & 2 & 2 & 0.690(15)  & 0.662(41)  & 0.654(99)  & 0.684(15)  & 0.663(41)  & 0.663(99) \\
RGEVP & 5 & 2 & 0.660(24)  & 0.667(80)  & 2.2(3.6)  & 0.656(24)  & 0.669(80)  & 2.2(3.6) \\
RGEVP & 8 & 2 & 0.696(22)  & 0.80(11)  & ---  & 0.691(22)  & 0.81(11)  & --- \\
RGEVP & 2 & 3& --- & 0.634(59)  & 0.72(18) & --- & 0.636(59)  & 0.73(18) \\
RGEVP & 5 & 3& --- & 0.69(13)  & --- & --- & 0.70(13)  & --- \\
RGEVP & 8 & 3& --- & 0.79(11)  & --- & --- & 0.79(11)  & --- \\
RGEVP & 2 & 4& --- & 0.595(97)  & 0.88(44) & --- & 0.596(97)  & 0.89(44) \\
RGEVP & 5 & 4& --- & 0.69(14)  & --- & --- & 0.69(14)  & --- \\
RGEVP & 8 & 4& --- & 0.755(70)  & 1.19(76) & --- & 0.757(70)  & 1.20(76) \\
\hline
\end{tabular}
\caption{Same as Table~\ref{tab:efm_I2_n0_L24} but for the $I=0$ two-pion second excited state on the $24^3$ lattice.  The re-basing matrix is calculated as: $5\times5\to 3\times3$ at $t_0 = 1$.}
\label{tab:efm_I0_n2_L24}
\end{table*}

\begin{table*}[tp]
\centering
\begin{tabular}{|ccc|cccccc|}
\hline
 GEVP type & $\delta_t$ & $t-t_0$ & $t = 3$ & $t = 5$ & $t = 7$ & $t = 3$ w DR & $t = 5$ w DR & $t = 7$ w DR\\
\hline
$4\times4$ & 2 & 1 & 0.788(49)  & 0.84(14)  & ---  & 0.782(49)  & 0.84(14)  & --- \\
$4\times4$ & 5 & 1 & 0.812(69)  & 1.7(1.9)  & ---  & 0.808(69)  & 1.7(1.9)  & --- \\
$4\times4$ & 8 & 1 & 0.93(19)  & ---  & -0.37(30)  & 0.93(19)  & ---  & -0.37(30) \\
$4\times4$ & 2 & 2 & 0.775(62)  & 0.84(15)  & ---  & 0.769(62)  & 0.84(15)  & --- \\
$4\times4$ & 5 & 2 & 0.806(79)  & 1.7(1.9)  & ---  & 0.802(79)  & 1.7(1.9)  & --- \\
$4\times4$ & 8 & 2 & 0.94(19)  & ---  & ---  & 0.93(19)  & ---  & --- \\
$4\times4$ & 2 & 3& --- & 0.84(16)  & --- & --- & 0.84(16)  & --- \\
$4\times4$ & 5 & 3& --- & 1.7(1.9)  & --- & --- & 1.7(1.9)  & --- \\
$4\times4$ & 8 & 3& --- & ---  & --- & --- & ---  & --- \\
$4\times4$ & 2 & 4& --- & 0.82(19)  & --- & --- & 0.82(19)  & --- \\
$4\times4$ & 5 & 4& --- & 1.7(1.9)  & --- & --- & 1.7(1.9)  & --- \\
$4\times4$ & 8 & 4& --- & ---  & --- & --- & ---  & --- \\
$5\times5$ & 2 & 1 & 0.767(60)  & 0.79(27)  & ---  & 0.761(60)  & 0.79(27)  & --- \\
$5\times5$ & 5 & 1 & 0.785(77)  & 1.8(2.4)  & ---  & 0.780(77)  & 1.8(2.4)  & --- \\
$5\times5$ & 8 & 1 & 0.92(24)  & ---  & 1.1(6.6)  & 0.92(24)  & ---  & 1.1(6.6) \\
$5\times5$ & 2 & 2 & 0.764(77)  & 0.81(18)  & ---  & 0.758(77)  & 0.81(18)  & --- \\
$5\times5$ & 5 & 2 & 0.784(99)  & 0.97(24)  & ---  & 0.779(99)  & 0.98(24)  & --- \\
$5\times5$ & 8 & 2 & 0.93(22)  & 1.01(14)  & 1.2(2.0)  & 0.93(22)  & 1.01(14)  & 1.2(2.0) \\
$5\times5$ & 2 & 3& --- & 0.81(19)  & --- & --- & 0.81(19)  & --- \\
$5\times5$ & 5 & 3& --- & 1.10(27)  & 1.5(1.6) & --- & 1.10(27)  & 1.5(1.6) \\
$5\times5$ & 8 & 3& --- & 1.00(14)  & 1.2(2.0) & --- & 1.01(14)  & 1.2(2.0) \\
$5\times5$ & 2 & 4& --- & 0.80(23)  & --- & --- & 0.80(23)  & --- \\
$5\times5$ & 5 & 4& --- & 1.19(34)  & 1.5(1.6) & --- & 1.20(34)  & 1.5(1.6) \\
$5\times5$ & 8 & 4& --- & 1.00(13)  & 1.2(2.0) & --- & 1.00(13)  & 1.2(2.0) \\
\hline
\end{tabular}
\caption{Same as Table~\ref{tab:efm_I2_n0_L24} but for the $I=0$ two-pion third excited state on the $24^3$ lattice.  The re-basing matrix is calculated as: $5\times5\to 3\times3$ at $t_0 = 1$.}
\label{tab:efm_I0_n3_L24}
\end{table*}

\begin{table*}[tp]
\centering
\begin{tabular}{|ccc|cccccc|}
\hline
 GEVP type & $\delta_t$ & $t-t_0$ & $t = 3$ & $t = 5$ & $t = 7$ & $t = 3$ w DR & $t = 5$ w DR & $t = 7$ w DR\\
\hline
$5\times5$ & 2 & 1 & 0.906(30)  & 0.81(21)  & ---  & 0.903(31)  & 0.82(21)  & --- \\
$5\times5$ & 5 & 1 & 0.879(22)  & 0.79(17)  & 1.5(1.6)  & 0.878(22)  & 0.80(17)  & 1.5(1.6) \\
$5\times5$ & 8 & 1 & 0.893(28)  & ---  & -0.2(6.3)  & 0.892(28)  & ---  & -0.2(6.3) \\
$5\times5$ & 2 & 2 & 0.891(38)  & 0.79(12)  & ---  & 0.887(39)  & 0.80(12)  & --- \\
$5\times5$ & 5 & 2 & 0.864(26)  & 1.6(2.6)  & 1.5(1.6)  & 0.863(26)  & 1.6(2.6)  & 1.5(1.6) \\
$5\times5$ & 8 & 2 & 0.881(36)  & ---  & ---  & 0.879(37)  & ---  & --- \\
$5\times5$ & 2 & 3& --- & 0.77(13)  & --- & --- & 0.78(13)  & --- \\
$5\times5$ & 5 & 3& --- & 1.5(2.6)  & --- & --- & 1.5(2.6)  & --- \\
$5\times5$ & 8 & 3& --- & ---  & --- & --- & ---  & --- \\
$5\times5$ & 2 & 4& --- & 0.75(15)  & --- & --- & 0.76(15)  & --- \\
$5\times5$ & 5 & 4& --- & 1.4(2.6)  & --- & --- & 1.4(2.6)  & --- \\
$5\times5$ & 8 & 4& --- & ---  & --- & --- & ---  & --- \\
\hline
\end{tabular}
\caption{Same as Table~\ref{tab:efm_I2_n0_L24} but for the $I=0$ two-pion fourth excited state on the $24^3$ lattice.  }
\label{tab:efm_I0_n4_L24}
\end{table*}

\begin{table*}[tp]
\centering
\begin{tabular}{|ccc|ccccc|}
\hline
 GEVP type & $\delta_t$ & $t-t_0$ & $t = 5$ & $t = 7$ & $t = 9$ & $t = 11$ & $t = 13$\\
\hline
$2\times2$ & 2 & 1 & 0.21049(69)  & 0.21075(52)  & 0.21076(65)  & 0.21030(64)  & 0.20982(62) \\
$2\times2$ & 5 & 1 & 0.21070(49)  & 0.21068(45)  & 0.21046(51)  & 0.21001(49)  & 0.20982(51) \\
$2\times2$ & 8 & 1 & 0.21060(46)  & 0.21055(41)  & 0.21032(47)  & 0.21011(45)  & 0.20993(45) \\
$2\times2$ & 11 & 1 & 0.21050(45)  & 0.21047(42)  & 0.21038(46) & ---& ---\\
$2\times2$ & 2 & 3 & 0.21049(69)  & 0.21074(52)  & 0.21077(65)  & 0.21031(64)  & 0.20985(63) \\
$2\times2$ & 5 & 3 & 0.21070(49)  & 0.21068(45)  & 0.21046(51)  & 0.21000(49)  & 0.20982(51) \\
$2\times2$ & 8 & 3 & 0.21060(46)  & 0.21055(41)  & 0.21032(47)  & 0.21011(45)  & 0.20994(45) \\
$2\times2$ & 11 & 3 & 0.21050(45)  & 0.21047(42)  & 0.21038(46) & ---& ---\\
$3\times3$ & 2 & 1 & 0.21051(69)  & 0.21075(52)  & 0.21073(65)  & 0.21030(65)  & 0.20990(66) \\
$3\times3$ & 5 & 1 & 0.21070(49)  & 0.21068(45)  & 0.21045(51)  & 0.21001(49)  & 0.20983(51) \\
$3\times3$ & 8 & 1 & 0.21060(46)  & 0.21055(41)  & 0.21031(47)  & 0.21011(45)  & 0.20992(45) \\
$3\times3$ & 11 & 1 & 0.21050(45)  & 0.21047(42)  & 0.21037(46) & ---& ---\\
$3\times3$ & 2 & 3 & 0.21051(69)  & 0.21075(52)  & 0.21075(65)  & 0.21031(65)  & 0.20986(64) \\
$3\times3$ & 5 & 3 & 0.21070(49)  & 0.21067(45)  & 0.21045(51)  & 0.21000(49)  & 0.20982(51) \\
$3\times3$ & 8 & 3 & 0.21060(46)  & 0.21055(41)  & 0.21031(47)  & 0.21011(45)  & 0.20993(45) \\
$3\times3$ & 11 & 3 & 0.21050(45)  & 0.21047(42)  & 0.21037(46) & ---& ---\\
$4\times4$ & 2 & 1 & 0.21056(69)  & 0.21075(52)  & 0.21081(67)  & 0.21010(82)  & 0.2089(21) \\
$4\times4$ & 5 & 1 & 0.21071(49)  & 0.21068(45)  & 0.21048(51)  & 0.21001(49)  & 0.2089(11) \\
$4\times4$ & 8 & 1 & 0.21061(46)  & 0.21055(41)  & 0.21033(47)  & 0.21011(45)  & 0.20952(72) \\
$4\times4$ & 11 & 1 & 0.21052(45)  & 0.21047(42)  & 0.21038(47) & ---& ---\\
$4\times4$ & 2 & 3 & 0.21052(69)  & 0.21076(53)  & 0.21077(65)  & 0.21020(67)  & 0.20995(84) \\
$4\times4$ & 5 & 3 & 0.21071(49)  & 0.21068(45)  & 0.21046(51)  & 0.20999(50)  & 0.20958(60) \\
$4\times4$ & 8 & 3 & 0.21061(46)  & 0.21055(41)  & 0.21032(47)  & 0.21011(45)  & 0.20977(48) \\
$4\times4$ & 11 & 3 & 0.21051(45)  & 0.21047(42)  & 0.21038(46) & ---& ---\\
RGEVP & 2 & 1 & 0.21051(69)  & 0.21075(52)  & 0.21073(65)  & 0.21029(65)  & 0.20991(66) \\
RGEVP & 5 & 1 & 0.21069(49)  & 0.21068(45)  & 0.21045(51)  & 0.21000(49)  & 0.20984(51) \\
RGEVP & 8 & 1 & 0.21059(46)  & 0.21055(41)  & 0.21031(47)  & 0.21011(45)  & 0.20993(46) \\
RGEVP & 11 & 1 & 0.21050(45)  & 0.21047(42)  & 0.21037(46) & ---& ---\\
RGEVP & 2 & 3 & 0.21050(69)  & 0.21075(52)  & 0.21076(65)  & 0.21029(65)  & 0.20987(63) \\
RGEVP & 5 & 3 & 0.21070(49)  & 0.21067(45)  & 0.21045(51)  & 0.21000(49)  & 0.20984(51) \\
RGEVP & 8 & 3 & 0.21060(46)  & 0.21055(41)  & 0.21032(47)  & 0.21011(45)  & 0.20994(45) \\
RGEVP & 11 & 3 & 0.21050(45)  & 0.21047(42)  & 0.21037(46) & ---& ---\\
\hline
\end{tabular}
\caption{Same as Table~\ref{tab:efm_I2_n0_L24} but for the $I=2$ two-pion ground state on the $32^3$ lattice.  Only results with DR are shown.  The re-basing matrix is calculated as: $4\times4\to 3\times3$ at $t_0 = 5$.}
\label{tab:efm_I2_n0_L32}
\end{table*}

\begin{table*}[tp]
\centering
\begin{tabular}{|ccc|ccccc|}
\hline
 GEVP type & $\delta_t$ & $t-t_0$ & $t = 5$ & $t = 7$ & $t = 9$ & $t = 11$ & $t = 13$\\
\hline
$2\times2$ & 2 & 1 & 0.45634(84)  & 0.4580(12)  & 0.4550(15)  & 0.4589(21)  & 0.4499(33) \\
$2\times2$ & 5 & 1 & 0.45660(55)  & 0.45714(76)  & 0.45536(100)  & 0.4559(14)  & 0.4524(21) \\
$2\times2$ & 8 & 1 & 0.45675(52)  & 0.45697(70)  & 0.45584(90)  & 0.4570(12)  & 0.4543(19) \\
$2\times2$ & 11 & 1 & 0.45663(51)  & 0.45697(68)  & 0.45585(85) & ---& ---\\
$2\times2$ & 2 & 3 & 0.45634(84)  & 0.4580(12)  & 0.4550(15)  & 0.4589(21)  & 0.4498(33) \\
$2\times2$ & 5 & 3 & 0.45660(55)  & 0.45715(76)  & 0.45535(100)  & 0.4559(14)  & 0.4524(21) \\
$2\times2$ & 8 & 3 & 0.45675(52)  & 0.45697(70)  & 0.45584(90)  & 0.4570(12)  & 0.4543(19) \\
$2\times2$ & 11 & 3 & 0.45662(51)  & 0.45697(68)  & 0.45584(85) & ---& ---\\
$3\times3$ & 2 & 1 & 0.45589(82)  & 0.4576(11)  & 0.4549(15)  & 0.4588(21)  & 0.4499(33) \\
$3\times3$ & 5 & 1 & 0.45624(54)  & 0.45690(75)  & 0.45523(97)  & 0.4558(14)  & 0.4522(21) \\
$3\times3$ & 8 & 1 & 0.45638(51)  & 0.45672(69)  & 0.45572(87)  & 0.4569(12)  & 0.4543(19) \\
$3\times3$ & 11 & 1 & 0.45626(50)  & 0.45671(67)  & 0.45572(82) & ---& ---\\
$3\times3$ & 2 & 3 & 0.45590(82)  & 0.4577(11)  & 0.4548(15)  & 0.4587(21)  & 0.4499(33) \\
$3\times3$ & 5 & 3 & 0.45625(54)  & 0.45691(75)  & 0.45522(96)  & 0.4558(14)  & 0.4522(21) \\
$3\times3$ & 8 & 3 & 0.45638(51)  & 0.45672(69)  & 0.45572(86)  & 0.4569(12)  & 0.4543(19) \\
$3\times3$ & 11 & 3 & 0.45626(50)  & 0.45672(67)  & 0.45572(82) & ---& ---\\
$4\times4$ & 2 & 1 & 0.45582(80)  & 0.4576(11)  & 0.4550(15)  & 0.4587(21)  & 0.4507(48) \\
$4\times4$ & 5 & 1 & 0.45619(54)  & 0.45689(74)  & 0.45529(95)  & 0.4555(14)  & 0.4501(29) \\
$4\times4$ & 8 & 1 & 0.45633(50)  & 0.45670(69)  & 0.45576(86)  & 0.4565(12)  & 0.4529(23) \\
$4\times4$ & 11 & 1 & 0.45621(49)  & 0.45669(67)  & 0.45574(81) & ---& ---\\
$4\times4$ & 2 & 3 & 0.45583(80)  & 0.4577(11)  & 0.4548(15)  & 0.4588(21)  & 0.4497(34) \\
$4\times4$ & 5 & 3 & 0.45620(53)  & 0.45690(74)  & 0.45521(96)  & 0.4557(14)  & 0.4509(23) \\
$4\times4$ & 8 & 3 & 0.45633(50)  & 0.45670(68)  & 0.45569(86)  & 0.4567(12)  & 0.4533(21) \\
$4\times4$ & 11 & 3 & 0.45621(49)  & 0.45669(66)  & 0.45568(81) & ---& ---\\
RGEVP & 2 & 1 & 0.45582(81)  & 0.4577(11)  & 0.4548(15)  & 0.4587(21)  & 0.4498(33) \\
RGEVP & 5 & 1 & 0.45619(54)  & 0.45689(74)  & 0.45516(96)  & 0.4557(14)  & 0.4520(21) \\
RGEVP & 8 & 1 & 0.45633(51)  & 0.45670(68)  & 0.45566(86)  & 0.4568(12)  & 0.4541(19) \\
RGEVP & 11 & 1 & 0.45620(50)  & 0.45669(66)  & 0.45564(82) & ---& ---\\
RGEVP & 2 & 3 & 0.45583(81)  & 0.4577(11)  & 0.4548(15)  & 0.4586(21)  & 0.4498(33) \\
RGEVP & 5 & 3 & 0.45620(54)  & 0.45689(74)  & 0.45517(96)  & 0.4557(14)  & 0.4521(21) \\
RGEVP & 8 & 3 & 0.45633(51)  & 0.45670(69)  & 0.45567(86)  & 0.4568(12)  & 0.4542(19) \\
RGEVP & 11 & 3 & 0.45621(50)  & 0.45669(66)  & 0.45565(82) & ---& ---\\
\hline
\end{tabular}
\caption{Same as Table~\ref{tab:efm_I2_n0_L24} but for the $I=2$ two-pion first excited state on the $32^3$ lattice.  Only results with DR are shown.  The re-basing matrix is calculated as: $4\times4\to 3\times3$ at $t_0 = 5$.}
\label{tab:efm_I2_n1_L32}
\end{table*}

\begin{table*}[tp]
\centering
\begin{tabular}{|ccc|ccccc|}
\hline
 GEVP type & $\delta_t$ & $t-t_0$ & $t = 5$ & $t = 7$ & $t = 9$ & $t = 11$ & $t = 13$\\
\hline
$3\times3$ & 2 & 1 & 0.6146(19)  & 0.6142(36)  & 0.6294(72)  & 0.601(18)  & 0.614(37) \\
$3\times3$ & 5 & 1 & 0.6154(14)  & 0.6185(27)  & 0.6249(52)  & 0.611(12)  & 0.635(28) \\
$3\times3$ & 8 & 1 & 0.6152(14)  & 0.6166(25)  & 0.6240(47)  & 0.608(11)  & 0.645(27) \\
$3\times3$ & 11 & 1 & 0.6154(13)  & 0.6171(25)  & 0.6253(47) & ---& ---\\
$3\times3$ & 2 & 3 & 0.6145(19)  & 0.6142(36)  & 0.6294(72)  & 0.601(18)  & 0.615(37) \\
$3\times3$ & 5 & 3 & 0.6154(14)  & 0.6185(27)  & 0.6249(52)  & 0.611(12)  & 0.635(28) \\
$3\times3$ & 8 & 3 & 0.6152(14)  & 0.6166(25)  & 0.6240(47)  & 0.608(11)  & 0.645(27) \\
$3\times3$ & 11 & 3 & 0.6154(13)  & 0.6171(25)  & 0.6253(47) & ---& ---\\
$4\times4$ & 2 & 1 & 0.6141(19)  & 0.6125(35)  & 0.6290(71)  & 0.601(18)  & 0.60(37) \\
$4\times4$ & 5 & 1 & 0.6147(14)  & 0.6173(26)  & 0.6248(53)  & 0.611(12)  & 0.638(28) \\
$4\times4$ & 8 & 1 & 0.6146(13)  & 0.6156(25)  & 0.6238(48)  & 0.608(11)  & 0.645(27) \\
$4\times4$ & 11 & 1 & 0.6147(13)  & 0.6159(24)  & 0.6250(48) & ---& ---\\
$4\times4$ & 2 & 3 & 0.6142(19)  & 0.6126(35)  & 0.6290(72)  & 0.601(18)  & 0.602(50) \\
$4\times4$ & 5 & 3 & 0.6147(14)  & 0.6173(26)  & 0.6249(53)  & 0.611(12)  & 0.635(28) \\
$4\times4$ & 8 & 3 & 0.6146(13)  & 0.6155(25)  & 0.6238(48)  & 0.608(11)  & 0.644(26) \\
$4\times4$ & 11 & 3 & 0.6148(13)  & 0.6159(24)  & 0.6250(48) & ---& ---\\
RGEVP & 2 & 1 & 0.6141(19)  & 0.6129(35)  & 0.6290(71)  & 0.601(18)  & 0.620(37) \\
RGEVP & 5 & 1 & 0.6147(14)  & 0.6174(26)  & 0.6249(53)  & 0.612(12)  & 0.636(27) \\
RGEVP & 8 & 1 & 0.6145(13)  & 0.6156(25)  & 0.6238(48)  & 0.608(11)  & 0.644(26) \\
RGEVP & 11 & 1 & 0.6147(13)  & 0.6160(24)  & 0.6250(48) & ---& ---\\
RGEVP & 2 & 3 & 0.6141(19)  & 0.6125(35)  & 0.6290(72)  & 0.602(18)  & 0.622(37) \\
RGEVP & 5 & 3 & 0.6147(14)  & 0.6173(26)  & 0.6250(54)  & 0.612(12)  & 0.636(27) \\
RGEVP & 8 & 3 & 0.6145(13)  & 0.6155(25)  & 0.6239(48)  & 0.608(11)  & 0.644(26) \\
RGEVP & 11 & 3 & 0.6147(13)  & 0.6159(24)  & 0.6251(48) & ---& ---\\
\hline
\end{tabular}
\caption{Same as Table~\ref{tab:efm_I2_n0_L24} but for the $I=2$ two-pion second excited state on the $32^3$ lattice.  Only results with DR are shown.  The re-basing matrix is calculated as: $4\times4\to 3\times3$ at $t_0 = 5$.}
\label{tab:efm_I2_n2_L32}
\end{table*}

\begin{table*}[tp]
\centering
\begin{tabular}{|ccc|ccccc|}
\hline
 GEVP type & $\delta_t$ & $t-t_0$ & $t = 5$ & $t = 7$ & $t = 9$ & $t = 11$ & $t = 13$\\
\hline
$4\times4$ & 2 & 1 & 0.7231(40)  & 0.7203(94)  & 0.759(23)  & 0.697(74)  & 0.64(45) \\
$4\times4$ & 5 & 1 & 0.7247(29)  & 0.7267(72)  & 0.736(17)  & 0.667(52)  & 0.635(96) \\
$4\times4$ & 8 & 1 & 0.7240(28)  & 0.7248(70)  & 0.738(17)  & 0.694(51)  & 0.70(11) \\
$4\times4$ & 11 & 1 & 0.7235(28)  & 0.7248(67)  & 0.745(17) & ---& ---\\
$4\times4$ & 2 & 3 & 0.7231(40)  & 0.7202(95)  & 0.759(23)  & 0.697(74)  & 0.64(13) \\
$4\times4$ & 5 & 3 & 0.7246(29)  & 0.7267(72)  & 0.736(17)  & 0.666(52)  & 0.636(98) \\
$4\times4$ & 8 & 3 & 0.7239(28)  & 0.7248(70)  & 0.738(17)  & 0.694(51)  & 0.70(11) \\
$4\times4$ & 11 & 3 & 0.7234(28)  & 0.7248(67)  & 0.745(17) & ---& ---\\
\hline
\end{tabular}
\caption{Same as Table~\ref{tab:efm_I2_n0_L24} but for the $I=2$ two-pion third excited state on the $32^3$ lattice.  Only results with DR are shown.  }
\label{tab:efm_I2_n3_L32}
\end{table*}

\begin{table*}[tp]
\centering
\begin{tabular}{|ccc|cccc|}
\hline
 GEVP type & $\delta_t$ & $t-t_0$ & $t = 4$ & $t = 7$ & $t = 9$ & $t = 11$\\
\hline
$3\times3$ & 2 & 1 & 0.20359(73)  & 0.20313(96)  & 0.2020(14)  & 0.20(11) \\
$3\times3$ & 5 & 1 & 0.20304(55)  & 0.20252(93)  & 0.2017(15)  & 0.2022(35) \\
$3\times3$ & 8 & 1 & 0.20287(49)  & 0.20269(60)  & 0.2020(98)  & 0.2013(16) \\
$3\times3$ & 11 & 1 & 0.20280(47)  & 0.20226(64)  & 0.2011(68) & ---\\
$3\times3$ & 2 & 3 & 0.20397(75)  & 0.20319(94)  & 0.2023(10)  & 0.201(50) \\
$3\times3$ & 5 & 3 & 0.20325(56)  & 0.20261(75)  & 0.2017(14)  & 0.202(18) \\
$3\times3$ & 8 & 3 & 0.20308(51)  & 0.20265(60)  & 0.2018(12)  & 0.2013(31) \\
$3\times3$ & 11 & 3 & 0.20291(48)  & 0.20228(61)  & 0.202(33) & ---\\
$4\times4$ & 2 & 1 & 0.20356(74)  & 0.20314(97)  & 0.201(19)  & 0.207(27) \\
$4\times4$ & 5 & 1 & 0.20302(56)  & 0.20255(100)  & 0.2015(24)  & 0.2025(16) \\
$4\times4$ & 8 & 1 & 0.20285(51)  & 0.20262(77)  & 0.203(12)  & 0.20(37) \\
$4\times4$ & 11 & 1 & 0.20274(49)  & 0.2022(27)  & 0.2012(19) & ---\\
$4\times4$ & 2 & 3 & 0.20395(75)  & 0.2033(10)  & 0.2022(12)  & 0.201(42) \\
$4\times4$ & 5 & 3 & 0.20322(57)  & 0.20264(78)  & 0.2016(12)  & 0.202(42) \\
$4\times4$ & 8 & 3 & 0.20305(52)  & 0.20260(63)  & 0.202(27)  & 0.2012(16) \\
$4\times4$ & 11 & 3 & 0.20287(49)  & 0.20234(67)  & 0.201(19) & ---\\
$5\times5$ & 2 & 1 & 0.20356(74)  & 0.20318(97)  & 0.205(77)  & --- \\
$5\times5$ & 5 & 1 & 0.20303(57)  & 0.2025(10)  & 0.20(16)  & --- \\
$5\times5$ & 8 & 1 & 0.20285(51)  & 0.2025(29)  & 0.2017(48)  & 0.2014(15) \\
$5\times5$ & 11 & 1 & 0.20273(50)  & 0.2022(58)  & 0.2011(12) & ---\\
$5\times5$ & 2 & 3 & 0.20395(75)  & 0.2033(10)  & 0.2022(13)  & 0.194(57) \\
$5\times5$ & 5 & 3 & 0.20322(57)  & 0.20263(79)  & 0.2013(17)  & 0.202(40) \\
$5\times5$ & 8 & 3 & 0.20305(52)  & 0.20253(71)  & 0.203(20)  & 0.2013(15) \\
$5\times5$ & 11 & 3 & 0.20286(49)  & 0.20236(72)  & 0.201(11) & ---\\
RGEVP & 2 & 1 & 0.20361(78)  & 0.20355(99)  & 0.2017(16)  & 0.2022(36) \\
RGEVP & 5 & 1 & 0.20307(59)  & 0.20277(81)  & 0.2015(14)  & 0.2021(18) \\
RGEVP & 8 & 1 & 0.20286(53)  & 0.20253(67)  & 0.2018(11)  & 0.2013(15) \\
RGEVP & 11 & 1 & 0.20273(49)  & 0.20235(66)  & 0.20118(100) & ---\\
RGEVP & 2 & 3 & 0.20374(78)  & 0.20328(97)  & 0.2021(11)  & 0.2015(60) \\
RGEVP & 5 & 3 & 0.20317(59)  & 0.20263(75)  & 0.2017(13)  & 0.2020(27) \\
RGEVP & 8 & 3 & 0.20297(55)  & 0.20261(60)  & 0.20184(99)  & 0.201(13) \\
RGEVP & 11 & 3 & 0.20273(51)  & 0.20232(66)  & 0.202(12) & ---\\
\hline
\end{tabular}
\caption{Same as Table~\ref{tab:efm_I2_n0_L24} but for the $I=0$ two-pion ground state on the $32^3$ lattice.  Only results with DR are shown.  The re-basing matrix is calculated as: $5\times5\to 4\times4$ at $t_0 = 1$, $4\times4\to 3\times3$ at $t_0 = 2$ and $3\times3\to 2\times2$ at $t_0 = 4$.}
\label{tab:efm_I0_n0_L32}
\end{table*}

\begin{table*}[tp]
\centering
\begin{tabular}{|ccc|cccc|}
\hline
 GEVP type & $\delta_t$ & $t-t_0$ & $t = 4$ & $t = 7$ & $t = 9$ & $t = 11$\\
\hline
$3\times3$ & 2 & 1 & 0.4057(70)  & 0.43(11)  & 0.27(77)  & 0.38(60) \\
$3\times3$ & 5 & 1 & 0.397(17)  & 0.31(12)  & 0.25(30)  & 0.15(56) \\
$3\times3$ & 8 & 1 & 0.379(29)  & 0.31(16)  & 0.09(21)  & 0.01(22) \\
$3\times3$ & 11 & 1 & 0.382(36)  & 0.28(14)  & 0.08(16) & ---\\
$3\times3$ & 2 & 3 & 0.4083(38)  & 0.413(71)  & 0.26(36)  & 0.31(84) \\
$3\times3$ & 5 & 3 & 0.4065(53)  & 0.32(12)  & 0.24(26)  & 0.15(36) \\
$3\times3$ & 8 & 3 & 0.3994(66)  & 0.32(16)  & 0.10(23)  & 0.01(22) \\
$3\times3$ & 11 & 3 & 0.4006(67)  & 0.28(15)  & 0.08(17) & ---\\
$4\times4$ & 2 & 1 & 0.4049(64)  & 0.46(11)  & 0.38(34)  & 0.0(2.1) \\
$4\times4$ & 5 & 1 & 0.396(19)  & 0.31(15)  & 0.24(40)  & 0.3(2.2) \\
$4\times4$ & 8 & 1 & 0.376(40)  & 0.27(25)  & -0.08(38)  & -0.04(46) \\
$4\times4$ & 11 & 1 & 0.38(18)  & 0.25(19)  & -0.02(26) & ---\\
$4\times4$ & 2 & 3 & 0.4060(43)  & 0.43(19)  & 0.22(56)  & 0.4(1.1) \\
$4\times4$ & 5 & 3 & 0.4062(53)  & 0.31(15)  & 0.22(35)  & 0.2(2.8) \\
$4\times4$ & 8 & 3 & 0.3993(63)  & 0.29(24)  & -0.03(36)  & -0.05(30) \\
$4\times4$ & 11 & 3 & 0.4005(61)  & 0.25(19)  & 0.00(24) & ---\\
$5\times5$ & 2 & 1 & 0.4054(70)  & 0.457(81)  & 0.39(52)  & --- \\
$5\times5$ & 5 & 1 & 0.396(22)  & 0.31(14)  & 0.29(47)  & --- \\
$5\times5$ & 8 & 1 & 0.38(12)  & 0.27(27)  & -0.2(3.9)  & -0.1(1.7) \\
$5\times5$ & 11 & 1 & 0.40(36)  & 0.25(44)  & -0.0(1.1) & ---\\
$5\times5$ & 2 & 3 & 0.4056(45)  & 0.43(19)  & 0.22(59)  & 0.4(1.6) \\
$5\times5$ & 5 & 3 & 0.4062(53)  & 0.31(14)  & 0.22(43)  & 0.3(2.6) \\
$5\times5$ & 8 & 3 & 0.3994(63)  & 0.29(26)  & -0.10(52)  & -0.08(55) \\
$5\times5$ & 11 & 3 & 0.4004(60)  & 0.24(21)  & -0.01(29) & ---\\
RGEVP & 2 & 1 & 0.4036(60)  & 0.436(19)  & 0.361(44)  & 0.357(87) \\
RGEVP & 5 & 1 & 0.4071(51)  & 0.401(17)  & 0.351(36)  & 0.397(92) \\
RGEVP & 8 & 1 & 0.3970(65)  & 0.397(21)  & 0.391(42)  & 0.418(88) \\
RGEVP & 11 & 1 & 0.3976(79)  & 0.404(22)  & 0.346(44) & ---\\
RGEVP & 2 & 3 & 0.441(35)  & 0.417(95)  & 0.26(17)  & 0.29(28) \\
RGEVP & 5 & 3 & 0.404(46)  & 0.32(11)  & 0.25(20)  & 0.22(27) \\
RGEVP & 8 & 3 & 0.374(57)  & 0.31(13)  & 0.13(18)  & 0.03(20) \\
RGEVP & 11 & 3 & 0.367(89)  & 0.25(17)  & 0.05(16) & ---\\
\hline
\end{tabular}
\caption{Same as Table~\ref{tab:efm_I2_n0_L24} but for the $I=0$ two-pion first excited state on the $32^3$ lattice.  Only results with DR are shown.  The re-basing matrix is calculated as: $5\times5\to 4\times4$ at $t_0 = 1$, $4\times4\to 3\times3$ at $t_0 = 2$ and $3\times3\to 2\times2$ at $t_0 = 4$.}
\label{tab:efm_I0_n1_L32}
\end{table*}

\begin{table*}[tp]
\centering
\begin{tabular}{|ccc|cccc|}
\hline
 GEVP type & $\delta_t$ & $t-t_0$ & $t = 4$ & $t = 7$ & $t = 9$ & $t = 11$\\
\hline
$3\times3$ & 2 & 1 & 0.489(54)  & 0.458(74)  & 0.29(75)  & 0.35(60) \\
$3\times3$ & 5 & 1 & 0.457(52)  & 0.432(49)  & 0.38(11)  & 0.39(49) \\
$3\times3$ & 8 & 1 & 0.436(44)  & 0.402(25)  & 0.400(31)  & 0.432(54) \\
$3\times3$ & 11 & 1 & 0.420(41)  & 0.419(18)  & 0.399(23) & ---\\
$3\times3$ & 2 & 3 & 0.486(57)  & 0.48(11)  & 0.30(25)  & 0.43(72) \\
$3\times3$ & 5 & 3 & 0.447(64)  & 0.426(64)  & 0.384(98)  & 0.389(88) \\
$3\times3$ & 8 & 3 & 0.415(68)  & 0.389(48)  & 0.391(55)  & 0.433(53) \\
$3\times3$ & 11 & 3 & 0.401(72)  & 0.416(24)  & 0.397(27) & ---\\
$4\times4$ & 2 & 1 & 0.50(48)  & 0.459(86)  & 0.47(77)  & 0.3(1.6) \\
$4\times4$ & 5 & 1 & 0.453(68)  & 0.433(54)  & 0.37(18)  & 0.3(3.1) \\
$4\times4$ & 8 & 1 & 0.428(59)  & 0.396(35)  & 0.398(40)  & 0.425(24) \\
$4\times4$ & 11 & 1 & 0.40(23)  & 0.420(18)  & 0.396(38) & ---\\
$4\times4$ & 2 & 3 & 0.519(22)  & 0.51(19)  & 0.30(26)  & 0.5(1.1) \\
$4\times4$ & 5 & 3 & 0.465(70)  & 0.432(71)  & 0.38(15)  & 0.4(4.4) \\
$4\times4$ & 8 & 3 & 0.431(93)  & 0.380(69)  & 0.389(73)  & 0.435(32) \\
$4\times4$ & 11 & 3 & 0.41(12)  & 0.421(25)  & 0.395(36) & ---\\
$5\times5$ & 2 & 1 & 0.57(58)  & 0.58(100)  & ---  & --- \\
$5\times5$ & 5 & 1 & 0.448(78)  & 0.430(43)  & 0.5(1.1)  & --- \\
$5\times5$ & 8 & 1 & 0.42(16)  & 0.396(38)  & 0.40(30)  & 0.4(1.1) \\
$5\times5$ & 11 & 1 & 0.37(44)  & 0.41(45)  & 0.4(1.0) & ---\\
$5\times5$ & 2 & 3 & 0.519(21)  & 0.51(19)  & 0.30(26)  & --- \\
$5\times5$ & 5 & 3 & 0.463(77)  & 0.430(62)  & 0.37(21)  & 0.4(6.4) \\
$5\times5$ & 8 & 3 & 0.43(10)  & 0.379(73)  & 0.391(80)  & 0.436(29) \\
$5\times5$ & 11 & 3 & 0.41(13)  & 0.422(23)  & 0.396(37) & ---\\
RGEVP & 2 & 1 & 0.533(10)  & 0.576(52)  & 0.41(12)  & 0.48(61) \\
RGEVP & 5 & 1 & 0.523(11)  & 0.475(36)  & 0.454(54)  & 0.42(17) \\
RGEVP & 8 & 1 & 0.487(19)  & 0.437(50)  & 0.35(36)  & 0.29(85) \\
RGEVP & 11 & 1 & 0.467(32)  & 0.418(19)  & 0.400(25) & ---\\
RGEVP & 2 & 3 & 0.485(61)  & 0.49(13)  & 0.27(29)  & 0.45(86) \\
RGEVP & 5 & 3 & 0.440(83)  & 0.436(93)  & 0.36(15)  & 0.4(1.1) \\
RGEVP & 8 & 3 & 0.403(89)  & 0.381(66)  & 0.391(65)  & 0.444(70) \\
RGEVP & 11 & 3 & 0.375(96)  & 0.421(30)  & 0.391(36) & ---\\
\hline
\end{tabular}
\caption{Same as Table~\ref{tab:efm_I2_n0_L24} but for the $I=0$ two-pion second excited state on the $32^3$ lattice.  Only results with DR are shown.  The re-basing matrix is calculated as: $5\times5\to 4\times4$ at $t_0 = 1$ and $4\times4\to 3\times3$ at $t_0 = 2$.}
\label{tab:efm_I0_n2_L32}
\end{table*}

\begin{table*}[tp]
\centering
\begin{tabular}{|ccc|cccc|}
\hline
 GEVP type & $\delta_t$ & $t-t_0$ & $t = 4$ & $t = 7$ & $t = 9$ & $t = 11$\\
\hline
$4\times4$ & 2 & 1 & 0.53(51)  & 0.78(35)  & -0.4(1.5)  & --- \\
$4\times4$ & 5 & 1 & 0.574(38)  & 0.68(29)  & 0.54(87)  & 0.6(3.7) \\
$4\times4$ & 8 & 1 & 0.570(36)  & 0.58(13)  & 0.55(17)  & --- \\
$4\times4$ & 11 & 1 & 0.554(22)  & 0.67(16)  & 0.53(18) & ---\\
$4\times4$ & 2 & 3 & 0.505(80)  & 0.76(33)  & -0.0(1.2)  & --- \\
$4\times4$ & 5 & 3 & 0.552(63)  & 0.68(28)  & 0.55(89)  & 0.6(5.5) \\
$4\times4$ & 8 & 3 & 0.544(52)  & 0.57(15)  & 0.51(27)  & --- \\
$4\times4$ & 11 & 3 & 0.519(41)  & 0.66(16)  & 0.50(23) & ---\\
$5\times5$ & 2 & 1 & 0.41(55)  & 0.5(1.0)  & 0.1(5.5)  & --- \\
$5\times5$ & 5 & 1 & 0.571(44)  & 0.630(94)  & ---  & --- \\
$5\times5$ & 8 & 1 & 0.567(42)  & 0.58(10)  & 0.5(1.4)  & --- \\
$5\times5$ & 11 & 1 & 0.549(27)  & 0.641(70)  & 0.56(16) & ---\\
$5\times5$ & 2 & 3 & 0.48(10)  & 0.71(14)  & 0.43(31)  & --- \\
$5\times5$ & 5 & 3 & 0.547(73)  & 0.64(11)  & 0.73(33)  & --- \\
$5\times5$ & 8 & 3 & 0.540(60)  & 0.57(10)  & 0.52(22)  & 1.16(84) \\
$5\times5$ & 11 & 3 & 0.514(47)  & 0.648(86)  & 0.54(17) & ---\\
RGEVP & 2 & 1 & 0.666(16)  & 0.686(88)  & 0.66(60)  & 0.3(8.2) \\
RGEVP & 5 & 1 & 0.605(29)  & 0.630(99)  & 0.67(35)  & 1.0(3.3) \\
RGEVP & 8 & 1 & 0.584(33)  & 0.577(80)  & 0.54(15)  & 1.4(2.4) \\
RGEVP & 11 & 1 & 0.563(25)  & 0.65(11)  & 0.53(15) & ---\\
RGEVP & 2 & 3 & 0.500(83)  & 0.78(36)  & -0.1(1.4)  & --- \\
RGEVP & 5 & 3 & 0.548(68)  & 0.71(36)  & 0.5(1.1)  & 0.5(5.1) \\
RGEVP & 8 & 3 & 0.541(54)  & 0.58(17)  & 0.51(30)  & --- \\
RGEVP & 11 & 3 & 0.516(44)  & 0.67(18)  & 0.50(24) & ---\\
\hline
\end{tabular}
\caption{Same as Table~\ref{tab:efm_I2_n0_L24} but for the $I=0$ two-pion third excited state on the $32^3$ lattice.  Only results with DR are shown.  The re-basing matrix is calculated as: $5\times5\to 4\times4$ at $t_0 = 1$.}
\label{tab:efm_I0_n3_L32}
\end{table*}

\begin{table*}[tp]
\centering
\begin{tabular}{|ccc|cccc|}
\hline
 GEVP type & $\delta_t$ & $t-t_0$ & $t = 4$ & $t = 7$ & $t = 9$ & $t = 11$\\
\hline
$5\times5$ & 2 & 1 & 0.670(16)  & 1.8(2.5)  & ---  & --- \\
$5\times5$ & 5 & 1 & 0.668(26)  & 2.1(8.2)  & ---  & --- \\
$5\times5$ & 8 & 1 & 0.685(30)  & 1.3(2.0)  & ---  & --- \\
$5\times5$ & 11 & 1 & 0.681(24)  & 1.4(1.7)  & --- & ---\\
$5\times5$ & 2 & 3 & 0.648(27)  & 1.7(2.4)  & ---  & --- \\
$5\times5$ & 5 & 3 & 0.666(31)  & 2.1(8.2)  & ---  & --- \\
$5\times5$ & 8 & 3 & 0.683(34)  & 1.3(2.1)  & ---  & --- \\
$5\times5$ & 11 & 3 & 0.675(29)  & 1.4(1.7)  & --- & ---\\
RGEVP & 2 & 1 & 0.670(16)  & 1.8(2.5)  & ---  & --- \\
RGEVP & 5 & 1 & 0.668(26)  & 2.1(8.2)  & ---  & --- \\
RGEVP & 8 & 1 & 0.685(30)  & 1.3(2.0)  & ---  & --- \\
RGEVP & 11 & 1 & 0.681(24)  & 1.4(1.7)  & --- & ---\\
RGEVP & 2 & 3 & 0.648(27)  & 1.7(2.4)  & ---  & --- \\
RGEVP & 5 & 3 & 0.666(31)  & 2.1(8.2)  & ---  & --- \\
RGEVP & 8 & 3 & 0.683(34)  & 1.3(2.1)  & ---  & --- \\
RGEVP & 11 & 3 & 0.675(29)  & 1.4(1.7)  & --- & ---\\
\hline
\end{tabular}
\caption{Same as Table~\ref{tab:efm_I2_n0_L24} but for the $I=0$ two-pion fourth excited state on the $32^3$ lattice.  Only results with DR are shown.  }
\label{tab:efm_I0_n4_L32}
\end{table*}

\end{document}